\documentclass[preprint,prd,tightenlines,superscriptaddress]{revtex4-1}
\RequirePackage{fix-cm}
\RequirePackage{graphicx}

\RequirePackage{lineno}
\RequirePackage[colorlinks,citecolor=blue,urlcolor=blue,linkcolor=blue]{hyperref}
\RequirePackage{tabularx}
\RequirePackage{multirow}
\RequirePackage{units}
\RequirePackage{siunitx}
\RequirePackage{hyphenat}
\RequirePackage{subfigure}
\RequirePackage[italic]{hepnames}
\RequirePackage{amssymb}
\RequirePackage{upgreek}
\RequirePackage{xspace}
\RequirePackage{color, colortbl}
\RequirePackage{bm}
\RequirePackage{mathtools}

\newcommand{\ee}{\text{e}}
\newcommand{\dd}{\text{d}}

\newcommand{\CP}{\ensuremath{C\hspace{-0.13em}P}\xspace}

\renewcommand{\Ph}{\HepParticle{h}{}{}\xspace}

\newcommand{\PX}{\HepParticle{X}{}{}\xspace}

\newcommand{\PKstarminus}{\HepParticle{K}{}{\ast-}\xspace}
\newcommand{\PDstarplus}{\HepParticle{D}{}{\ast+}\xspace}

\newcommand{\phione}{\ensuremath{\upphi_1}\xspace}
\newcommand{\phitwo}{\ensuremath{\upphi_2}\xspace}

\newcommand{\PBtag}{\HepParticle{\PB}{}{}^{0}_{\rm tag}\xspace}

\renewcommand{\PUpsilonFourS}{\HepParticleResonance{\PgU}{\mathrm{4}S}{}{}\xspace}
\newcommand{\Pa}{\HepParticle{a}{}{}\xspace}
\newcommand{\Pb}{\HepParticle{b}{}{}\xspace}

\renewcommand{\PBzero}{\ensuremath{\HepParticle{\PB}{}{}^0}\xspace}
\renewcommand{\APBzero}{\ensuremath{\HepParticle{\APB}{}{}^0}\xspace}
\renewcommand{\APDzero}{\ensuremath{\HepParticle{\APD}{}{}^0}\xspace}
\renewcommand{\Pgpz}{\ensuremath{\HepParticle{\Pgp}{}{}^0}\xspace}
\renewcommand{\PDzero}{\ensuremath{\HepParticle{\PD}{}{}^0}\xspace}
\renewcommand{\PKzS}{\ensuremath{\HepParticle{\PK}{}{}^0_{\rm S}}\xspace}
\renewcommand{\PKzL}{\ensuremath{\HepParticle{\PK}{}{}^0_{\rm L}}\xspace}

\begin{document}

%\linenumbers

%% Avoid "orphans" and "widows"
\clubpenalty = 10000  % no orphans
\widowpenalty = 10000 % no widows

\vspace*{-3\baselineskip}
\resizebox{!}{3cm}{\includegraphics{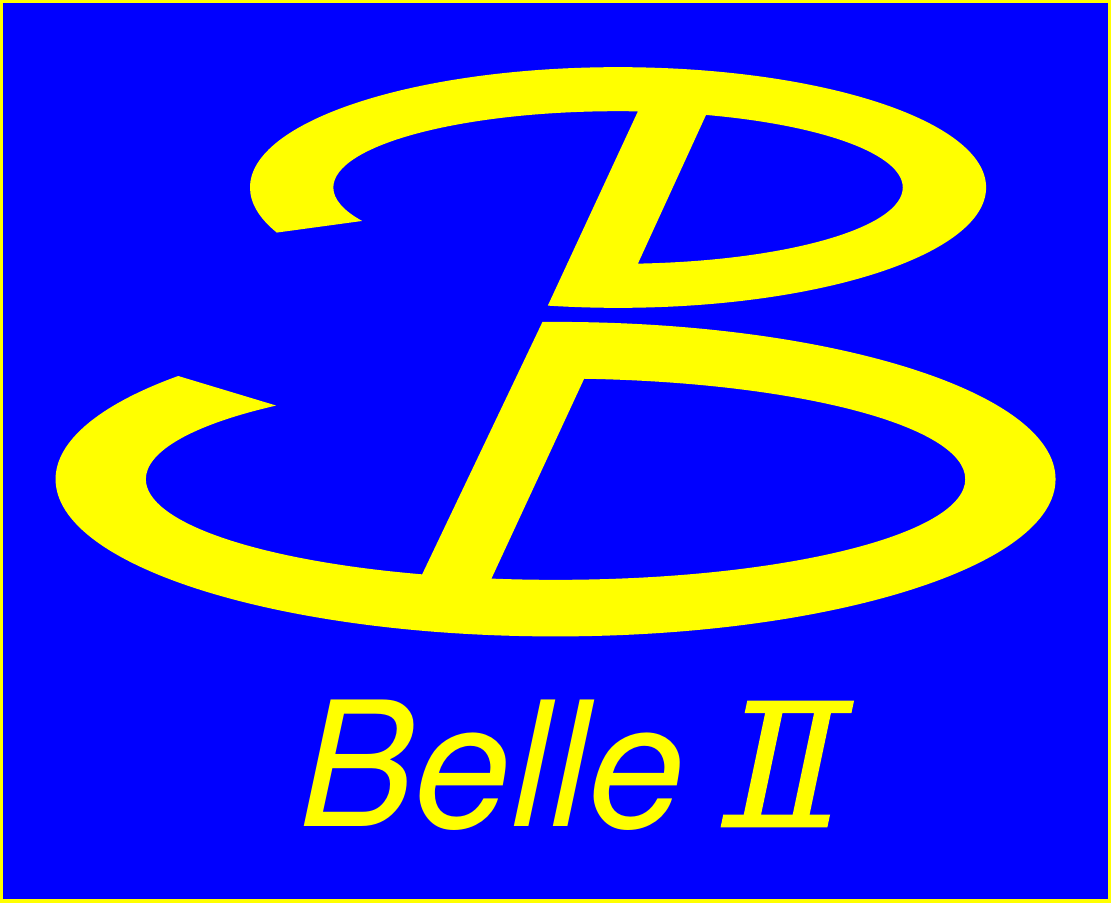}}

\vspace*{-5\baselineskip}
\begin{flushright}
BELLE2-PUB-TE-2021-002\\
EPJC-21-09-075\\
%Version 3.0 \\
\today
\end{flushright}

\quad\\[0.5cm]

\title{\PB-flavor tagging at \mbox{Belle II}%\thanksref{t1}
}

%%% Paper:    (2020  conference papers)
%%% Journal:  (2020 conferences)
%%% ====================================================================
%%% Use \input{authors-conf2020} to insert this material into your latex file.
\newcommand{\instSinica}{Academia Sinica, Taipei 11529, Taiwan}
\newcommand{\instCPPM}{Aix Marseille Universit\'{e}, CNRS/IN2P3, CPPM, 13288 Marseille, France}
\newcommand{\instBeihang}{Beihang University, Beijing 100191, China}
\newcommand{\instBUAP}{Benemerita Universidad Autonoma de Puebla, Puebla 72570, Mexico}
\newcommand{\instBNL}{Brookhaven National Laboratory, Upton, New York 11973, U.S.A.}
\newcommand{\instBINP}{Budker Institute of Nuclear Physics SB RAS, Novosibirsk 630090, Russian Federation}
\newcommand{\instCMU}{Carnegie Mellon University, Pittsburgh, Pennsylvania 15213, U.S.A.}
\newcommand{\instCinvestavIPN}{Centro de Investigacion y de Estudios Avanzados del Instituto Politecnico Nacional, Mexico City 07360, Mexico}
\newcommand{\instPrague}{Faculty of Mathematics and Physics, Charles University, 121 16 Prague, Czech Republic}
\newcommand{\instChiangMai}{Chiang Mai University, Chiang Mai 50202, Thailand}
\newcommand{\instChiba}{Chiba University, Chiba 263-8522, Japan}
\newcommand{\instChonnam}{Chonnam National University, Gwangju 61186, South Korea}
\newcommand{\instConacyt}{Consejo Nacional de Ciencia y Tecnolog\'{\i}a, Mexico City 03940, Mexico}
\newcommand{\instDESY}{Deutsches Elektronen--Synchrotron, 22607 Hamburg, Germany}
\newcommand{\instDuke}{Duke University, Durham, North Carolina 27708, U.S.A.}
\newcommand{\instITAR}{Institute of Theoretical and Applied Research (ITAR), Duy Tan University, Hanoi 100000, Vietnam}
\newcommand{\instENEA}{ENEA Casaccia, I-00123 Roma, Italy}
\newcommand{\instEri}{Earthquake Research Institute, University of Tokyo, Tokyo 113-0032, Japan}
\newcommand{\instJuelich}{Forschungszentrum J\"{u}lich, 52425 J\"{u}lich, Germany}
\newcommand{\instFuJen}{Department of Physics, Fu Jen Catholic University, Taipei 24205, Taiwan}
\newcommand{\instFudan}{Key Laboratory of Nuclear Physics and Ion-beam Application (MOE) and Institute of Modern Physics, Fudan University, Shanghai 200443, China}
\newcommand{\instGoettingen}{II. Physikalisches Institut, Georg-August-Universit\"{a}t G\"{o}ttingen, 37073 G\"{o}ttingen, Germany}
\newcommand{\instGifu}{Gifu University, Gifu 501-1193, Japan}
\newcommand{\instSOKENDAI}{The Graduate University for Advanced Studies (SOKENDAI), Hayama 240-0193, Japan}
\newcommand{\instGyeongsang}{Gyeongsang National University, Jinju 52828, South Korea}
\newcommand{\instHanyang}{Department of Physics and Institute of Natural Sciences, Hanyang University, Seoul 04763, South Korea}
\newcommand{\instKEK}{High Energy Accelerator Research Organization (KEK), Tsukuba 305-0801, Japan}
\newcommand{\instJPARC}{J-PARC Branch, KEK Theory Center, High Energy Accelerator Research Organization (KEK), Tsukuba 305-0801, Japan}
\newcommand{\instHSE}{Higher School of Economics (HSE), Moscow 101000, Russian Federation}
\newcommand{\instIISER}{Indian Institute of Science Education and Research Mohali, SAS Nagar, 140306, India}
\newcommand{\instIITBhubaneswar}{Indian Institute of Technology Bhubaneswar, Satya Nagar 751007, India}
\newcommand{\instIITGuwahati}{Indian Institute of Technology Guwahati, Assam 781039, India}
\newcommand{\instIITHyderabad}{Indian Institute of Technology Hyderabad, Telangana 502285, India}
\newcommand{\instIITMadras}{Indian Institute of Technology Madras, Chennai 600036, India}
\newcommand{\instIndiana}{Indiana University, Bloomington, Indiana 47408, U.S.A.}
\newcommand{\instIHEPRussia}{Institute for High Energy Physics, Protvino 142281, Russian Federation}
\newcommand{\instHEPHYVienna}{Institute of High Energy Physics, Vienna 1050, Austria}
\newcommand{\instIHEPChina}{Institute of High Energy Physics, Chinese Academy of Sciences, Beijing 100049, China}
\newcommand{\instChennai}{Institute of Mathematical Sciences, Chennai 600113, India}
\newcommand{\instIPP}{Institute of Particle Physics (Canada), Victoria, British Columbia V8W 2Y2, Canada}
\newcommand{\instIOP}{Institute of Physics, Vietnam Academy of Science and Technology (VAST), Hanoi, Vietnam}
\newcommand{\instIFIC}{Instituto de Fisica Corpuscular, Paterna 46980, Spain}
\newcommand{\instFrascati}{INFN Laboratori Nazionali di Frascati, I-00044 Frascati, Italy}
\newcommand{\instNapoliINFN}{INFN Sezione di Napoli, I-80126 Napoli, Italy}
\newcommand{\instPadovaINFN}{INFN Sezione di Padova, I-35131 Padova, Italy}
\newcommand{\instPerugiaINFN}{INFN Sezione di Perugia, I-06123 Perugia, Italy}
\newcommand{\instPisaINFN}{INFN Sezione di Pisa, I-56127 Pisa, Italy}
\newcommand{\instRomaINFN}{INFN Sezione di Roma, I-00185 Roma, Italy}
\newcommand{\instRomaTreINFN}{INFN Sezione di Roma Tre, I-00146 Roma, Italy}
\newcommand{\instTorinoINFN}{INFN Sezione di Torino, I-10125 Torino, Italy}
\newcommand{\instTriesteINFN}{INFN Sezione di Trieste, I-34127 Trieste, Italy}
\newcommand{\instJAEA}{Advanced Science Research Center, Japan Atomic Energy Agency, Naka 319-1195, Japan}
\newcommand{\instMainz}{Johannes Gutenberg-Universit\"{a}t Mainz, Institut f\"{u}r Kernphysik, D-55099 Mainz, Germany}
\newcommand{\instGiessen}{Justus-Liebig-Universit\"{a}t Gie\ss{}en, 35392 Gie\ss{}en, Germany}
\newcommand{\instKarlsruhe}{Institut f\"{u}r Experimentelle Teilchenphysik, Karlsruher Institut f\"{u}r Technologie, 76131 Karlsruhe, Germany}
\newcommand{\instKennesaw}{Kennesaw State University, Kennesaw, Georgia 30144, U.S.A.}
\newcommand{\instKitasato}{Kitasato University, Sagamihara 252-0373, Japan}
\newcommand{\instKISTI}{Korea Institute of Science and Technology Information, Daejeon 34141, South Korea}
\newcommand{\instKorea}{Korea University, Seoul 02841, South Korea}
\newcommand{\instKSU}{Kyoto Sangyo University, Kyoto 603-8555, Japan}
\newcommand{\instKyotoU}{Kyoto University, Kyoto 606-8501, Japan}
\newcommand{\instKyungpook}{Kyungpook National University, Daegu 41566, South Korea}
\newcommand{\instLPI}{P.N. Lebedev Physical Institute of the Russian Academy of Sciences, Moscow 119991, Russian Federation}
\newcommand{\instLNNU}{Liaoning Normal University, Dalian 116029, China}
\newcommand{\instLMU}{Ludwig Maximilians University, 80539 Munich, Germany}
\newcommand{\instISUni}{Iowa State University,  Ames, Iowa 50011, U.S.A.}
\newcommand{\instLuther}{Luther College, Decorah, Iowa 52101, U.S.A.}
\newcommand{\instMNITJaipur}{Malaviya National Institute of Technology Jaipur, Jaipur 302017, India}
\newcommand{\instMPP}{Max-Planck-Institut f\"{u}r Physik, 80805 M\"{u}nchen, Germany}
\newcommand{\instMPGHLL}{Semiconductor Laboratory of the Max Planck Society, 81739 M\"{u}nchen, Germany}
\newcommand{\instMcGill}{McGill University, Montr\'{e}al, Qu\'{e}bec, H3A 2T8, Canada}
\newcommand{\instMETU}{Middle East Technical University, 06531 Ankara, Turkey}
\newcommand{\instMEPhI}{Moscow Physical Engineering Institute, Moscow 115409, Russian Federation}
\newcommand{\instNagoya}{Graduate School of Science, Nagoya University, Nagoya 464-8602, Japan}
\newcommand{\instNagoyaKMI}{Kobayashi-Maskawa Institute, Nagoya University, Nagoya 464-8602, Japan}
\newcommand{\instNagoyaIAR}{Institute for Advanced Research, Nagoya University, Nagoya 464-8602, Japan}
\newcommand{\instNaraWu}{Nara Women's University, Nara 630-8506, Japan}
\newcommand{\instUNAM}{National Autonomous University of Mexico, Mexico City, Mexico}
\newcommand{\instNTUTaiwan}{Department of Physics, National Taiwan University, Taipei 10617, Taiwan}
\newcommand{\instNUUTaiwan}{National United University, Miao Li 36003, Taiwan}
\newcommand{\instKrakow}{H. Niewodniczanski Institute of Nuclear Physics, Krakow 31-342, Poland}
\newcommand{\instNiigata}{Niigata University, Niigata 950-2181, Japan}
\newcommand{\instNSU}{Novosibirsk State University, Novosibirsk 630090, Russian Federation}
\newcommand{\instOkinawa}{Okinawa Institute of Science and Technology, Okinawa 904-0495, Japan}
\newcommand{\instOsakaCity}{Osaka City University, Osaka 558-8585, Japan}
\newcommand{\instRCNP}{Research Center for Nuclear Physics, Osaka University, Osaka 567-0047, Japan}
\newcommand{\instPNNL}{Pacific Northwest National Laboratory, Richland, Washington 99352, U.S.A.}
\newcommand{\instPanjab}{Panjab University, Chandigarh 160014, India}
\newcommand{\instPeking}{Peking University, Beijing 100871, China}
\newcommand{\instPanjabPAU}{Punjab Agricultural University, Ludhiana 141004, India}
\newcommand{\instRIKENMSL}{Meson Science Laboratory, Cluster for Pioneering Research, RIKEN, Saitama 351-0198, Japan}
\newcommand{\instRIKEN}{Theoretical Research Division, Nishina Center, RIKEN, Saitama 351-0198, Japan}
\newcommand{\instXavier}{St. Francis Xavier University, Antigonish, Nova Scotia, B2G 2W5, Canada}
\newcommand{\instSeoul}{Seoul National University, Seoul 08826, South Korea}
\newcommand{\instShandong}{Shandong University, Jinan 250100, China}
\newcommand{\instSPU}{Showa Pharmaceutical University, Tokyo 194-8543, Japan}
\newcommand{\instSoochow}{Soochow University, Suzhou 215006, China}
\newcommand{\instSoongsil}{Soongsil University, Seoul 06978, South Korea}
\newcommand{\instLjubljanaJSI}{J. Stefan Institute, 1000 Ljubljana, Slovenia}
\newcommand{\instKyiv}{Taras Shevchenko National Univ. of Kiev, Kiev, Ukraine}
\newcommand{\instTata}{Tata Institute of Fundamental Research, Mumbai 400005, India}
\newcommand{\instTUM}{Department of Physics, Technische Universit\"{a}t M\"{u}nchen, 85748 Garching, Germany}
\newcommand{\instECUTUM}{Excellence Cluster Universe, Technische Universit\"{a}t M\"{u}nchen, 85748 Garching, Germany}
\newcommand{\instTelAviv}{Tel Aviv University, School of Physics and Astronomy, Tel Aviv, 69978, Israel}
\newcommand{\instToho}{Toho University, Funabashi 274-8510, Japan}
\newcommand{\instTohoku}{Department of Physics, Tohoku University, Sendai 980-8578, Japan}
\newcommand{\instTitech}{Tokyo Institute of Technology, Tokyo 152-8550, Japan}
\newcommand{\instTokyoMetropolitan}{Tokyo Metropolitan University, Tokyo 192-0397, Japan}
\newcommand{\instUAS}{Universidad Autonoma de Sinaloa, Sinaloa 80000, Mexico}
\newcommand{\instNapoliUNIV}{Dipartimento di Scienze Fisiche, Universit\`{a} di Napoli Federico II, I-80126 Napoli, Italy}
\newcommand{\instNapoliUNIVA}{Dipartimento di Agraria, Universit\`{a} di Napoli Federico II, I-80055 Portici (NA), Italy}
\newcommand{\instPadovaUNIV}{Dipartimento di Fisica e Astronomia, Universit\`{a} di Padova, I-35131 Padova, Italy}
\newcommand{\instPerugiaUNIV}{Dipartimento di Fisica, Universit\`{a} di Perugia, I-06123 Perugia, Italy}
\newcommand{\instPisaUNIV}{Dipartimento di Fisica, Universit\`{a} di Pisa, I-56127 Pisa, Italy}
\newcommand{\instRomaUNIV}{Universit\`{a} di Roma ``La Sapienza,'' I-00185 Roma, Italy}
\newcommand{\instRomaTreUNIV}{Dipartimento di Matematica e Fisica, Universit\`{a} di Roma Tre, I-00146 Roma, Italy}
\newcommand{\instTorinoUNIV}{Dipartimento di Fisica, Universit\`{a} di Torino, I-10125 Torino, Italy}
\newcommand{\instTriesteUNIV}{Dipartimento di Fisica, Universit\`{a} di Trieste, I-34127 Trieste, Italy}
\newcommand{\instMontreal}{Universit\'{e} de Montr\'{e}al, Physique des Particules, Montr\'{e}al, Qu\'{e}bec, H3C 3J7, Canada}
\newcommand{\instIJCLab}{Universit\'{e} Paris-Saclay, CNRS/IN2P3, IJCLab, 91405 Orsay, France}
\newcommand{\instIPHC}{Universit\'{e} de Strasbourg, CNRS, IPHC, UMR 7178, 67037 Strasbourg, France}
\newcommand{\instAdelaide}{Department of Physics, University of Adelaide, Adelaide, South Australia 5005, Australia}
\newcommand{\instBonn}{University of Bonn, 53115 Bonn, Germany}
\newcommand{\instUBC}{University of British Columbia, Vancouver, British Columbia, V6T 1Z1, Canada}
\newcommand{\instCincinnati}{University of Cincinnati, Cincinnati, Ohio 45221, U.S.A.}
\newcommand{\instFlorida}{University of Florida, Gainesville, Florida 32611, U.S.A.}
\newcommand{\instHamburg}{University of Hamburg, 20148 Hamburg, Germany}
\newcommand{\instHawaii}{University of Hawaii, Honolulu, Hawaii 96822, U.S.A.}
\newcommand{\instHeidelberg}{University of Heidelberg, 68131 Mannheim, Germany}
\newcommand{\instLjubljanaUniLJ}{Faculty of Mathematics and Physics, University of Ljubljana, 1000 Ljubljana, Slovenia}
\newcommand{\instLouisville}{University of Louisville, Louisville, Kentucky 40292, U.S.A.}
\newcommand{\instMalaya}{National Centre for Particle Physics, University Malaya, 50603 Kuala Lumpur, Malaysia}
\newcommand{\instLjubljanaUM}{University of Maribor, 2000 Maribor, Slovenia}
\newcommand{\instMelbourne}{School of Physics, University of Melbourne, Victoria 3010, Australia}
\newcommand{\instMississippi}{University of Mississippi, University, Mississippi 38677, U.S.A.}
\newcommand{\instUOM}{University of Miyazaki, Miyazaki 889-2192, Japan}
\newcommand{\instNovaGorica}{University of Nova Gorica, 5000 Nova Gorica, Slovenia}
\newcommand{\instPittsburgh}{University of Pittsburgh, Pittsburgh, Pennsylvania 15260, U.S.A.}
\newcommand{\instUSTC}{University of Science and Technology of China, Hefei 230026, China}
\newcommand{\instSAlabama}{University of South Alabama, Mobile, Alabama 36688, U.S.A.}
\newcommand{\instSCarolina}{University of South Carolina, Columbia, South Carolina 29208, U.S.A.}
\newcommand{\instSydney}{School of Physics, University of Sydney, New South Wales 2006, Australia}
\newcommand{\instTabuk}{Department of Physics, Faculty of Science, University of Tabuk, Tabuk 71451, Saudi Arabia}
\newcommand{\instUTokyo}{Department of Physics, University of Tokyo, Tokyo 113-0033, Japan}
\newcommand{\instIPMU}{Kavli Institute for the Physics and Mathematics of the Universe (WPI), University of Tokyo, Kashiwa 277-8583, Japan}
\newcommand{\instVictoria}{University of Victoria, Victoria, British Columbia, V8W 3P6, Canada}
\newcommand{\instVPI}{Virginia Polytechnic Institute and State University, Blacksburg, Virginia 24061, U.S.A.}
\newcommand{\instWayneState}{Wayne State University, Detroit, Michigan 48202, U.S.A.}
\newcommand{\instYamagata}{Yamagata University, Yamagata 990-8560, Japan}
\newcommand{\instYerevan}{Alikhanyan National Science Laboratory, Yerevan 0036, Armenia}
\newcommand{\instYonsei}{Yonsei University, Seoul 03722, South Korea}
\newcommand{\HUNNU}{Hunan Normal University, 410081 Changsha, China}

\affiliation{\instCPPM}
\affiliation{\instBNL}
\affiliation{\instCMU}
\affiliation{\instPrague}
\affiliation{\instDESY}
\affiliation{\instDuke}
\affiliation{\instITAR}
\affiliation{\instHanyang}
\affiliation{\instKEK}
\affiliation{\instJPARC}
\affiliation{\instHSE}
\affiliation{\HUNNU}
\affiliation{\instIITGuwahati}
\affiliation{\instIITHyderabad}
\affiliation{\instHEPHYVienna}
\affiliation{\instIFIC}
\affiliation{\instFrascati}
\affiliation{\instNapoliINFN}
\affiliation{\instPadovaINFN}
\affiliation{\instPerugiaINFN}
\affiliation{\instPisaINFN}
\affiliation{\instRomaTreINFN}
\affiliation{\instTorinoINFN}
\affiliation{\instTriesteINFN}
\affiliation{\instKarlsruhe}
\affiliation{\instKyungpook}
\affiliation{\instLPI}
\affiliation{\instLMU}
\affiliation{\instISUni}
\affiliation{\instMPP}
\affiliation{\instNagoya}
\affiliation{\instNagoyaKMI}
\affiliation{\instNagoyaIAR}
\affiliation{\instNaraWu}
\affiliation{\instNiigata}
\affiliation{\instRIKENMSL}
\affiliation{\instSPU}
\affiliation{\instLjubljanaJSI}
\affiliation{\instTata}
\affiliation{\instTelAviv}
\affiliation{\instTohoku}
\affiliation{\instNapoliUNIV}
\affiliation{\instPadovaUNIV}
\affiliation{\instPerugiaUNIV}
\affiliation{\instPisaUNIV}
\affiliation{\instRomaTreUNIV}
\affiliation{\instTorinoUNIV}
\affiliation{\instTriesteUNIV}
\affiliation{\instIPHC}
\affiliation{\instBonn}
\affiliation{\instUBC}
\affiliation{\instCincinnati}
\affiliation{\instHawaii}
\affiliation{\instLjubljanaUniLJ}
\affiliation{\instLouisville}
\affiliation{\instLjubljanaUM}
\affiliation{\instMelbourne}
\affiliation{\instMississippi}
\affiliation{\instSAlabama}
\affiliation{\instIPMU}
\affiliation{\instVPI}
\affiliation{\instWayneState}
\affiliation{\instYerevan}

\author{F.~Abudin{\'e}n}\affiliation{\instTriesteINFN}
\author{N.~Akopov}\affiliation{\instYerevan} % 9443
\author{A.~Aloisio}\affiliation{\instNapoliINFN,\instNapoliUNIV}% 2194
\author{V.~Babu}\affiliation{\instDESY} % 5623
\author{Sw.~Banerjee}\affiliation{\instLouisville}% 8603
\author{M.~Bauer}\affiliation{\instKarlsruhe}
\author{J.~V.~Bennett}\affiliation{\instMississippi} % 2454
\author{F.~U.~Bernlochner}\affiliation{\instBonn} % 2282
\author{M.~Bessner}\affiliation{\instHawaii} % 3783
\author{S.~Bettarini}\affiliation{\instPisaINFN,\instPisaUNIV} % 2350
% ...

\begin{abstract}
We report on new flavor tagging algorithms developed to determine the quark-flavor content of bottom (\PB) mesons at Belle~II. The algorithms provide essential inputs for measurements of quark-flavor mixing and charge-parity violation. 
We validate and evaluate the performance of the algorithms using hadronic \PB~decays with flavor-specific final states reconstructed in a data set corresponding to an integrated luminosity of $62.8$\,fb$^{-1}$, collected at the $\PUpsilonFourS$~resonance with the Belle~II detector at the SuperKEKB collider.
We measure the total effective tagging efficiency to be
\begin{center}
\mbox{$\varepsilon_{\rm eff} = \big(30.0 \pm 1.2(\text{stat}) \pm 0.4(\text{syst})\big)\%$}
\end{center}
for a category-based algorithm and
\begin{center}
\mbox{$\varepsilon_{\rm eff} = \big(28.8 \pm 1.2(\text{stat}) \pm 0.4(\text{syst})\big)\%$}
\end{center}
for a deep-learning-based algorithm.
\end{abstract}


\maketitle

\section{Introduction}
\label{intro}
Determining the quark-flavor content of heavy-flavored hadrons is
essential in many measurements of quark-flavor mixing and \CP~violation. A keystone of the Belle~II physics program is the study of $\PBzero-\APBzero$~mixing and \CP~violation in decays of neutral \PB~mesons~\cite{Kou:2018nap,Belle:2004hwe,BaBar:2002jxa,LHCb:2016gsk,Adachi:2012et,Aubert:2009aw,Aaij:2015vza,Aubert:2004cp,Adachi:2018jqe,Adachi:2013mae,Lees:2012mma,Vanhoefer:2015ijw,Aubert:2007nua,Kusaka:2007dv,Lees:2013nwa,Chen:2006nk,Lees:2012kxa,Fujikawa:2008pk,Aubert:2008ad,Aubert:2008gy,Ushiroda:2006fi}. The study of these processes is key to constrain the Cabibbo-Kobayashi-Maskawa~(CKM) angles $\phione/\upbeta$
%~\cite{Adachi:2012et,Aubert:2009aw,Aaij:2015vza,Aubert:2004cp,Adachi:2018jqe} 
and $\phitwo/\upalpha$~\cite{Cabibbo:1963yz,Kobayashi:1973fv,Carter:1980tk,Dib:1989uz,Gronau:1990ka,Lipkin:1991st},
%~\cite{Adachi:2013mae,Lees:2012mma,LHCb:2017ood,Vanhoefer:2015ijw,Aubert:2007nua,Kusaka:2007dv,Lees:2013nwa}
as well as to study flavor anomalies
%~\cite{Chen:2006nk,Lees:2012kxa,Fujikawa:2008pk,Aubert:2008ad,Aubert:2008gy,Ushiroda:2006fi}
that could ultimately reveal possible deviations from the Standard Model expectations~\cite{Atwood:1997zr,Gronau:2005kz,Beneke:2005pu}.  

%Kou:2018nap,Abe:2004mz,Aubert:2002rg,Aaij:2013gja,Adachi:2012et,Aubert:2009aw,Aaij:2015vza,Aubert:2004cp,Adachi:2018jqe,Adachi:2013mae,Lees:2012mma,LHCb:2017ood,Vanhoefer:2015ijw,Aubert:2007nua,Kusaka:2007dv,Lees:2013nwa,Chen:2006nk,Lees:2012kxa,Fujikawa:2008pk,Aubert:2008ad,Aubert:2008gy,Ushiroda:2006fi

At Belle~II, $\PB$~mesons are produced in $\PB\APB$~pairs at the $\PUpsilonFourS$ resonance, which decays almost half of the time into a pair of neutral $\PB\APB$~mesons.
Most measurements of \CP~violation and $\PBzero-\APBzero$~mixing require the full reconstruction of a signal $\PB$~meson (signal side), and to determine the quark-flavor content of the accompanying \PB~meson (tag side) at the time of its decay, a task referred to as flavor tagging. 

Flavor tagging is possible because 
many decay modes of neutral $\PB$~mesons provide flavor signatures through flavor-specific final states. Flavor signatures are characteristics of the \PB-decay products that are correlated with the flavor of the neutral $\PB$~meson, which is the charge sign of the $\Pqb$ quark or antiquark that it contains. For example, in semileptonic decays such as 
\mbox{$\APBzero\to D^{*+} \Plm \Pagnl$}~(charge-conjugate processes are implied everywhere in this paper),
 a negatively charged lepton tags unambiguously a $\overline{B}^0$, which
 contains a negatively charged $\Pqb$, while a positively charged lepton tags a ${B}^0$, which
 contains a positively charged $\Paqb$.

To determine the quark-flavor of \PB~mesons, Belle~II exploits the information associated with the \PB-decay products using multivariate machine-learning algorithms. We develop two algorithms. The first one is a category-based flavor tagger~\cite{Abudinen:2018}, which is inspired by previous flavor taggers developed by the Belle and the BaBar collaborations~\cite{Kakuno:2004cf,Bevan:2014iga}. The category-based flavor tagger first identifies \PBzero-decay products and then combines all information to determine the \PBzero~flavor. The second algorithm is a deep-learning neural network~(DNN) flavor tagger~\cite{Gemmler:2020}, that determines the \PBzero flavor in a single step without pre-identifying \PBzero-decay products. 

In the following, we focus on the description of the algorithms and their training procedure. We evaluate the performance of the algorithms by measuring the $\PBzero-\APBzero$~mixing probability from the signal yield integrated in decay-time (time-integrated analysis). 
% Time-dependent measurements are currently being prepared and will be published in the future.

To evaluate the performance, we reconstruct signal $\PB$~decays with final states that allow us to unambiguously identify the flavor of the signal side and determine the flavor of the tag side using the flavor taggers. We
reconstruct signal $\PB$~decays into hadronic final states with branching fractions of $10^{-5}$ or greater to obtain a sufficiently large signal sample in the used data set. 
We evaluate the tagging performance on neutral \PB mesons and, as a cross check, on charged \PB~mesons.
The training of the flavor taggers, the signal \PB~reconstruction procedure and the event selection criteria  are developed and optimized using Monte~Carlo~(MC)~simulation before application to experimental data. 

The remainder of this paper is organized as follows. Section~\ref{sec:belle2} describes the Belle~II detector, followed by a description of the data sets and analysis framework in Sec.~\ref{sec:data}. The category-based flavor tagger is described in Sec.~\ref{sec:categoryTagger} and the DNN tagger in Sec.~\ref{sec:DNNTagger}. The training of the flavor taggers is detailed in Sec.~\ref{sec:training}. We describe the reconstruction of the calibration samples in Sec.~\ref{sec:reco} and the determination of efficiencies and wrong-tag fractions in Sec.~\ref{sec:fit}. We then compare the performance of the flavor taggers in data and in simulation in Sec.~\ref{sec:splot} and present the results of the calibration in Secs.~\ref{sec:results} and~\ref{sec:linearity}. A comparison with the Belle algorithm is provided in Sec.~\ref{sec:compBelle}, followed by a summary of the paper in Sec.~\ref{sec:summary}.

\section{The Belle II detector}

\label{sec:belle2}

Belle~II is a particle-physics spectrometer with a solid-angle acceptance of almost $4\pi$~\cite{Kou:2018nap, Abe:2010gxa}. 
It is designed to reconstruct the products of electron-positron collisions produced by the SuperKEKB asymmetric-energy collider~\cite{Akai:2018mbz}, 
located at the KEK laboratory in Tsukuba, Japan. 
Belle~II comprises several subdetectors arranged around the interaction point in a cylindrical geometry. 
The innermost subdetector is the vertex detector, 
which uses position-sensitive silicon layers to sample the trajectories of charged particles (tracks) in the vicinity of the interaction region to determine the decay vertex of their long-lived parent particles.
The vertex detector includes two inner layers of silicon pixel sensors~(PXD) and four outer layers of double-sided silicon microstrip sensors~(SVD). 
The second pixel layer is currently incomplete and covers only one sixth of azimuthal angle. 
Charged-particle momenta and charges are measured by a large-radius, helium-ethane, small-cell central drift chamber~(CDC), which also offers charged-particle-identification information via the measurement of particles' energy-loss~${\dd E}/{\dd x}$ by specific ionization. 
A Cherenkov-light angle and time-of-propagation detector~(TOP) surrounding the chamber provides charged-particle identification in the central detector volume, supplemented by proximity-focusing, aerogel,
ring-imaging Cherenkov detectors~(ARICH) in the forward regions. 
A CsI(Tl)-crystal electromagnetic
calorimeter~(ECL) enables energy measurements of electrons and photons. 
A solenoid surrounding the 
calorimeter generates a uniform axial 1.5\,T magnetic field filling its inner volume. 
Layers of plastic scintillator and resistive-plate chambers~(KLM), interspersed between the
magnetic flux-return iron plates, enable identification of \PKzL~mesons and muons. 
We employ all subdetectors in this work.

\section{Framework and data}

\label{sec:data}

Both flavor taggers are part of the Belle~II analysis software framework~\cite{Kuhr:2018lps}, which is used to process all data. We train the flavor taggers using a sample of 20~million signal-only MC~events~\cite{Ryd:2005zz,GEANT4:2002zbu}, where the signal $\PB$~meson decays to the invisible final state $\PBzero\to\Pgngt\Pagngt$ and the tag-side $\PB$~meson decays to any possible final state according to known branching fractions~\cite{Zyla:2020zbs}. To perform different tests after training, we use similar signal-only MC~samples where the signal $\PB$~meson decays to benchmark decay modes such as \mbox{\small$\PBzero\to\PKp\Pgpm$},  
{\small$\PBzero\to\PJpsi(\to\Pgmp\Pgmm)\,\PKzS(\to\Pgpp\Pgpm)$}, and 
{\small$\PBzero\to\Peta^\prime(\to\Pgpp\Pgpm\Peta(\to\Pgpp\Pgpm\Pgpz))\,\PKzS(\to\Pgpp\Pgpm)$}.

We evaluate the performance of the flavor taggers using \textit{generic MC~simulation}. 
The generic MC~simulation consists of samples that include $\Pep\Pem\to\PBzero\APBzero$, $\PBplus\PBminus$, $\Pqu\Paqu$, $\Pqd\Paqd$, $\Pqc\Paqc$, and $\Pqs\APqs$~processes~\cite{Ryd:2005zz,GEANT4:2002zbu,Zyla:2020zbs, Sjostrand:2014zea} in proportions representing their different production cross~sections and correspond to an integrated luminosity of 700\,fb$^{-1}$, about eleven times the data sample used in the measurement. 
We use these samples to optimize the event selection and to compare the flavor distributions and fit results obtained from the experimental data with the expectations.

Signal-only and generic MC simulation include the effect of simulated beam-induced backgrounds~\cite{Kou:2018nap,Lewis:2018ayu}, caused by the Touschek effect (scattering and
loss of beam particles), 
by beam-gas scattering and by
synchrotron radiation, 
as well as simulated
luminosity-dependent backgrounds, caused by Bhabha scattering~\cite{Kleiss:1994wq,CarloniCalame:2000pz} and by two-photon quantum electrodynamical processes~\cite{Berends:1984gf}.

As for experimental data, we use all good-quality data collected at the \PUpsilonFourS~resonance between March~11$^\text{th}$, 2019 and July~1$^\text{st}$, 2020; this sample corresponds to an integrated luminosity of $62.8\,\si{fb^{-1}}$~\cite{Abudinen:2019osb}. To reduce the data sample size to a manageable level, all events are required to meet loose data-skim selection criteria, based on total energy and charged-particle multiplicity in the event. Almost $100\%$ of the $\PB\APB$~events meet the data-skim selection criteria.

\section{The category-based flavor tagger}

\label{sec:categoryTagger}

The Belle~II flavor taggers are multivariate algorithms that receive as input kinematic, track-hit, and charged-particle identification~(PID) information about the particles on the tag side and provide as output the product $q\cdot r$, where $q$ is the flavor of the tag-side $\PB$~meson, and $r$ the dilution factor. By definition, the dilution factor $r$ is equal to $1-2w$, where $w$ corresponds to the fraction of wrongly tagged events. A dilution factor $r=0$ indicates a
fully diluted flavor (no possible distinction between $\PBzero$ and $\APBzero$), whereas a dilution factor $r=1$ indicates a perfectly tagged flavor. By convention, $q = +1$ corresponds to a tag-side \PBzero, and $q=-1$ corresponds to a tag-side \APBzero.

The new category-based algorithm relies on flavor-specific decay modes. Each decay mode has a particular decay topology and provides a flavor-specific signature. Similar or complementary decay modes are combined to obtain additional flavor signatures. The different flavor signatures are sorted into thirteen \emph{tagging categories}, which are described in detail in Sec.~\ref{sec:categories}. Table~\ref{table:targets} shows an overview of all thirteen categories together with the underlying decay modes and the respective flavor-specific final state particles, which we call \emph{target particles}.

{\renewcommand{\arraystretch}{1.1}
\begin{table}[pt]
\caption{Tagging categories and their targets~(left) with examples of the considered decay modes~(right). The target particles for each category are shown using the same colors on the left and on the right. Here, $p^*$ stands for momentum in the center-of-mass frame, $\Plpm$ for charged leptons (\Pmuon or \Pelectron) and $\PX$ for other possible particles in the decays. }
\begin{center}
    \label{table:targets} \vspace{0.4cm}
  \begin{minipage}[ht]{0.55\linewidth}
  \begin{tabular}{ @{}c@{} @{}l@{} }
    \begin{tabular}{ l  @{}c@{} }
    \hline
    Categories & Targets for \APBzero \\ \hline\hline
    Electron & \color[rgb]{0.000,0.3,0} $\Pelectron$\\ %   
    Intermediate Electron & \color[rgb]{0.8,0.1,0.5}\ $\APelectron$ \\
    Muon &\color[rgb]{0.000,0.3,0} $\Pmuon$  \\
    Intermediate Muon & \color[rgb]{0.8,0.1,0.5}\ $\APmuon$\\
    Kinetic Lepton & \color[rgb]{0.000,0.3,0} $\Pleptonminus$ \\
    Intermediate Kinetic Lepton& \color[rgb]{0.8,0.1,0.5}\ $\Pleptonplus$\\
    Kaon & \color[rgb]{0.8,0,0} $\PKm$ \\
    Kaon-Pion & \thinspace \thinspace {\color[rgb]{0.8,0,0} $\PKm$}, \color[rgb]{0,0.44,0.75} $\Pgpp$ \thinspace \\
    Slow Pion & \color[rgb]{0,0.44,0.75} $\Pgpp$  \\
    Maximum $p^*$ & {\color[rgb]{0.000,0.3,0} $\Plm$}, \color[rgb]{1,0.4,0} $\Pgpm$  \\
    Fast-Slow-Correlated (FSC)  & {\color[rgb]{0.000,0.3,0} $\Pleptonminus$}, \color[rgb]{0,0.44,0.75} $\Pgpp$\\
    Fast Hadron & \color[rgb]{1,0.4,0} $\Pgpm$, $\PKm$ \\
    Lambda & \color[rgb]{0.44,0.18,0.63}$\PLambda$ \\
    \hline
%    \hline\multicolumn{2}{|c|}{Total = 13} \\ \hline
    \end{tabular}
    \end{tabular}
    \end{minipage}
    \begin{minipage}[ht]{0.3\linewidth}
     \vspace{-0.21cm}
     \begin{tabular}{ c }
     Underlying decay modes
     \end{tabular}
     \begin{center}
    \includegraphics{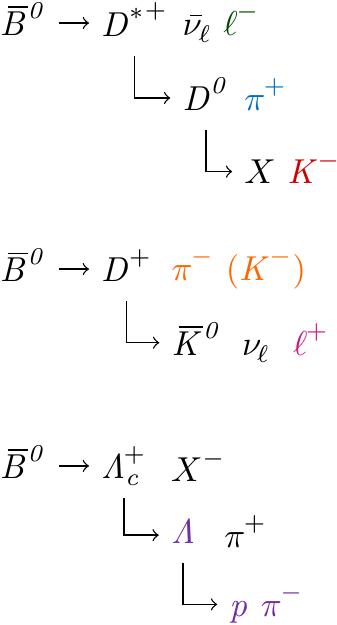}
    \end{center}
     \end{minipage}
\end{center}
\end{table}}

We identify the target particles among all available particle candidates on the tag side using discriminating input variables. Some input variables require information from all reconstructed tracks~(charged candidates)~\cite{BelleIITrackingGroup:2020hpx} and all neutral clusters~(neutral candidates) on the tag side. Neutral clusters are clusters in the electromagnetic calorimeter~(reconstructed photons) and in the KLM detector~(reconstructed \PKzL particles) that are not associated with a reconstructed track. Summing the input variables for all categories yields a total of $186$ inputs in the current configuration of the algorithm~(see Sec.~\ref{sec:categories}). Some variables are used multiple times for the same candidates in different categories. To save computing time, each variable is calculated only once for each candidate.

We adopted the useful concept of tagging categories from the previous Belle and BaBar flavor taggers~\cite{Kakuno:2004cf,Bevan:2014iga}. However, the new Belle II category-based flavor tagger includes more categories and more input variables than previously implemented algorithms.
% and is optimized to cope with the new high-luminosity background conditions at SuperKEKB. 

%,  reducing the number of variables to be calculated to $n_\text{unique} = 108$. 

%\clearpage

\subsection{Categories and input variables}

\label{sec:categories}

In the following, we describe the flavor signatures and the input variables. 
Table~\ref{table:variables} shows an overview of the input variables for each category.  
Except for the 
Maximum-$p^*$ category, PID variables are used for all categories. 
The PID variables correspond to PID likelihoods $\mathcal{L}$~\cite{Kou:2018nap}, which can be either combined likelihoods considering 6 possible long-lived charged-particle hypotheses (\Pe, \Pmu, \PK, \Pgp, \Pp, and deuteron), or binary  likelihoods considering only two of the hypotheses. 
The PID likelihoods can be calculated using all sub-detectors providing particle identification, or single ones (TOP, ARICH, ECL, KLM, or ${\dd E}/{\dd x}$ from CDC). For example, $\mathcal{L}_{\Pgp/\Pe}^{{\dd E}/{\dd x}}$~stands for binary $\Pgp/\Pe$ PID using only CDC information.  \vspace{0.1cm}

{\renewcommand{\arraystretch}{1.4}
\begin{table*}
    \centering
        \caption{Discriminating input variables for each category and for the DNN flavor tagger. For some of the categories the $p$-value of the track fit is taken into account. For the Lambda category, the $p$-value of the reconstructed $\PLambda$-decay vertex is used.  All variables are calculated for every particle candidate.} 
    \label{table:variables}\vspace{0.4cm}        
    % \begin{tabularx}{1\textwidth}{ l@{\hskip 4pt}c }
    \begin{tabular}{ l@{\hskip 20pt} c }
    \hline\hline 
    Categories & Discriminating input variables \\ \hline
    Electron & $\mathcal{L}_{\Pe}$,$\ \mathcal{L}_{\Pe}^\text{TOP}$,$\ \mathcal{L}_{\Pe}^\text{ARICH}$,
    $\ \mathcal{L}_{\Pe}^\text{ECL}$,$\ p^*$,$\ p_\text{t}^*$,$\ p$,$\ p_\text{t}$,$\, \cos{\theta}$,\\ %   
    Int. Electron &  $M_\text{rec}^2$,$\, E_{90}^W$,$\, p^*_\text{miss}$,$\, \cos{\theta^*_\text{miss}}$,$\, \vert\cos{\theta^*_{\text{T}}}\vert$,$\, p$-val.\\ \hline
    Muon & $\mathcal{L}_{\Pgm}$,$\ \mathcal{L}_{\Pgm}^\text{TOP}$,$\ \mathcal{L}_{\Pgm}^\text{ARICH}$,
    $\ \mathcal{L}_{\Pgm}^\text{KLM}$,$\ p^*$,$\ p_\text{t}^*$,$\ p$,$\, p_\text{t}$,$\, \cos{\theta}$,\\ %  
    Int. Muon &  $M_\text{rec}^2$,$\, E_{90}^W$,$\, p^*_\text{miss}$,$\, \cos{\theta^*_\text{miss}}$,$\, \vert\cos{\theta^*_{\text{T}}}\vert$,$\, p$-val.(only for Int.)\\ \hline
    Kin. Lepton & $\mathcal{L}_{\Pe}$,$\ \mathcal{L}_{\Pe}^\text{TOP}$,$\ \mathcal{L}_{\Pe}^\text{ARICH}$,
    $\ \mathcal{L}_{\Pe}^\text{ECL}$,$\ \mathcal{L}_{\Pgm}$,$\ \mathcal{L}_{\Pgm}^\text{TOP}$,$\ \mathcal{L}_{\Pgm}^\text{ARICH}$,
    $\ \mathcal{L}_{\Pgm}^\text{KLM}$,$\, p^*$,$\ p_\text{t}^*$,$\ p$,$\ p_\text{t}$,\\ 
    Int. Kin. Lep.& $\, \cos{\theta}$,$\ M_\text{rec}^2$,$\, E_{90}^W$,$\, p^*_\text{miss}$,$\, \cos{\theta^*_\text{miss}}$,$\, \vert\cos{\theta^*_{\text{T}}}\vert$,$\, p$-val.(only for Int.)\\\hline    
    \multirow{2}{*}{Kaon} &  $\mathcal{L}_{\PK}$,$\ \mathcal{L}_{\PK}^{{\dd E}/{\dd x}}$,$\ \mathcal{L}_{\PK}^\text{TOP}$,$\ \mathcal{L}_{\PK}^\text{ARICH}$,$\, p^*$,$\, p_\text{t}^*$,$\, p$,$\, p_\text{t}$,$\, \cos{\theta}$,\\ 
    & $n_{\PKs}$,$\, \sum p_\text{t}^2$,$\ M_\text{rec}^2$,$\, E_{90}^W$,$\, p^*_\text{miss}$,$\, \cos{\theta^*_\text{miss}}$, $\vert\cos{\theta^*_{\text{T}}}\vert$,$\, p$-val.\\ \hline
    Slow Pion & $\mathcal{L}_{\Pgp}$,$\ \mathcal{L}_{\Pgp}^\text{TOP}$,$\ \mathcal{L}_{\Pgp}^\text{ARICH}$,$\ \mathcal{L}_{\Pgp/\Pe}^{{\dd E}/{\dd x}}$,$\ \mathcal{L}_\text{\Pe}$,$\ \mathcal{L}_{\PK}$,$\ \mathcal{L}_{\PK}^{{\dd E}/{\dd x}}$,$\ \mathcal{L}_{\PK}^\text{TOP}$,$\ \mathcal{L}_{\PK}^\text{ARICH}$,\\
    Fast Hadron & $p^*$,$\ p_\text{t}^*$,$\ p$,$\ p_\text{t}$,$\cos{\theta}$,$\ n_{\PKs}$,$\, \sum p_\text{t}^2$,$\ M_\text{rec}^2$,$ E_{90}^W$,$\ p^*_\text{miss}$,$\ \cos{\theta^*_\text{miss}}$,$\,\vert\cos{\theta^*_{\text{T}}}\vert$ \\ \hline  
    Kaon-Pion & $\ \mathcal{L}_{\PK}$,$\ y_\text{Kaon}$,$\ y_\text{SlowPion}$,$\ \cos{\theta^*_{\PK\pi}}$,$\ q_{\PK} \cdot q_{\Pgp} $ \\ \hline   
    Maximum $p^*$ & $ p^*$,$\ p_\text{t}^*$,$\ p$,$\ p_\text{t}$,$\ \vert\cos{\theta^*_{\text{T}}}\vert $  \\ \hline
    FSC  & $\mathcal{L}_{\PK}$,$\ p^*_\text{Slow}$,$\ p^*_\text{Fast}$,$\ \vert\cos{\theta^*_\text{T, Slow}}\vert$, $\vert\cos{\theta^*_\text{T, Fast}}\vert$, $\cos{\theta^*_{\text{SlowFast}}}$,$\ q_\text{Slow} \cdot q_\text{Fast} $\\ \hline
  
    Lambda & $\mathcal{L}_\text{\Pp}$,$\, \mathcal{L}_{\Pgp}$,
    $\ p^*_{\PLambda}$,$\, p_{\PLambda}$,$\ p^*_{\Pgp}$,$\, p_{\Pp}$,
    $\ p^*_{\Pgp}$,$\, p_{\Pgp}$,$\, q_{\PLambda}$,$\, M_{\PLambda}$,
    $\, n_{\PKs}$,$\, \cos{\theta_{\bm{x}_{\PLambda},\,\bm{p}_{\PLambda}}}$,
    $\, \sigma_{\PLambda}^{zz}$, $\ p$-val. \\\hline\hline
    
    DNN &  $\mathcal{L}_{\Pe}$, $\mathcal{L}_{\Pmu}$, $\mathcal{L}_{\PK}$,  $\mathcal{L}_{\Ppi}$, $\mathcal{L}_{\Pproton}$, $p^*$, $\cos(\theta^*)$, $\phi^*$,  $N_{\rm PXD}$, $N_{\rm SVD}$.\\ \hline    
    \end{tabular}
\end{table*}}

\textbf{Electron, Muon, and Kinetic lepton:} these categories exploit the signatures provided by primary leptons from $\PB$~decays occurring via transitions $\Pbottom\to~\Pcharm~\Pleptonminus\APnulepton$, or $\Pbottom\to~\Pup~\Pleptonminus\APnulepton$, where $\Plepton$ corresponds to an electron, muon or both depending on the category. 
    Useful variables to identify primary leptons are the momentum $p$, the transverse momentum $p_\text{t}$, and the cosine of the polar angle $\cos{\theta}$, which can be calculated in the lab frame, or in the \PUpsilonFourS frame (denoted with a $^*$ superscript). We consider the following variables calculated only in the \PUpsilonFourS frame:
 
  \vspace{0.2cm}
 
  \begin{itemize}
\item[$\bullet$] $M_\text{rec}^2 = m_{\PX}^2 = g_{\mu,\nu}p_{\PX}^{*\mu} p_{\PX}^{*\nu}$, the squared invariant mass of the recoiling system $\PX$ whose four-momentum is defined by
\begin{eqnarray*}
\hspace{2.9cm}  p_{\PX}^{*\mu} & = & \sum_{i\neq \Plepton} p_i^{*\mu}\text{,}
\end{eqnarray*}
where the index $i$ goes over all charged and neutral candidates on the tag side and $\Pl$ corresponds to the index of the lepton candidate. 

 \vspace{0.2cm}

\item[$\bullet$] $p_\text{miss}^* = \vert\bm{p}_\text{miss}^*\vert =\vert \bm{p}^*_{\PBtag} - \bm{p}^*_{\PX} - \bm{p}^*_{\Plepton}\vert$, the absolute value of the missing momentum.

 \vspace{0.2cm}

\item[$\bullet$] $\cos{\theta^*_\text{miss}}$, the cosine of the angle between the momentum $\bm{p}^*_{\Plepton}$ of the lepton candidate and the missing momentum $\bm{p}_\text{miss}^*$. 

 \vspace{0.2cm}

\item[$\bullet$] $E_{90}^{\PW}$, the energy in the hemisphere defined by the direction of the virtual \PWpm in the \PB~meson decay,
\begin{eqnarray*}
\hspace{1.8cm}  E_{90}^{\PW}=\ \sum_{i\in \PX,\ \bm{p}^*_i\cdot \bm{p}^*_{\PW} > 0} E_i^{\rm ECL} \text{,}
\end{eqnarray*}
where the sum extends over all charged and neutral candidates in the recoiling system \PX that are in the hemisphere of the \PWpm, and $E^{\rm ECL}$ corresponds to the energy deposited in the ECL.  
The momentum of the virtual \PWpm is calculated as
\begin{eqnarray*}
\hspace{0.9cm} \bm{p}^*_{\PW}\ =\ \bm{p}^*_{\Plepton} + \bm{p}^*_{\Pgn} \approx\ \bm{p}^*_{\Plepton} + \bm{p}_\text{miss}^* = -\bm{p}^*_{\PX} \text{,}
\end{eqnarray*}
where the momentum $\bm{p}_{\Pgn}$ of the neutrino is estimated using the missing momentum $\bm{p}_\text{miss}^*$. 
In the equation above we assume the \PB~meson to be almost at rest in the \PUpsilonFourS frame, that is
$\bm{p}^*_{\PBtag} \approx  \mathbf{0}$ and thus $\bm{p}_\text{miss}^* \approx - \left(\bm{p}^*_{\PX} + \bm{p}^*_{\Plepton}\right)$.

 \vspace{0.2cm}

\item[$\bullet$] $\vert\cos{\theta^*_{\text{T}}}\vert$, the absolute value of the cosine of the angle between the momentum $\bm{p}^*_{\Plepton}$ of the lepton and the thrust axis of the tag-side \PB~meson in the \PUpsilonFourS frame. 
In general, a thrust axis $\bm{T}$ can be defined as the unit vector that maximizes the thrust 
\begin{eqnarray*}
\hspace{2.6cm} T = \frac{ \sum_{i} \left| \bm{T} \cdot \bm{p}_i \right| }{ \sum_{i} \left| \bm{p}_i \right| }\text{,}
\label{EQ:TOOLS-BKGD:THRUST}
\end{eqnarray*}
where the sum extends over a group of particles. 
For the thrust axis of the tag-side \PB~meson, the sum extends over all charged and neutral candidates on the tag side.

\end{itemize}

\textbf{Intermediate Electron, Intermediate Muon,
    and Intermediate Kinetic Lepton:} 
    these categories exploit flavor signatures from  secondary leptons produced through the decay of charmed mesons and baryons occurring via transitions \Pbottom$\to$~\Pcharm$\to$~\Pstrange~(\Pdown)~\Pleptonplus\Pnulepton. 
    In this case the charge-flavor correspondence is reversed with respect to primary leptons: 
    a positively charged secondary lepton tags a \APBzero meson, and a negatively charged one a \PBzero meson. 
    Since their momentum spectrum is much softer in comparison with primary leptons, we refer to secondary leptons as intermediate leptons. 
    %Intermediate electrons and intermediate muons are the targets of the \textit{Intermediate Electron} and the \textit{Intermediate Muon} categories. Both are considered as targets in the \textit{Intermediate Kinetic Lepton} category.

\vspace{0.2cm}
\textbf{Kaon:} this category exploits the signature from
    kaons originating from decays of charmed mesons and baryons produced via $\Pbottom\to~\Pcharm\to~\Pstrange$ transitions. 
    Such kaons are referred to as \textit{right-sign} kaons. 
    They tag a \APBzero if they are negatively charged, and a \PBzero if they are positively charged. 

The kaon category provides the largest tagging power due to the high abundance of charged kaons (around $80\%$ of the $\PB$~decays contain one) 
    and because the fraction of right-sign kaons~(around $70\%$) 
    is much larger than the fraction of wrong-sign kaons~(around $10\%$) produced through processes of the kind
    $
\APbottom\to\PWp\left(\to\Pcharm\APstrange/\Pcharm\APdown\right)\,\PX$, with $\Pcharm\to\Pstrange\to\PKm$.

To identify target kaons, we include the following input variables:
\begin{itemize}
\item[$\bullet$] $n_{\PKzS}$, the number of reconstructed \PKzS~candidates on the tag side. 
Charged kaons originating from $\Pbottom\to~\Pcharm\to~\Pstrange$ transitions are usually not accompanied by \PKzS candidates, 
while wrong-sign kaons or charged kaons originating from $\Pstrange\APstrange$ pairs out of the vacuum are usually accompanied by one or more \PKzS~candidates. 

\item[$\bullet$] $\sum p_\text{t}^2$, the sum of the squared transverse momentum of all tracks on the tag side in the lab frame. 
\item[$\bullet$] $M_\text{rec}^2$, $E_{90}^{\PW}$, $p^*_\text{miss}$, $\cos{\theta^*_\text{miss}}$, and $\vert\cos{\theta^*_{\text{T}}}\vert$, 
the variables in the \PUpsilonFourS frame that discriminate against the lepton background.   
\end{itemize}

\textbf{Slow Pion:} the target particles of this category are secondary pions from decays \mbox{$\APBzero\to\PDstarplus(\to\PDzero\Pgpp)\PX^{-}$}. 
Due to the small mass difference between \PDstarplus and \PDzero, the secondary pions have a soft momentum spectrum and are therefore called slow pions. 
To identify slow pions we include some variables of the Kinetic Lepton and the kaon category, which help distinguish the background from slow leptons and kaons. 
    
%Slow pions are produced nearly at rest in the \PDstarplus frame together with the \PDzero meson. In the \PUpsilonFourS frame, they fly close to the \PDzero decay products and opposite to the other tag-side \PB-decay products, while in the lab frame they travel close to the beam boost direction. Thus, the polar angle  $\theta$ in the lab frame, and the angle to the thrust axis $\theta^*_{\text{T}}$ in the \PUpsilonFourS frame are useful variables. 
\vspace{0.2cm}
\textbf{Kaon-Pion:} this category exploits the flavor signatures of decays containing both a right-sign kaon and a slow pion.
We use the following input variables to identify both target particles:
\begin{itemize}
\item[$\bullet$] $y_\text{Kaon}$, the probability of being a target kaon obtained from the individual Kaon category~(see
Sec.~\ref{sec:CatBasedAlgo}).
\item[$\bullet$] $y_\text{SlowPion}$, the probability of being a target slow pion obtained from the individual Slow Pion category~(see
Sec.~\ref{sec:CatBasedAlgo}).
\item[$\bullet$] $\cos{\theta^*_{\PK\Pgp}}$, the cosine of the angle between the kaon and the slow-pion momentum in the \PUpsilonFourS frame. 
\item[$\bullet$] $q_{\PK} \cdot q_{\Pgp}$, the charge product of the kaon and the slow-pion candidates. 
\end{itemize} 

\textbf{Fast Hadron:} the targets of this category are kaons and pions from the \PW~boson in \mbox{$\Pbottom\to\,\Pcharm\,\PWm$} or  \mbox{$\Pbottom\to\Pup\,\PWm$} decays, 
    and from \textit{one-prong} decays of primary tau leptons from  $\Pb\to~\Pgtm(\to~\Ph^-~\Pgngt)~\Pagngt~\PX$ transitions,
where $\Ph^-$ stands for a \Pgpm or a \PKminus. 
This category considers as targets also those kaons and pions produced 
through intermediate resonances that decay via strong
processes conserving the flavor information, 
for example 
$\APBzero\to\PKstarminus(\to\PKm\Pgpz)\,\PX^+$.
The target kaons and pions are referred to as fast hadrons because of their hard momentum spectrum. 
% A negatively (positively) charged fast hadron indicates a \APBzero (\PBzero) meson. 
To identify them we use the same set of variables as the Slow Pion category, which also distinguish fast kaons and pions among the background of slow particles.  

\vspace{0.2cm}
\textbf{Maximum $p^*$:}  this category is a very inclusive tag based on selecting the charged particle with the highest momentum in the \PUpsilonFourS frame and using its charge as a flavor tag. 
In this way we give a higher weight to primary particles that may have not been selected either as a primary lepton or as a fast hadron. 
Primary hadrons and leptons from the $\PWpm$~boson in $\Pbottom\to~\Pqc~\PWm$ or in $\Pbottom\to~\Pqu~\PWm$ transitions have a very hard momentum spectrum and are most likely to be the tag-side particles with the largest momenta in a given event. 

\vspace{0.2cm}    
\textbf{Fast-Slow-Correlated (FSC):} the targets of this category are both slow pions and high-momentum primary particles. To identify them,  we use the following input variables:
\begin{itemize}
\item[$\bullet$] $p^*_\text{Slow}$, the momentum of the slow pion candidate in the \PUpsilonFourS frame.
\item[$\bullet$] $p^*_\text{Fast}$, the momentum of the fast candidate in the \PUpsilonFourS frame.
\item[$\bullet$] $\vert\cos{\theta^*_\text{T, Slow}}\vert$, the absolute value of the cosine of the angle between the thrust axis and the momentum of the slow pion candidate.
\item[$\bullet$] $\vert\cos{\theta^*_\text{T, Fast}}\vert$, the absolute value of the cosine of the angle between the thrust axis and the momentum of the fast candidate.
\item[$\bullet$] $\cos{\theta^*_{\text{SlowFast}}}$, the cosine of the angle between the momenta of the slow and the fast candidate.
\item[$\bullet$] $q_\text{Slow} \cdot q_\text{Fast}$, the charge product of the slow pion and the fast candidate.  
\end{itemize} 
    
\textbf{Lambda:} this category exploits the additional flavor signatures provided by $\PLambda$ baryons from $b\to c\to s$ transitions. 
A $\PLambda$ baryon indicates a \APBzero, and a $\bar{\PLambda}$ a \PBzero. 
%Although the abundance of decays with target $\PLambda$~particles is rather small, they provide complementary flavor tagging information.
    Here, \PLambda candidates are reconstructed from pairs of proton and pion candidates. To identify target $\PLambda$~particles, we use the momentum of the reconstructed $\PLambda$, the momenta of the proton and the pion, and also the following input variables: 
\begin{itemize}
\item[$\bullet$] $q_{\PLambda}$, the flavor of the $\PLambda$ baryon.
\item[$\bullet$] $M_{\PLambda}$, the reconstructed mass of the $\PLambda$.
\item[$\bullet$] $n_{\PKzS}$, the number of reconstructed \PKzS~candidates on the tag side.
\item[$\bullet$] $\cos{\theta_{\bm{x}_{\PLambda}\bm{p}_{\PLambda}}}$, the cosine of the angle between the $\PLambda$ momentum $\bm{p}_{\PLambda}$ and the direction from the interaction point to the reconstructed $\PLambda$ vertex $\bm{x}_{\PLambda}$ in the lab frame.
\item[$\bullet$] $|\bm{x}_{\PLambda}|$, the absolute distance between the $\PLambda$ vertex and the interaction point. 
\item[$\bullet$] $\sigma_{\PLambda}^{zz}$, the uncertainty on the $\PLambda$ vertex fit in the direction along the beam ($z$ direction). 
\item[$\bullet$] $\chi^2_{\PLambda}$, the $\chi^2$ probability of the vertex fit of the reconstructed $\PLambda$ decay vertex. 
\end{itemize}

% \clearpage

In comparison with previous versions of the Belle~II flavor taggers~\cite{Abudinen:2018,Gemmler:2020,Kou:2018nap}, for the current version of the algorithms we exclude track impact parameters~(displacement from nominal interaction point), because they are not yet well simulated for small displacements below $0.1\,\si{cm}$. Track impact parameters provide additional separation power between primary particles produced at the \PB-decay vertex (and thus with small track impact parameters) and secondary particles with decay vertices displaced from the interaction point. Thus, we will consider to use them again in the future. 

For the current version of the algorithms, we also exclude the $p$-value of the track fit for the Muon and the Kinetic Lepton categories since we observe discrepancies between data and simulation in the $p$-value distribution of particles identified as primary muons.

\subsection{Algorithm}

\label{sec:CatBasedAlgo}

\begin{figure}[hb]
\centering
\includegraphics[width=0.7\textwidth]{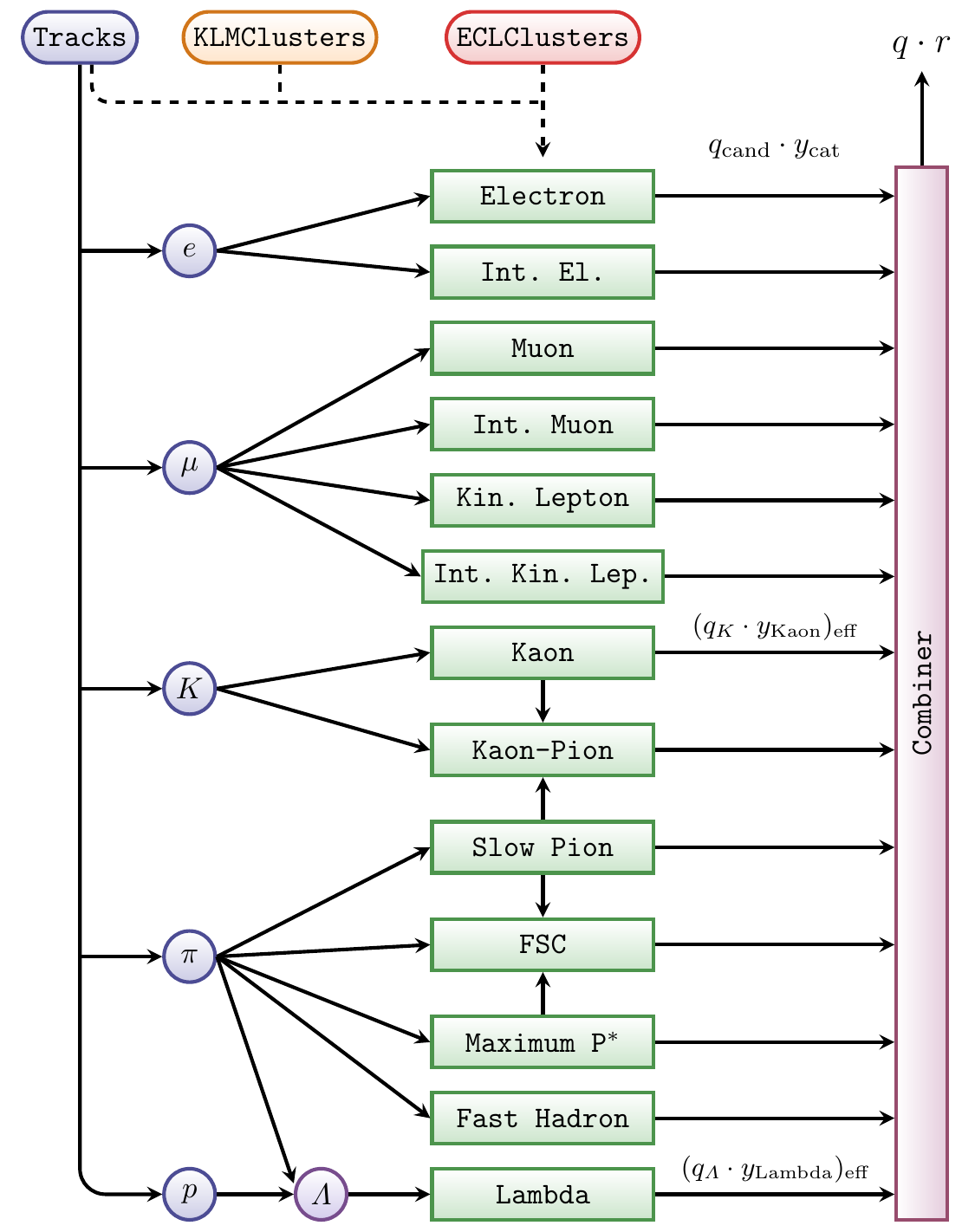}
\vspace{0.5cm}
\caption{Schematic overview of the category-based flavor tagger. The tracks on the tag side are used to build five different lists of candidates: \Pe, \Pmu, \PK, \Pgp, and \Pp. Each category considers the list of candidates belonging to its own targets. The different categories are represented by green boxes, and the combiner by a magenta box.}
  \label{fig:FlavTagScheme}
\end{figure}

The category-based flavor tagger performs a two-level procedure with an event level for each category
followed by a combiner level. Figure~\ref{fig:FlavTagScheme} shows a schematic overview.
The algorithm is based on Fast Boosted Decision Tree (FBDT)~\cite{Keck:2017gsv} classifiers, 
which are stochastic gradient-boosted decision trees that incorporate several mechanisms for regularization and are optimized to save computing resources during training and application. 

At the event~level, the flavor tagger identifies decay products providing flavor signatures among the $\Pepm$, $\Pmupm$, $\PKpm$, $\Pgppm$, and $\PLambda$ candidates.  Each category considers the list of particle candidates corresponding to its target particles. The event-level process is performed for each category, which corresponds to an FBDT classifier that receives the input variables associated with the category. 

\begin{figure*}[h]
\centering
\includegraphics[width=\textwidth]{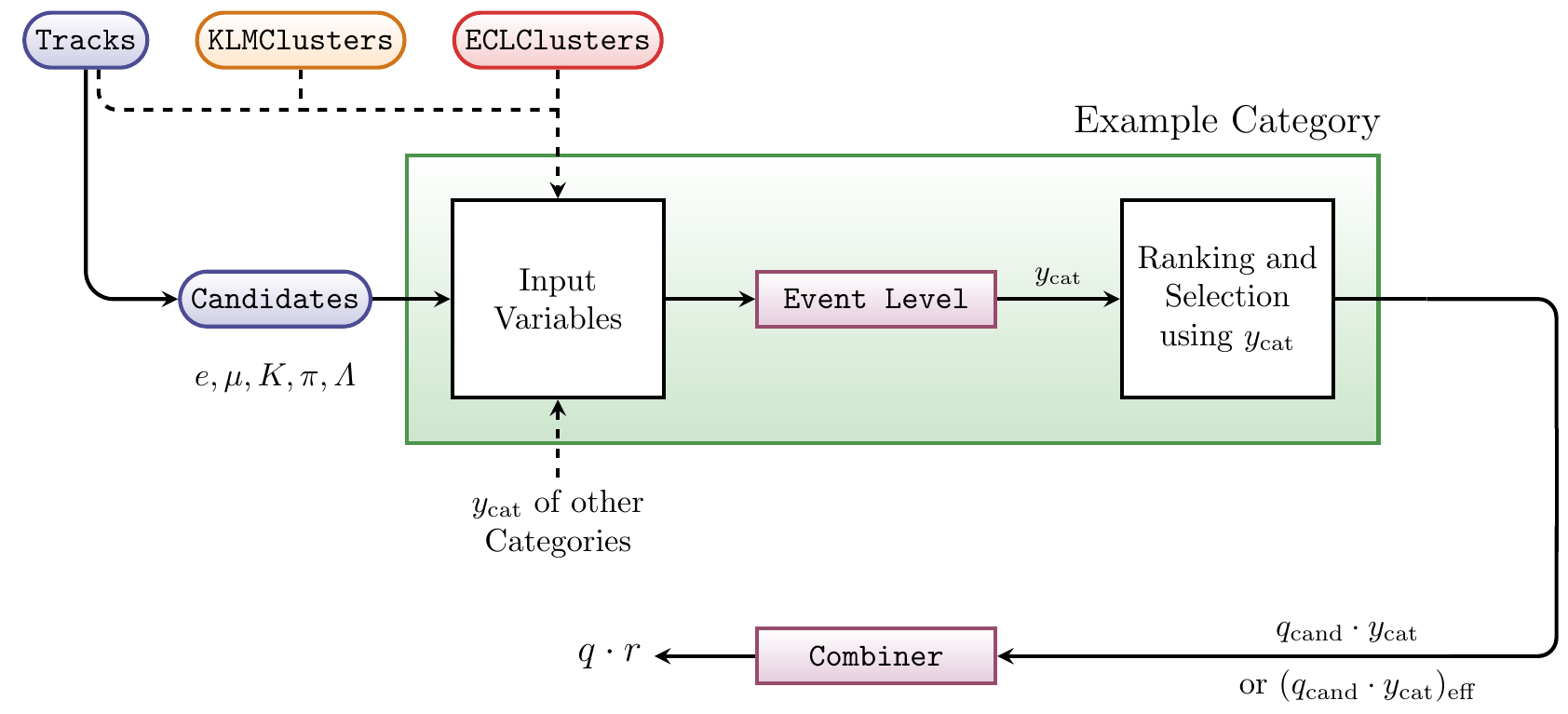}
\caption{Procedure for each single category (green box): the candidates correspond to the reconstructed tracks for a specific mass hypothesis. Some of the input variables consider all reconstructed tracks and all neutral ECL and KLM clusters on the tag side. The magenta boxes represent multivariate methods: $y_{\text{cat}}$ is the output of the event level. The output of the combiner is equivalent to the product $q\cdot r$. Each box corresponds to an FBDT classifier.}
  \label{fig:FlavTagSingleCat}
\end{figure*}

The event-level multivariate method assigns to each particle candidate a real-valued output \mbox{$y_\text{cat}\in[0, 1]$} corresponding to the probability of being the target of the
corresponding category providing the right flavor tag.
%To determine $y_\text{cat}$, the event-level multivariate methods
Within each category, the particle candidates are ranked according to the values of $y_\text{cat}$. The candidate with the highest $y_\text{cat}$ is identified as flavor-specific decay product. Figure~\ref{fig:FlavTagSingleCat} illustrates the procedure.
Only for the Maximum $p^*$ category, the candidates are ranked according to their momenta in the \PUpsilonFourS frame.  Two special categories get information from other categories: the Kaon-Pion category and the Fast-Slow-Correlated (FSC) category.

At the combiner level, the algorithm combines the information provided by all categories into the final product $q\cdot r$ using a combiner-level FBDT. Combining the information provided by all categories improves the performance of the flavor tagger as the \PB decays possibly offer more than one flavor-specific signature. 
The combiner receives an input from each category corresponding to the product $q_{\rm cand}\cdot y_{\rm cat}$, where $q_{\rm cand}$ is the charge (or $q_{\PLambda}$ flavor) of the candidate identified as flavor-specific decay product, and $y_{\rm cat}$ is the probability provided by the event-level FBDT. 
Only for the Kaon and the
Lambda categories the input is the effective product 
\begin{align*}
\label{eqn:qpProduct}
\small
(q_\text{cand}&\cdot y_\text{cat})_\text{eff} = 
 \frac{\prod_i\big(1 + \left(q_\text{cand}\cdot y_\text{cat}\right)_i\big) - \prod_i\big(1 - \left(q_\text{cand}\cdot y_\text{cat}\right)_i\big)}{\prod_i\big(1 + \left(q_\text{cand}\cdot y_\text{cat}\right)_i\big) + \prod_i\big(1 - \left(q_\text{cand}\cdot y_\text{cat}\right)_i\big)}\text{,} 
\end{align*}
where the products extend over the three particles with the highest $y_\text{cat}$ value. The use of $(q_\text{cand}\cdot y_\text{cat})_\text{eff}$ for the Kaon and the Lambda categories slightly improves the tagging performance. We find no significant improvement when we use it for the other categories. 

The structure of the FBDT classifiers, the learning procedure, and the preprocessing of the input data is controlled with different so-called \emph{hyper-parameters}. 
We use the default hyper-parameter values optimized for the Full-Event-Interpretation algorithm~\cite{Keck:2017mui,Keck:2018lcd}, 
which performs similarly complex classifications to identify \PB~mesons and other intermediate particles. 
The \emph{number of levels} in each tree is three; the \emph{number of cuts} for the cumulative probability histograms of each input variable is eight; 
the fraction of the sample to train each tree (\emph{sampling rate}) is 0.5; the \emph{learning rate} to regulate the training is 0.1. For the flavor tagger, only the \emph{number of trees} was optimized to $500$. 
For the training procedure, the FBDT algorithm transforms the distribution of the input variables to a uniform distribution and uses a negative binomial log-likelihood loss function. 

%\clearpage

The FBDT algorithm provides an internal ranking of input variables by counting how often the variables are used to split decision tree nodes and by weighting each split according to the separation gain and the number of events in the node~\cite{Keck:2017gsv, Keck:2017mui}. Based on this ranking~\cite{Abudinen:2018}, we generally observe that the input variables with largest separation power at the event level are the PID variables followed by the particle momenta. Variables requiring information from all tracks and neutral clusters, for example $M_\text{rec}^2$, $E_{90}^{\PW}$, and $\cos{\theta^*_\text{miss}}$, provide marginal additional separation power. At the combiner level, the categories with largest separation power are the Kaon and the Kinetic Lepton categories followed by the Maximum~$p^*$, Slow Pion, FSC, and Fast Hadron categories. The other categories provide marginal additional separation power.

\section{The deep-learning flavor tagger}

\label{sec:DNNTagger}

To explore the advantages of deep-learning multivariate methods, we developed a DNN flavor tagger based on a deep-learning multi-layer perceptron~(MLP). 
The algorithm is designed to learn the correlations between the characteristics of the tag-side tracks and the flavor of the tag-side \PB~meson avoiding any pre-selection of decay products. 
The algorithm provides as output the product $q\cdot r$.
The implementation of the algorithm is based on the machine-learning library Tensor-flow~\cite{Zadeh:2018mki}.   

\begin{figure}[hb]
\centering
    \includegraphics[width=0.6\textwidth]{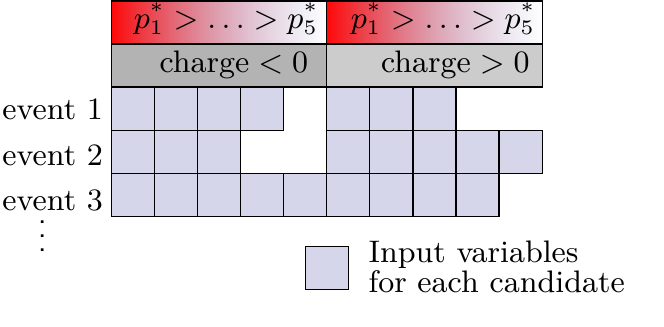}
\caption{\label{fig:DNNVarStr}Scheme of the input information for the MLP in the deep-learning flavor tagger. The tracks on the tag side are sorted into two groups according to their electric charge and ranked according to their momenta in the \PUpsilonFourS frame. The top $5$ tracks in each group are selected as candidates. The MLP receives input variables for each candidate. 
% If an event contains fewer than $5$ positive tracks or fewer than $5$ negative tracks, the input variables of the missing candidates are set to zero.
}
\end{figure}

The DNN flavor tagger sorts the tracks on the tag side into two groups, a positive and a negative one, depending on the electric charge of the particle~(see Fig.~\ref{fig:DNNVarStr}). 
The algorithm ranks the tracks in each group according to their momenta in the \PUpsilonFourS frame, and selects the top five tracks in each group. 
We find on average around six tag-side tracks per event in simulation with about equal fraction of positive and negative tracks. 
About $96\%$ of the events have fewer than ten tag-side tracks.
If an event contains fewer than five positive or fewer than five negative tracks, the algorithm sets the input variables for the missing candidates to zero.

For each charged-particle candidate, the
deep-learning MLP receives ten input variables in the current configuration: 
five PID likelihoods $\mathcal{L}_{\Pe}$, $\mathcal{L}_{\Pmu}$, $\mathcal{L}_{\PK}$, $\mathcal{L}_{\Ppi}$, and $\mathcal{L}_{\Pproton}$, 
the magnitude of the momentum $p^*$,  the cosine of the polar angle $\cos\theta^*$, the azimuth angle $\phi^*$, and the number of hits in the vertex detectors $N_{\rm PXD}$ and $N_{\rm SVD}$. 
Multiplying the number of input variables by the number of candidates yields $100$, corresponding to the number of input nodes. 

We optimize the hyper-parameters of the MLP by performing various scans. The optimized MLP contains eight \emph{hidden layers} with $300$ nodes each. 
Based on previous studies on similarly complex classification tasks~\cite{Baldi:2014kfa}, we employ the $\tanh$ function as \emph{activation function} to describe possible non-linear dependences between the inputs and the \PB-meson flavor. 
The inputs are transformed to be uniformly distributed in the range $[-1,1]$ corresponding to the range of the activation function. Zero, the default value for missing tracks, corresponds to the mean of the transformed distribution.

For the training procedure, we use a binary cross-entropy loss function with regularization terms~\cite{Gemmler:2020}. 
The loss function is minimized using a mini-batch
stochastic gradient-descent algorithm based on backpropagation~\cite{rumelhart1986learning}. 

% \clearpage

\section{Training procedure}

\label{sec:training}

Both flavor taggers are trained using $\PBzero\to\Pgngt\Pagngt$ samples~(see Sec.~\ref{sec:data}). In this way, we avoid possible bias due to \CP~asymmetries or reconstruction performance since these samples are generated without built-in \CP violation,  and all reconstructed objects~(tracks, photons, and KLM clusters) can be used to form the tag side without passing through reconstruction of the signal side. 

The algorithms are trained with a sample of about ten million MC~events and tested afterwards with an independent sample of the same size to exclude overtraining. We find no significant improvement in tagging performance using two to five times larger training samples. We train the algorithms for each MC campaign to optimize them for the most up-to-date data processing and background expectation. 

For the category-based algorithm, the training sample is divided into two statistically independent MC~samples of the same size: one sample for the event level, and one sample for the combiner level. The event level is trained first and each category is trained independently. The combiner is trained afterwards.

For the DNN algorithm, we take about $10\%$ of the training sample as an independent validation sample. We monitor the training procedure by calculating the value of the loss function on the validation sample at each training epoch and stop the training procedure if the value starts increasing for a fixed number of 100 epochs. We then save the MLP configuration at the epoch leading to the best performance on the validation sample. Typically, the training is stopped after about 500 epochs. We train 10 different MLPs with different initial random weights and keep only the one leading to the best performance. 
% However, we find no significant difference for different initial random weights.

Over-fitting is checked for each of the multivariate methods in both flavor taggers by comparing the distribution of the output on the training sample with the output on the testing sample.
The output on the training and on the testing sample have to be statistically compatible.

For the DNN tagger, the MLP complexity calls for significant computing resources to train the algorithm. We use GPUs to train the deep-learning MLP to exploit their parallel computation capabilities.
On a GTX~970 GPU~\cite{GPUGTX}, the training procedure for the eight-layer~MLP takes about $48$ hours. In comparison, the training procedure for the category-based flavor tagger takes about five hours running on a single~CPU core.

\begin{figure}[hb]
\centering
    \includegraphics[width=0.8\textwidth]{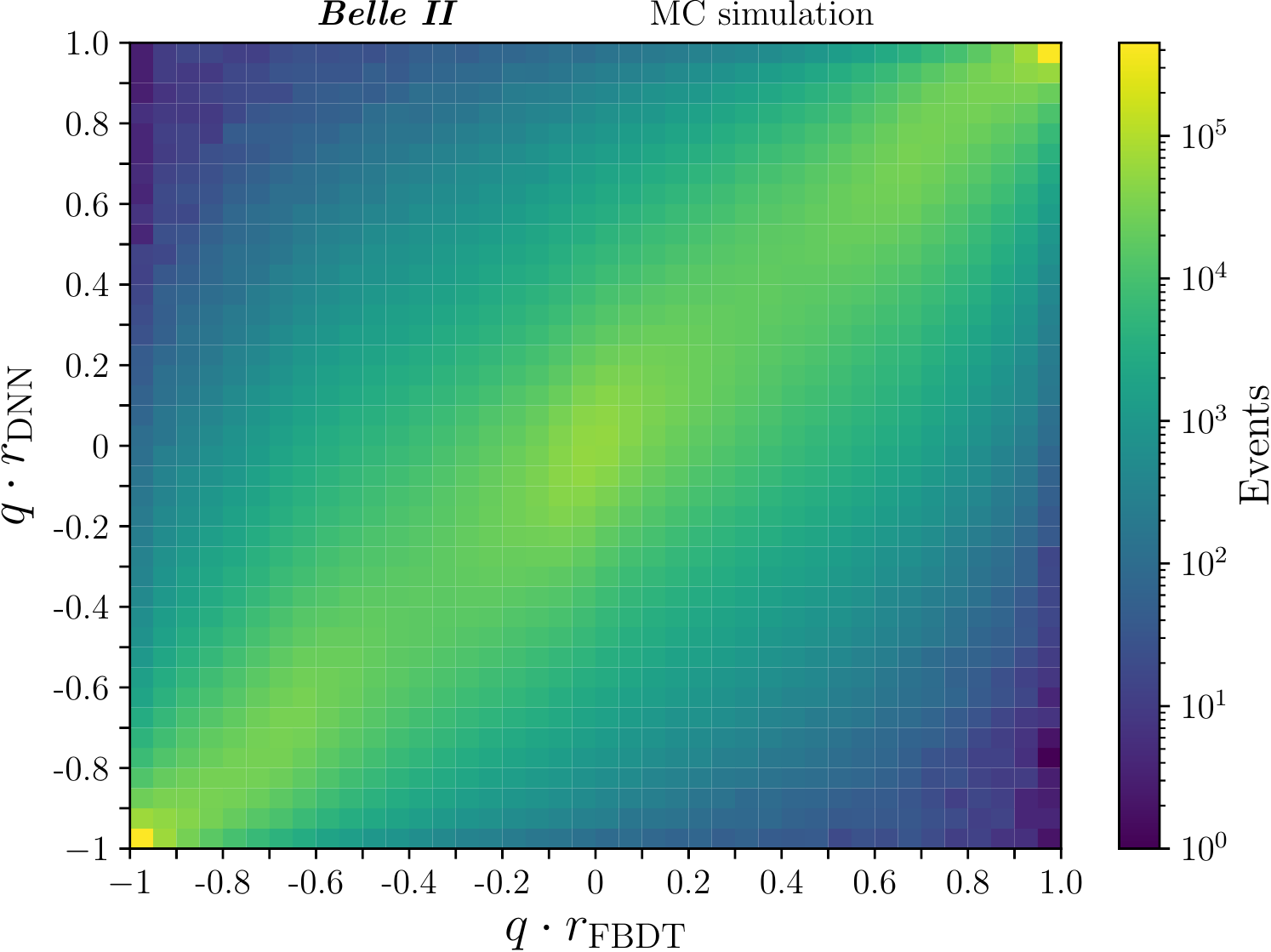}
\caption{\label{fig:FBDTDNNscatter} Distribution of the DNN tagger output vs. the category-based~(FBDT) tagger output in the testing \mbox{$\PBzero\to\Pgngt\Pagngt$} simulation sample. 
% If an event contains fewer than $5$ positive tracks or fewer than $5$ negative tracks, the input variables of the missing candidates are set to zero.
}
\end{figure}

\begin{figure*}[htb]
 \raggedright
 \includegraphics[width=0.475\textwidth]{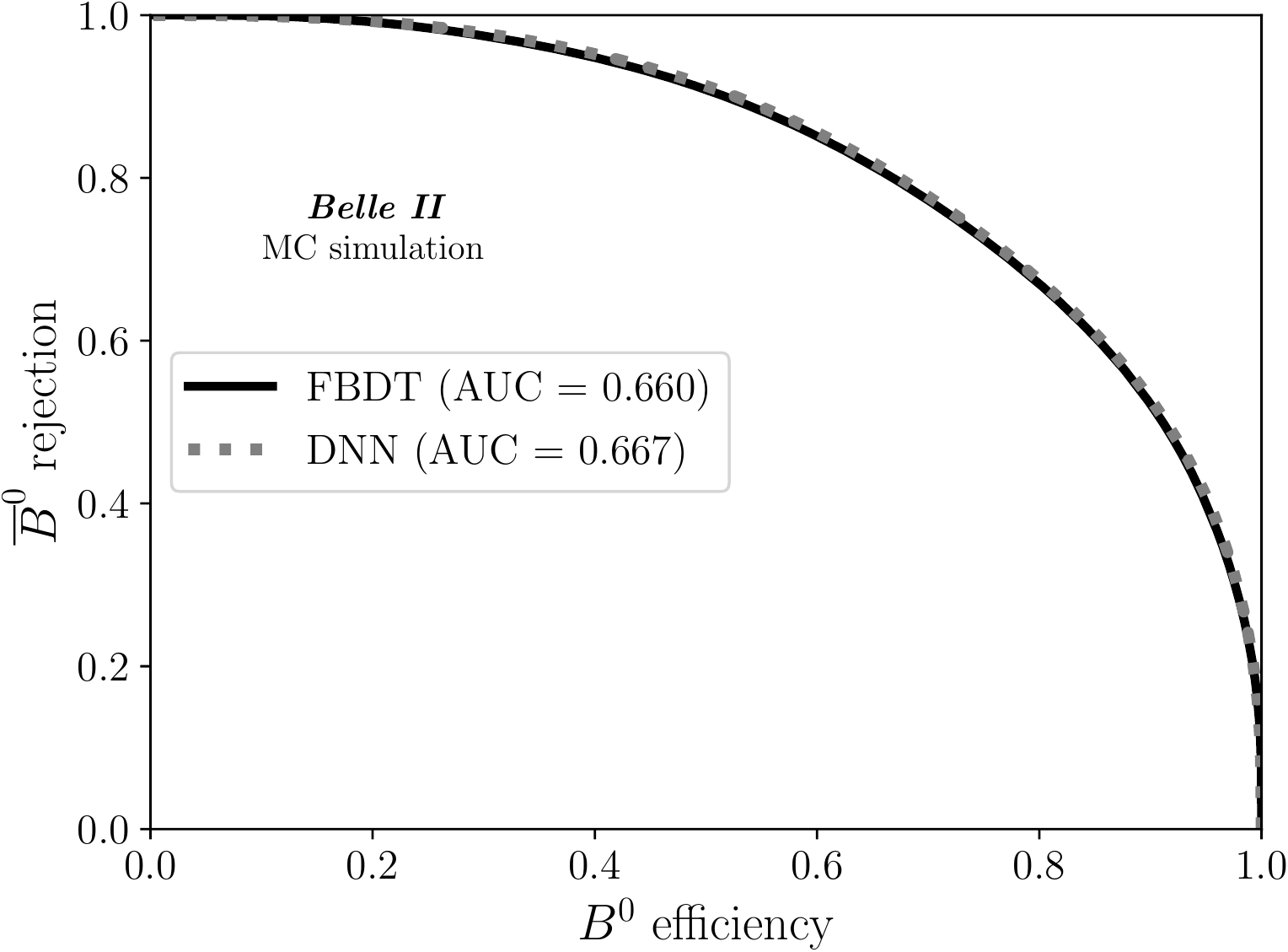}\hfill
 \includegraphics[width=0.475\textwidth]{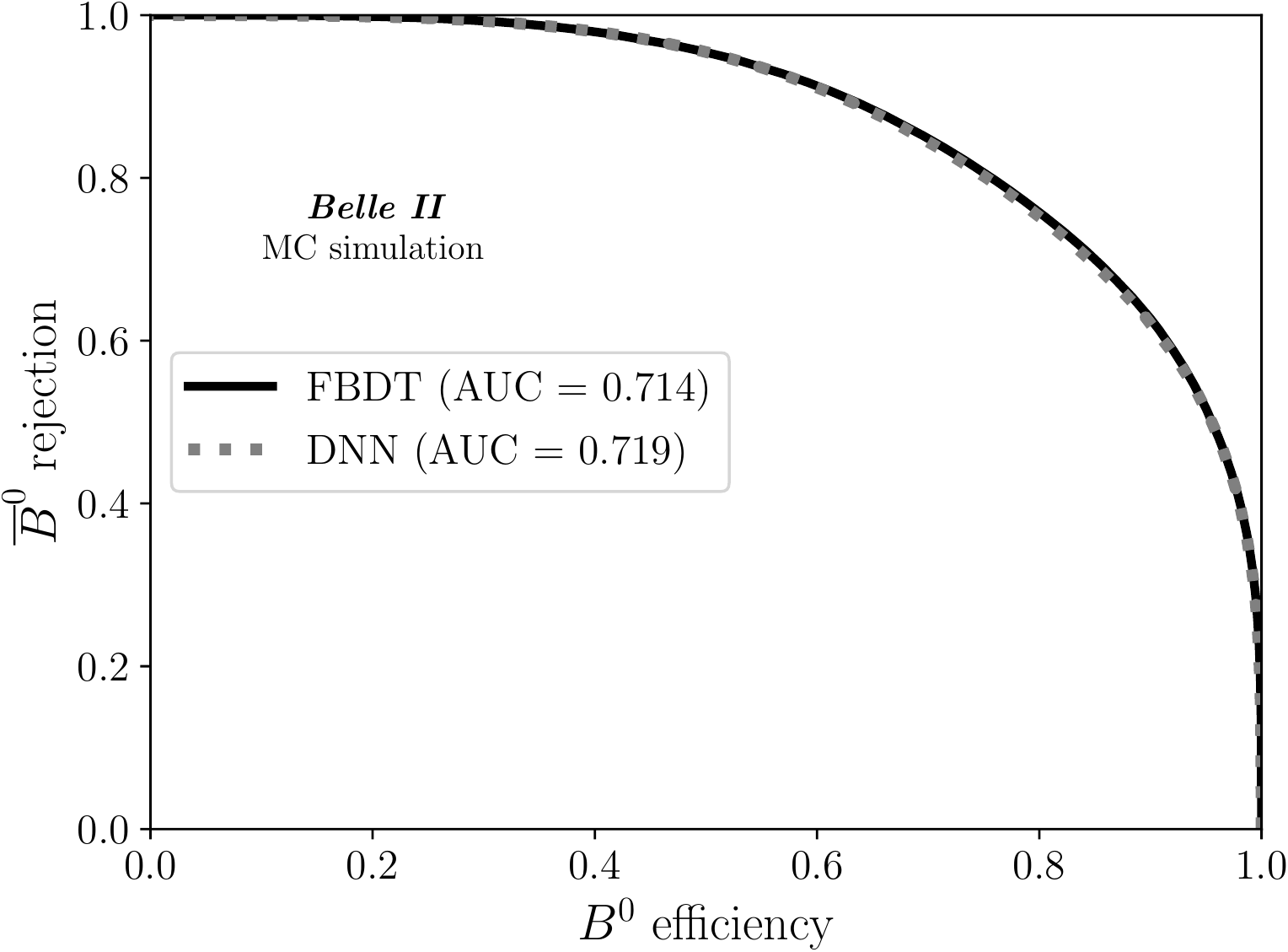}\\\vspace{0.3cm} 
 
 \text{ \hspace{2.5cm} (a) All events  \hspace{4.5cm} (d) Events with less than 5 tracks}\\\vspace{0.5cm}
 
 \includegraphics[width=0.475\textwidth]{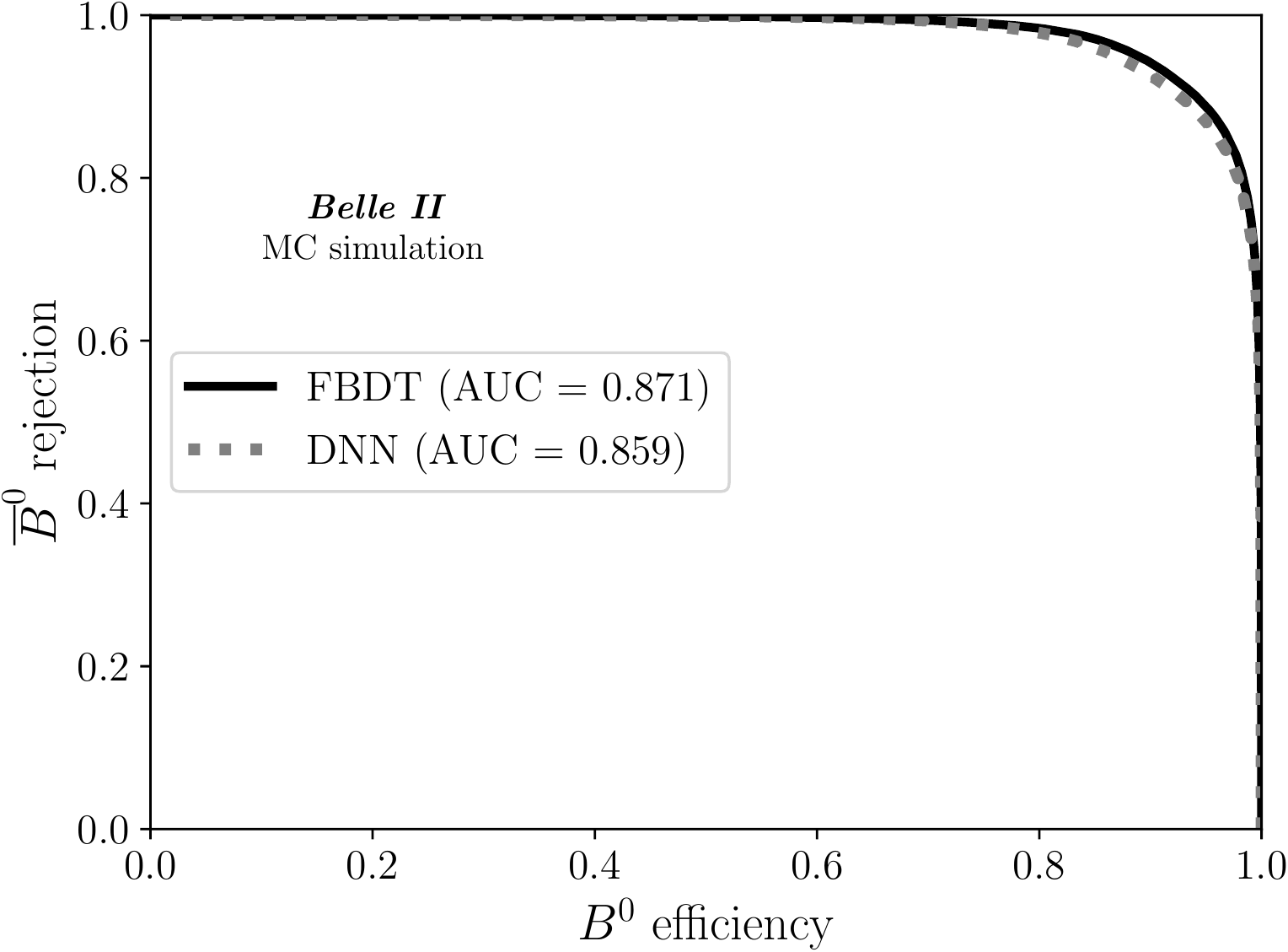}\hfill
 \includegraphics[width=0.475\textwidth]{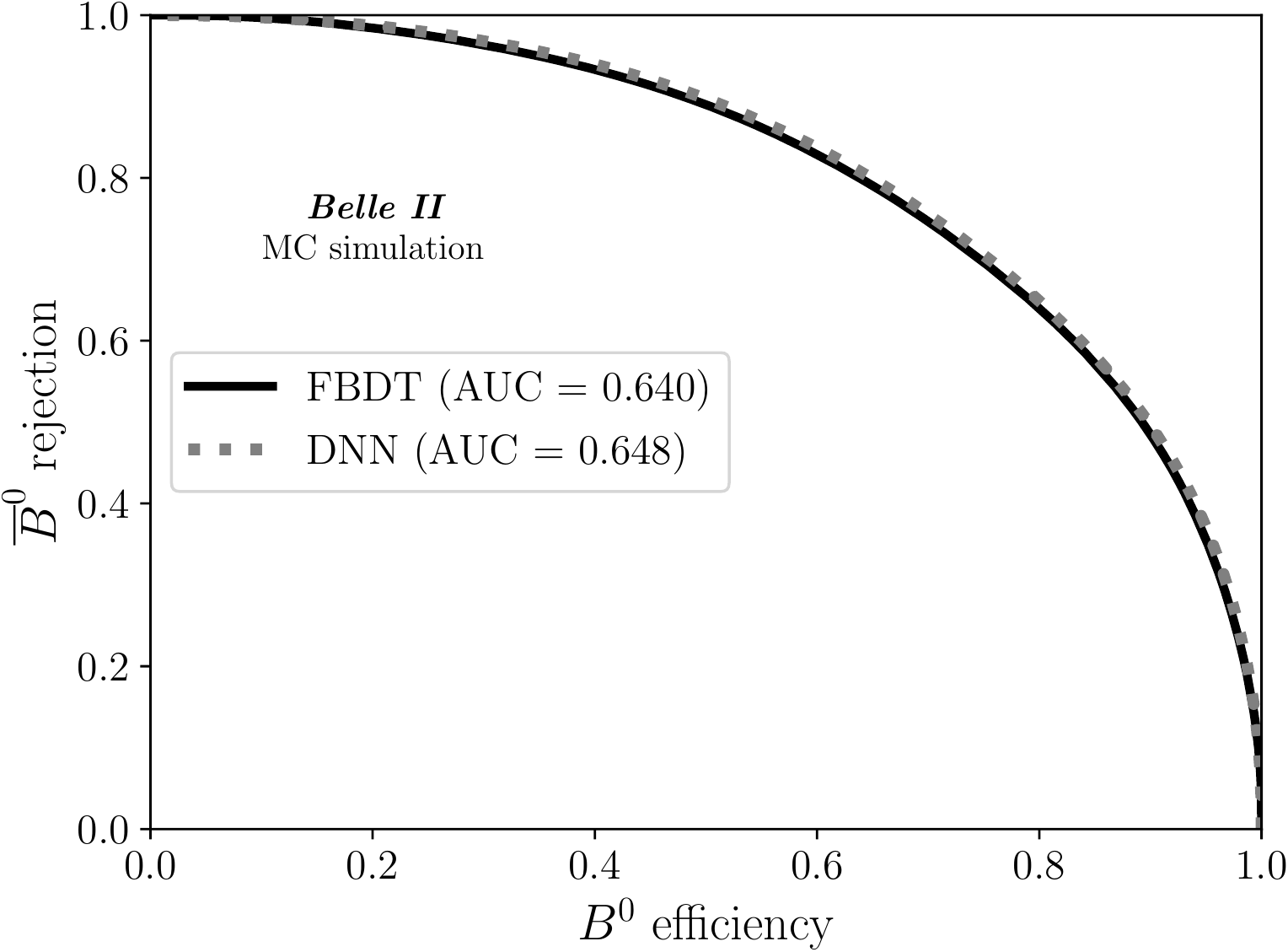}\\
 \vspace{0.3cm} 
 
  \text{ \hspace{0.5cm} (b) Events with one Kinetic Lepton target \hspace{2.1cm} (e) Events with 5 to 10 tracks}\\\vspace{0.5cm}
 
  \includegraphics[width=0.475\textwidth]{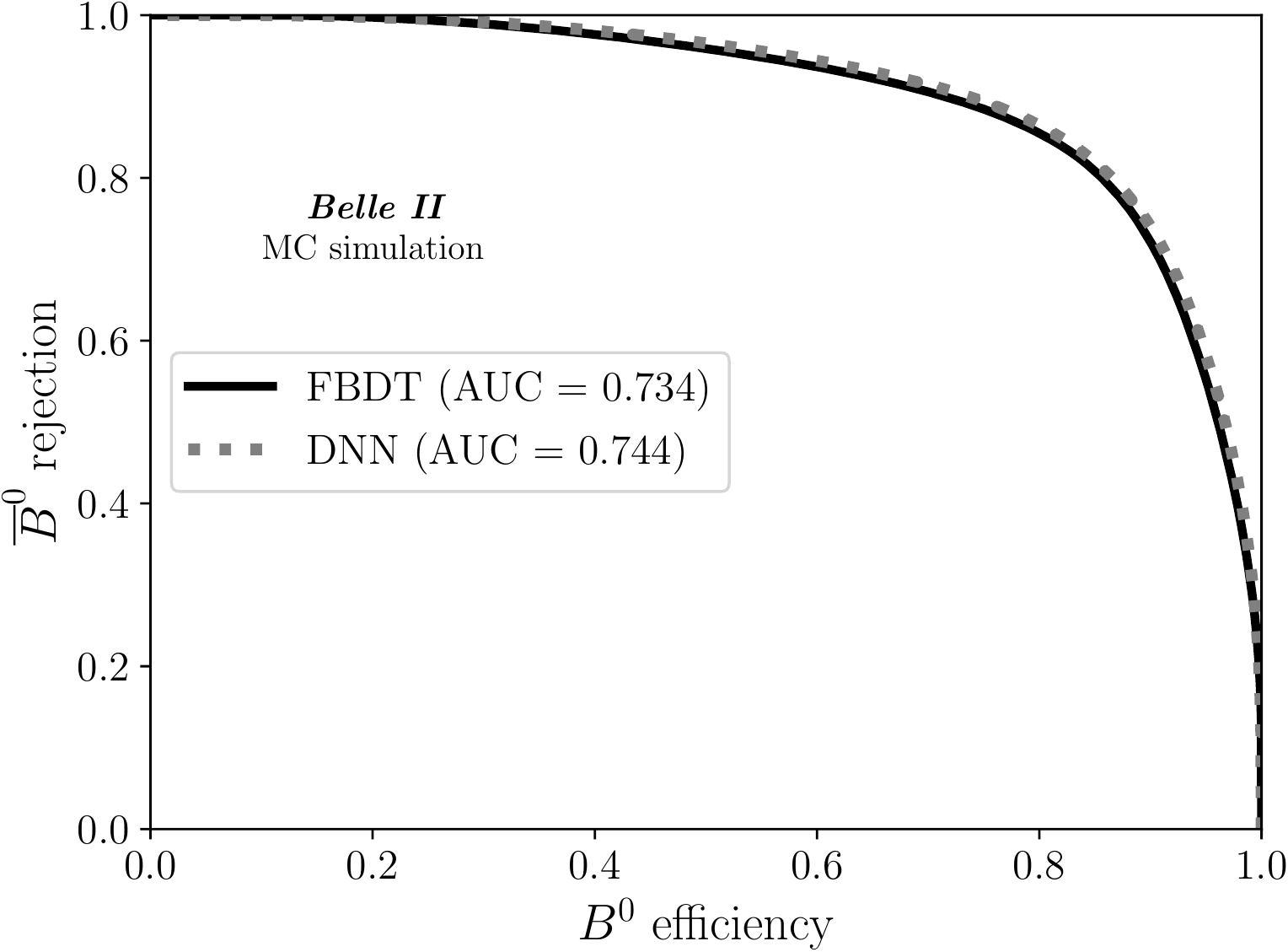}\hfill
 \includegraphics[width=0.475\textwidth]{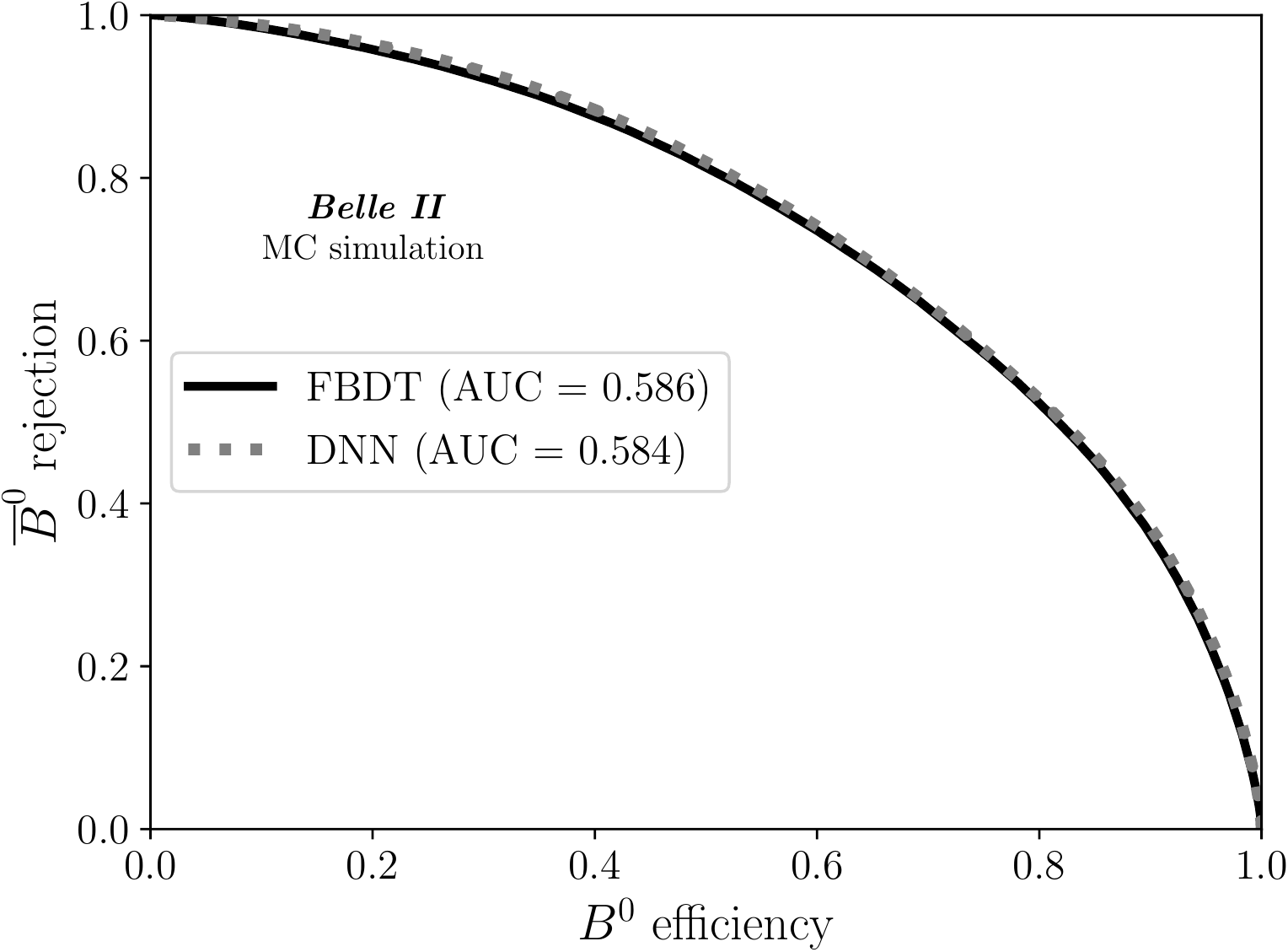}\\
 
   \text{ \hspace{1.1cm} (c) Events with one Kaon target \hspace{2.5cm} (f) Events with more than 10 tracks}\\\vspace{0.1cm}
 
 \caption{Receiver operating characteristic curves comparing the performance of the DNN tagger and the category-based~(FBDT) tagger in the testing $\PBzero\to\Pgngt\Pagngt$ simulation sample. Curves and values of area under the curve~(AUC) are shown for (left) all events, events with one target particle of the Kinetic Lepton category and events with one target particle of the Kaon category, and (right) events with less than 5 tracks, with 5 to 10 tracks and with more than 10 tracks.}
 \label{fig:ROCsFBDTDNN}
\end{figure*}

We compare the performance of both flavor taggers using the testing  $\PBzero\to\Pgngt\Pagngt$ sample.
Figure~\ref{fig:FBDTDNNscatter} shows the 2D distribution of the DNN output vs. the combiner FBDT output. From the sample we estimate a Pearson correlation coefficient around $90\%$. Figure~\ref{fig:ROCsFBDTDNN} shows the receiver operating characteristics~(ROC) and the area under the ROC curve~(AUC) for all events, for events containing a target particle of the Kinetic Lepton or Kaon categories, and for events containing less than 5~tracks, 5 to 10~tracks, and more than 10~tracks. The category-based tagger reaches a slightly better performance for events with a target of the Kinetic Lepton category and for events with more than ten tracks. On the other hand, the DNN tagger reaches a slightly better performance for events with one target of the Kaon category and for events with less than 10 tracks. However, in general, both algorithms reach about the same performance for all events and for the various sub-samples.        

After the training, we perform checks using signal-only MC samples, where the signal $\PB$~meson decays to one benchmark mode such as 
{$\PBzero\to\PKp\Pgpm$},   
{\small$\PBzero\to\PJpsi(\to\Pgmp\Pgmm)\,\PKzS( \to\Pgpp\Pgpm)$}, 
\mbox{\small$\PBzero\to\Peta^\prime(\to\Pgpp\Pgpm$\break
$\Peta(\to\Pgpp\Pgpm\Pgpz))\,\PKzS(\to\Pgpp\Pgpm)$}, or one of the neutral \PB~decays listed in the following section. We reconstruct the signal \PB~decay in each event and use the tag-side objects as input for the flavor taggers. For correctly associated MC~events, we verify that the tagging performance is consistent with the one obtained using the $\PBzero\to\Pgngt\Pagngt$ sample.

\vspace{1cm}

\clearpage

\section{Reconstruction of calibration samples}

\label{sec:reco}

To evaluate the performance of the flavor taggers, we reconstruct the following signal \PB~decays,
\begin{center}
\begin{tabular}{r l r l}
$\bullet$ & $\PBplus \to \APDzero\, \Pgpp$,  &
\hspace{0.2cm} $\bullet$ & $\PBzero \to \PDminus\, \Pgpp$,\\
$\bullet$ & $\PBplus \to \APDzero\, \Prhoplus$, &
\hspace{0.2cm} $\bullet$ & $\PBzero \to \PDminus\, \Prhoplus$, \\
$\bullet$ & $\PBplus \to \APD^{*0}(\to\APDzero\,\Pgpz)\, \Pgpp$,  &
\hspace{0.2cm} $\bullet$ & $\PBzero \to \PD^{*-}(\to\APDzero\,\Pgpm)\, \Pgpp$,\\
$\bullet$ & $\PBplus \to \APD^{*0}(\to\APDzero\,\Pgpz)\, \Prhoplus$, &
\hspace{0.2cm} $\bullet$ & $\PBzero \to \PD^{*-}(\to\APDzero\,\Pgpm)\, \Prhoplus$, \\
$\bullet$ & $\PBplus \to \APD^{*0}(\to\APDzero\,\Pgpz)\,  \Pa_1^+$, &
\hspace{0.2cm} $\bullet$ & $\PBzero \to \PD^{*-}(\to\APDzero\,\Pgpm)\, \Pa_1^+$, \\
\end{tabular}\vspace{0.5cm}
\end{center}
for which we reconstruct the following \PD~decays,
\begin{center}
\begin{tabular}{r l r l}
$\bullet$ & $\APDzero\to\PKplus\Pgpm$, \hspace{1.5cm}  &
\hspace{0.3cm} $\bullet$ & $\PDminus\to\PKp\Pgpm\Pgpm$, \hspace{1.4cm} \\
$\bullet$ & $\APDzero\to\PKplus\Pgpm\Pgpp\Pgpm$, &
\hspace{0.3cm} $\bullet$ & $\PDminus\to\PKzS\,\Pgpm$, \\
$\bullet$ & $\APDzero\to\PKplus\Pgpm\Pgpz$, &
\hspace{0.3cm} $\bullet$ & $\PDminus\to\PKzS\,\Pgpm\Pgpz$,\\
$\bullet$ & $\APDzero\to\PKzS\,\Pgpp\Pgpm$, &
\hspace{0.3cm} $\bullet$ & $\PDminus\to\PKp\Pgpm\Pgpm\Pgpz$. \\
\end{tabular}    
\end{center}

\subsection{Reconstruction and baseline selection}

We reconstruct charged-pion and charged-kaon candidates by starting from the most inclusive charged-particle selections. To reduce the background from tracks that do not originate from the interaction region, we require fiducial criteria that restrict the candidates to loose ranges of displacement from the nominal interaction point
($|dr|<\SI{0.5}{cm}$ radial and $|dz|<\SI{3}{cm}$ longitudinal) and to the full polar-acceptance in the central drift chamber~\mbox{($\SI{17}{\degree}<\theta<\SI{150}{\degree}$)}. Additionally, we use PID information to identify kaon candidates by requiring the likelihood $\mathcal{L}_{\PK}$ to be larger than $0.4$.

We reconstruct neutral pion candidates by requiring photons to exceed energies of $\SI{80}{MeV}$ in the forward region, $\SI{30}{MeV}$ in the central volume, and $\SI{60}{MeV}$ in the backward region. We restrict the diphoton mass to be in the range \mbox{$\SI{120} <\; M(\gamma\gamma) < \SI{145}{MeV}/c^2$}. The mass of the $\pi^0$ candidates is constrained to its known value~\cite{Zyla:2020zbs} in subsequent kinematic fits. 

For $K_{\rm S}^0$ reconstruction, we use pairs of oppositely charged particles that originate from a common decay vertex and have a dipion mass in the range 
\mbox{$\SI{450} <\; M(\pi^+\pi^-) < \SI{550}{MeV}/c^2$}. 
To reduce combinatorial background, we apply additional requirements, dependent on $K_{\rm S}^0$~momentum, on the distance between trajectories of the two charged-pion candidates, the $K^0_{\rm S}$ flight distance, and the angle between the pion-pair momentum and the $K^0_{\rm S}$~flight direction.

The resulting $K^\pm$, $\pi^\pm$, $\pi^0$, and $\PKzS$ candidates are combined to form  $\PD^{(*)}$~candidates in the various final states, by requiring their invariant masses to satisfy
\begin{itemize}
\item[$\bullet$] $\SI{1.84} <\; M\big(\small\PKp\Pgpm,\, \PKp\Pgpm\Pgpp\Pgpm,\,\PKp\Pgpm\Pgpz,\, \PKzS\,\Pgpp\Pgpm\big) < \SI{1.89}{GeV}/c^2$,
\item[$\bullet$] $\SI{1.844} <\; M\big(\PKp\Pgpm\Pgpm,\, \PKzS\,\Pgpm,\,\PKzS\,\Pgpm\Pgpz,\, \PKp\Pgpm\Pgpm\Pgpz\big) < \SI{1.894}{GeV}/c^2$,
\item[$\bullet$] $\SI{0.14} <\; M(\PDzero\Pgpz) - M(\PDzero) < \SI{0.144}{GeV}/c^2$,
\item[$\bullet$] $\SI{0.143} <\; M(\PDzero\Pgpp) - M(\PDzero)< \SI{0.147}{GeV}/c^2$,
\end{itemize}
where $M(\PDzero)$ and $M(\PDzero\Pgpz,\,\PDzero\Pgpp)$ are the invariant masses of the reconstructed \PDzero and \PDstar candidates.
We reconstruct $\Prhopm$ candidates from pairs of charged and neutral pions, and $\Pa_1^{\pm}$~candidates from three charged pions with the following requirements:
\begin{itemize}
\item[$\bullet$] $ \vert  M(\Pgpp\Pgpz) - M_{\Prho} \vert < \SI{0.15}{GeV}/c^2$,
\item[$\bullet$]  $ \vert M(\Pgpp\Pgpm\Pgpp) - M_{\Pa_1}\vert < \SI{0.4}{GeV}/c^2$,
\end{itemize}
where $M_{\Prho}$ and $M_{\Pa_1}$ are the known masses~\cite{Zyla:2020zbs} of the $\Prho$~and $\Pa_1$~mesons.
To identify primary~$\Pgppm$ and $\Pgppm$ candidates used to reconstruct $\Prhopm$ and $\Pa_1^{\pm}$~candidates, we also require the likelihood $\mathcal{L}_{\Pgp}$ to be larger than $0.1$ and the $\Pgppm$~momentum in the $\PUpsilonFourS$ frame to be larger than $\SI{0.2}{GeV}/c$.

To reconstruct the signal $\PB$~candidates, we combine the $\PD^{(*)}$ candidates with appropriate additional candidate
particles, $\Pgppm$, $\Prhopm$ or $\Pa_1^{\pm}$, by performing simultaneous kinematic-vertex fits of the entire decay chain~\cite{Belle-IIanalysissoftwareGroup:2019dlq} into each of our signal channels. 
We perform the kinematic-vertex fits without constraining the decay-vertex position or the invariant mass of the decaying particles and require the fit to converge. Requiring the kinematic-vertex fit to converge keeps about $96\%$ of the correctly associated MC~events. 

We use the following kinematic variables to distinguish $\PB$~signals from the dominant continuum background from $\Pep\Pem\to \Pqu\Paqu$, $\Pqd\Paqd$, $\Pqs\Paqs$, and $\Pqc\Paqc$~processes:
\begin{itemize}
    \item[$\bullet$] $M_{\rm bc} \equiv \sqrt{s/(4c^4) - (p^{*}_B/c)^2}$, the beam-energy constrained mass, which is the invariant mass of the \PB~candidate calculated with the \PB~energy replaced by half the collision energy $\sqrt{s}$, which is more precisely known; 
    \vspace{0.2cm}
    
    \item[$\bullet$] $\Delta E \equiv E^{*}_{B} - \sqrt{s}/2$, the energy difference  between the energy $E^{*}_{B}$ of the reconstructed \PB~candidate and half of the collision energy, both measured in the $\PUpsilonFourS$ frame.   
\end{itemize}
We retain $\PB$~candidates that have~$M_{\rm bc} > \SI{5.27}{GeV}/c^2$ and $\vert \Delta E\vert < \SI{0.12}{GeV}$. 
Additionally, for channels with \Prhopm candidates, we remove combinatorial background from soft $\Pgpz$~mesons collinear with the \Prhopm, by requiring the cosine of the helicity angle $\theta_{\rm H}$ between the \PB and the \Pgpp~momenta in the \Prho frame to satisfy \mbox{ $\cos{\theta_{\rm H}}< 0.8$}. 

We form the tag side of the signal $\PB$~candidates using all remaining tracks and photons that fulfill the loose fiducial criteria, and KLM clusters. 
The category-based and the DNN taggers receive the tag-side objects and run independently of each other.

\subsection{Continuum suppression and final selection}

To suppress continuum background, we apply requirements on the two topological variables with the highest discrimination power between signal from hadronic \PB~decays and continuum background: 
$\cos{\theta_{\rm T}^{\rm sig, tag}}$, the cosine of the angle between the thrust axis of the signal~$\PB$~(reconstructed) and the thrust axis of the tag-side \PB~(remaining tracks and clusters), and $R_2$, the ratio between the second and zeroth Fox-Wolfram moments~\cite{Fox:1978vw} calculated using the full event information.

We vary the selections on  $\cos{\theta_{\rm T}^{\rm sig, tag}}$ and $R_2$ to maximize the figure of merit ${\rm S}/\sqrt{{\rm S}+{\rm B}}$, 
where ${\rm S}$ and ${\rm B}$ are the number of signal and 
background \PB~candidates
in the signal-enriched range \mbox{$M_{\rm bc}>5.27\,\si{GeV}/c^2$} and $\SI{-0.12} < \Delta E <\SI{0.09}{GeV}$.
Both $\cos{\theta_{\rm T}^{\rm sig, tag}}$ and $R_2$ requirements are optimized simultaneously using simulation.  
We optimize the requirements for charged and for neutral candidates independently. The optimized requirements are found to be $\cos{\theta_{\rm T}^{\rm sig, tag}} < 0.87$ and $ R_2 < 0.43$ for charged $\PB$~candidates, and $\cos{\theta_{\rm T}^{\rm sig, tag}} < 0.95$ and $ R_2 < 0.35$ for neutral $\PB$~candidates.

Applying the optimized $R_2$ and $\cos{\theta_{\rm T}^{\rm sig, tag}}$ requirements keeps about $81\%$ of the charged signal \PB~candidates and about $77\%$ of the neutral ones, and improves the figure of merit by about $12\%$ for charged \PB~candidates and by about $14\%$ for neutral ones. We observe no significant difference in tagging performance before and after the $R_2$ and $\cos{\theta_{\rm T}^{\rm sig, tag}}$ requirements. 

After applying the $\cos{\theta_{\rm T}^{\rm sig, tag}}$ and $R_2$ requirements, more than one candidate per event populate the resulting $\Delta E$~distributions, with average multiplicities for the various channels ranging from 1.00 to 3.00 (about $75\%$ of the channels have multiplicities between 1.00 and 2.00). We select a single $\PB$~candidate per event by selecting the one with the highest $p$-value of the kinematic-vertex fit.  
The analyses of charged and neutral $\PB$~channels are independent: we select one candidate among the charged and one among the neutral channels independently.

\section{Determination of efficiencies and wrong-tag fractions}

\label{sec:fit}

The tagging efficiency $\varepsilon$ corresponds to the fraction of events to which a flavor tag can be assigned. Since the category-based and the DNN algorithms need only one charged track on the tag side to provide a tag, the tagging efficiency is close to $100\%$ for both, with good consistency between data and simulation as Table~\ref{tab:tagging_efficiencies} shows. 

\begin{table}[h!]
    \centering
    \caption{Tagging efficiencies $\varepsilon\pm\delta\varepsilon$ for charged and neutral \mbox{$\PB\to\PD^{(*)}h^{+}$}~candidates in data and in simulation. All values are given in percent. The uncertainties are only statistical.}
    \begin{tabular}{l c c}\hline\hline
     Channel & MC & Data \\ \hline
      $\PBzero\to\PD^{(*)-}h^{+}$    &  $99.77 \pm 0.01$ & $99.75 \pm 0.02$ \\
      $\PBplus\to\APD^{(*)0}h^{+}$   &  $99.80 \pm 0.01$ & $99.73 \pm 0.02$ \\      
      \hline
    \end{tabular}
    \label{tab:tagging_efficiencies}
\end{table}{}

To estimate the fraction of wrongly tagged events $w$, we fit the time-integrated $\PBzero-\APBzero$ mixing probability to the data. We take into account that $\varepsilon$ and $w$ can be slightly different for $\PBzero$ and $\APBzero$~mesons due to charge-asymmetries in detection and reconstruction. We express $\varepsilon$ and $w$ as
\begin{equation*}
\hspace{1.45cm} \varepsilon=\ \frac{\varepsilon_{\PBzero} + \varepsilon_{\APBzero}}{2}\text{,}\qquad\enskip w =\, \frac{w_{\PBzero} + w_{\APBzero}}{2}\text{,}
\label{eqs:AvrEpsAvrW}
\end{equation*}
and introduce the differences 
\begin{equation*}
\hspace{1.2cm} \Delta\varepsilon =\ \varepsilon_{\PBzero} - \varepsilon_{\APBzero}\text{,}\qquad  \Delta w =\ w_{\PBzero} - w_{\APBzero}\text{,}
\label{eqs:DeltaEpsDeltaW}
\end{equation*}
where the subscript corresponds to the true flavor, for example $w_{\PBzero}$ is the fraction of true $\PBzero$ mesons that are wrongly classified as $\APBzero$.

 For neutral $\PB\APB$~pairs produced at the \mbox{$\PUpsilonFourS$}, 
 the time-integrated probability for an event with
 a signal \PB~flavor \mbox{$q_{\rm sig}\in\{-1,+1\}$}
 and tag-side \PB~flavor
 \mbox{$q_{\rm tag}\in\{-1,+1\}$} is given by
\begin{equation*}
\phantom{10pt}\mathcal{P}(q_\text{sig}, q_\text{tag})=\frac{1}{2}\bigg[1 -q_\text{sig}\cdot q_\text{tag}\cdot (1-2\cdot\chi_d)\bigg]\text{,}
 \label{eqn:PQMix}
 \end{equation*}
 where $\chi_d$ is the time-integrated $\PBzero-\APBzero$ mixing 
 probability, whose current world average is
 \mbox{$\chi_d = 0.1858 \pm 0.0011$}~\cite{Amhis:2016xyh}. 
 The equation above assumes that for any event the signal and the tag-side \PB~flavor are correctly identified. 
 To include the effect of the flavor tagging algorithms, 
 one can express the observed probability 
 $\mathcal{P}(q_\text{sig}, q_\text{tag})^\text{obs}$ in terms of the efficiencies $\varepsilon_{\PBzero}$ and $\varepsilon_{\APBzero}$, and the wrong tag fractions $w_{\PBzero}$ and 
$w_{\APBzero}$. 
The probability becomes
\begin{align*}
\small\mathcal{P}^\text{obs}(q_\text{sig}, q_\text{tag} = +1) =\,
&\varepsilon_{\PBzero} (1-w_{\PBzero})\cdot \mathcal{P}(q_\text{sig}, q_\text{tag} = +1) + \varepsilon_{\APBzero}w_{\APBzero}\cdot \mathcal{P}(q_\text{sig}, q_\text{tag} = -1)\text{,} \nonumber \\ 
\mathcal{P}^\text{obs}(q_\text{sig}, q_\text{tag} = -1) =\,
&\varepsilon_{\APBzero}(1-w_{\APBzero})\cdot\mathcal{P}(q_\text{sig}, q_\text{tag} = -1) + \varepsilon_{\PBzero}w_{\PBzero}\cdot\mathcal{P}(q_\text{sig}, q_\text{tag} = +1)\text{,}  
\end{align*}
which can be written in terms of $\varepsilon$, $w$, $\mu = \Delta \varepsilon/(2\varepsilon)$ and $\Delta w$ as
{\small
 \begin{align}
\mathcal{P}(q_\text{sig}, q_\text{tag})^\text{obs}=
\frac{1}{2}\varepsilon\bigg[1 &-
q_\text{tag}\cdot\Delta w 
+ q_\text{tag}\cdot\mu\cdot(1-2w) \nonumber\\
&-\big[q_\text{tag}\cdot(1-2w) + \mu\cdot(1-q_\text{tag}\cdot\Delta w)\big]
\cdot q_\text{sig}\cdot(1-2\cdot\chi_d)\bigg]\text{.}
\label{eqn:PQMixFlavT}
\end{align}}

We sort the events in bins of the dilution factor $r$ provided by the flavor taggers and measure the value of $\varepsilon$, $w$, $\mu$, and $\Delta w$ in each $r$~bin~(7~bins in total). To compare with our predecessor experiment, we use the binning introduced by Belle~\cite{Kakuno:2004cf,Bevan:2014iga}. 

Since we need to consider the background, we develop a statistical model with a signal and a background component. We determine the signal yield~$N_{\rm sig}$, the background yield~$N_{\rm bkg}$, the partial efficiencies~$\varepsilon_i$, the wrong-tag fractions~$w_i$, and the asymmetries $\mu_i$ and $\Delta w_i$ in each $r$-bin $i$ from an extended maximum likelihood fit to the unbinned distributions of~$\Delta E$, $q_{\rm sig}$, and $q_{\rm tag}$. We check that the $\Delta E$ distribution is statistically independent from those of $q_{\rm sig}$ and $q_{\rm tag}$ with Pearson correlation coefficients below $2\%$. 

In the fit model, the probability density
func-
tion~(PDF) for each component~$j$ is given by
\begin{equation*}
   \phantom{10pt} \mathcal{P}_j(\Delta E, q_\text{sig}, q_\text{tag}) \equiv \mathcal{P}_j(\Delta E)\cdot \mathcal{P}_j^\text{obs}(q_\text{sig}, q_\text{tag})\text{.} 
\end{equation*}
We model the signal $\Delta E$ PDF using a Gaussian plus a Crystal Ball function~\cite{Skwarnicki:1986xj} determined empirically using signal MC~events obtained  from the generic simulation~(see Sec.~\ref{sec:data}), with the additional flexibility of a global shift of peak position and a global scaling factor for the width as suggested by a likelihood-ratio test. The background $\Delta E$ PDF is modeled using an exponential function with a floating exponent. Residual peaking backgrounds in generic simulation have expected yields below $0.5\%$ of the signal one and are thus neglected. 

The flavor PDF $\mathcal{P}(q_\text{sig}, q_\text{tag})^\text{obs}$ has the same form for signal and background~(Eq.~\ref{eqn:PQMixFlavT}) with independent $\varepsilon_i$,  $w_i$, $\Delta w_i$, $\mu_i$, and $\chi_d$ parameters for signal and background. We fix the background $\chi_d^{\rm bkg}$ parameter to zero as we obtain values compatible with zero when we let it float. We find, on the other hand, that the background parameters $\varepsilon_i^{\rm bkg}$,  $w_i^{\rm bkg}$, $\Delta w_i^{\rm bkg}$, and $\mu_i^{\rm bkg}$ have to be free to obtain unbiased results for the signal ones.

The total extended likelihood is given by
\begin{equation*}
\mathcal{L} \enskip \equiv \prod_i \enskip \frac{\ee^{-\sum_j N_j\cdot \varepsilon_i}}{N^i!} \prod_{k=1}^{N^i} \enskip\sum_{j} N_j \cdot \mathcal{P}_j^i(\Delta E^k, q_\text{sig}^k, q_\text{tag}^k)\text{,}
\end{equation*}
where $i$ extends over the $r$~bins, $k$ extends over the events in the $r$~bin~$i$, and $j$ over the two components: signal and background.  The PDFs for the different components have no common parameters. Here, $N_j$ denotes the yield for the component $j$, and $N^i$ denotes the total number of events in the $i$-th $r$~bin. The partial efficiencies~$\varepsilon_i$ are included in the flavor part of $\mathcal{P}_j$.  Since we can fit only to events with flavor information, the sum of all $\varepsilon_i$ must be one. We therefore replace the epsilon for the first bin (with lowest $r$) with 
\begin{equation*}
  \hspace{2.9cm}  \varepsilon_{1} = 1 - \sum_{i=2}^{7}\varepsilon_{i}\text{,}
    \label{eq:epsilon1}
\end{equation*} 
and obtain its uncertainty $\delta\varepsilon_1$ from the width of the residuals of simplified simulated experiments. 

\begin{figure*}[htb]
 \centering
 \includegraphics[width=0.475\textwidth]{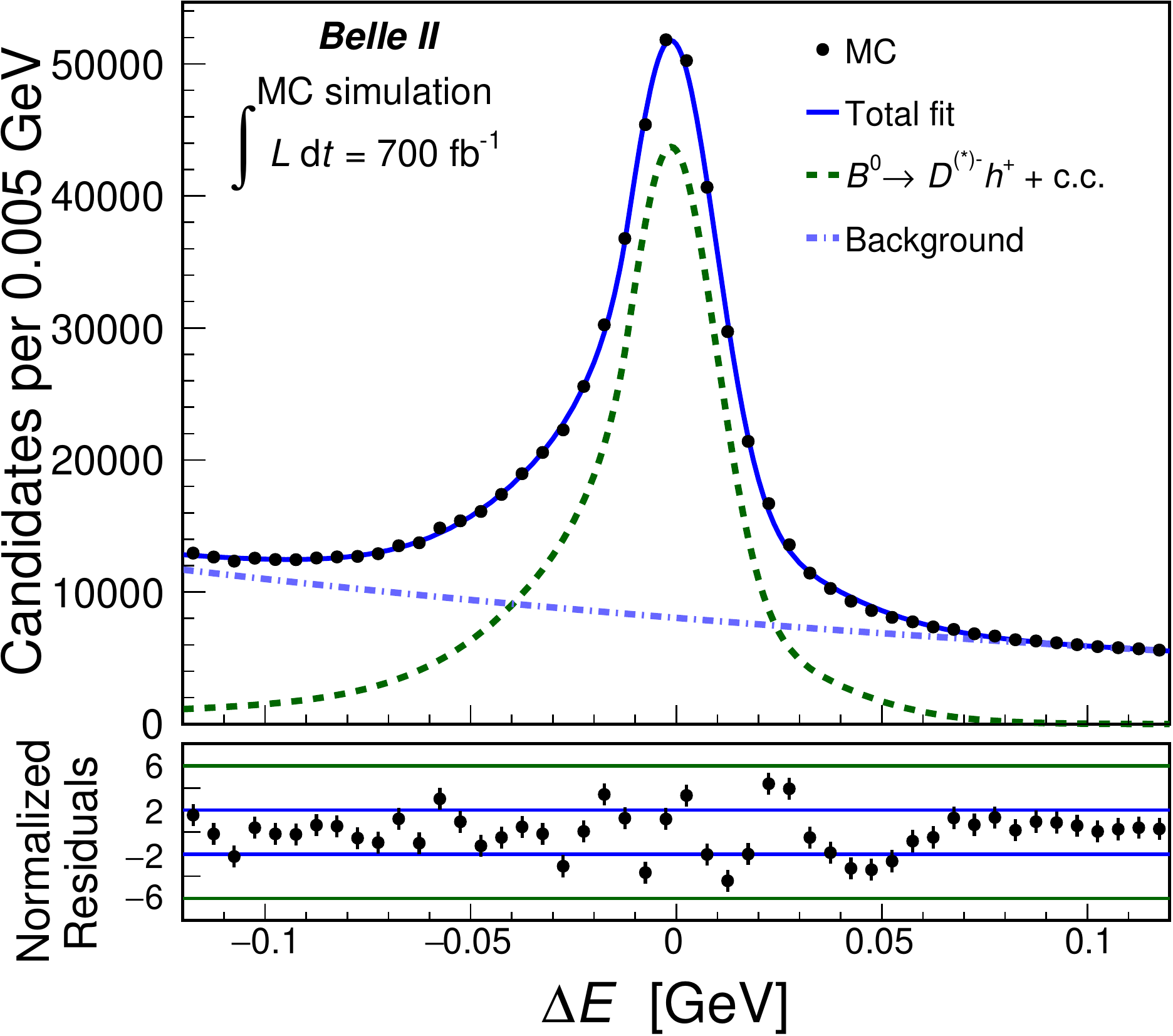}\hfill
 \includegraphics[width=0.475\textwidth]{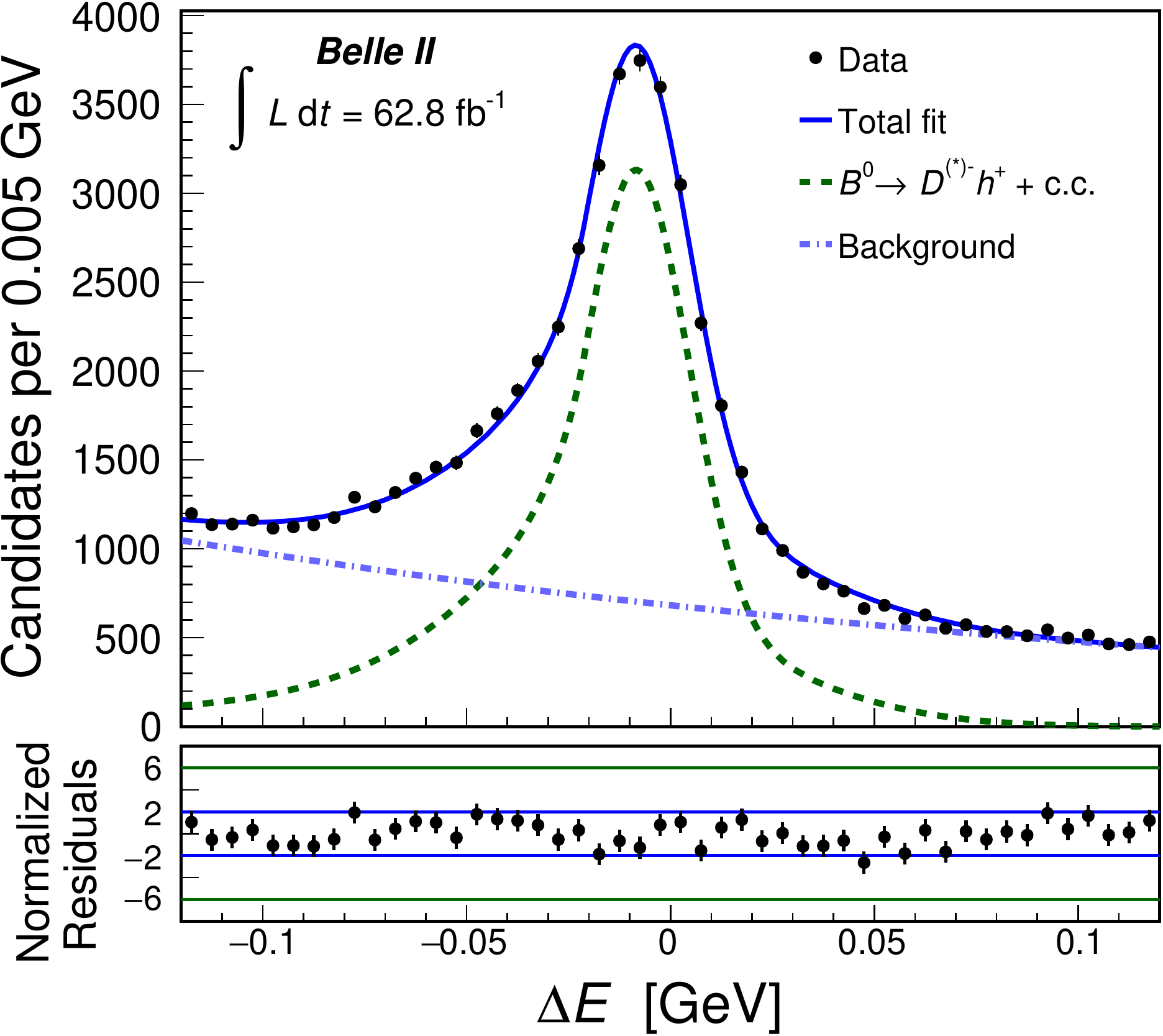}\\
    
\vspace{0.3cm} \text{(a) $\PBzero\to\PD^{(*)-}h^{+}$.}\\\vspace{1.0cm}
 
  \includegraphics[width=0.475\textwidth]{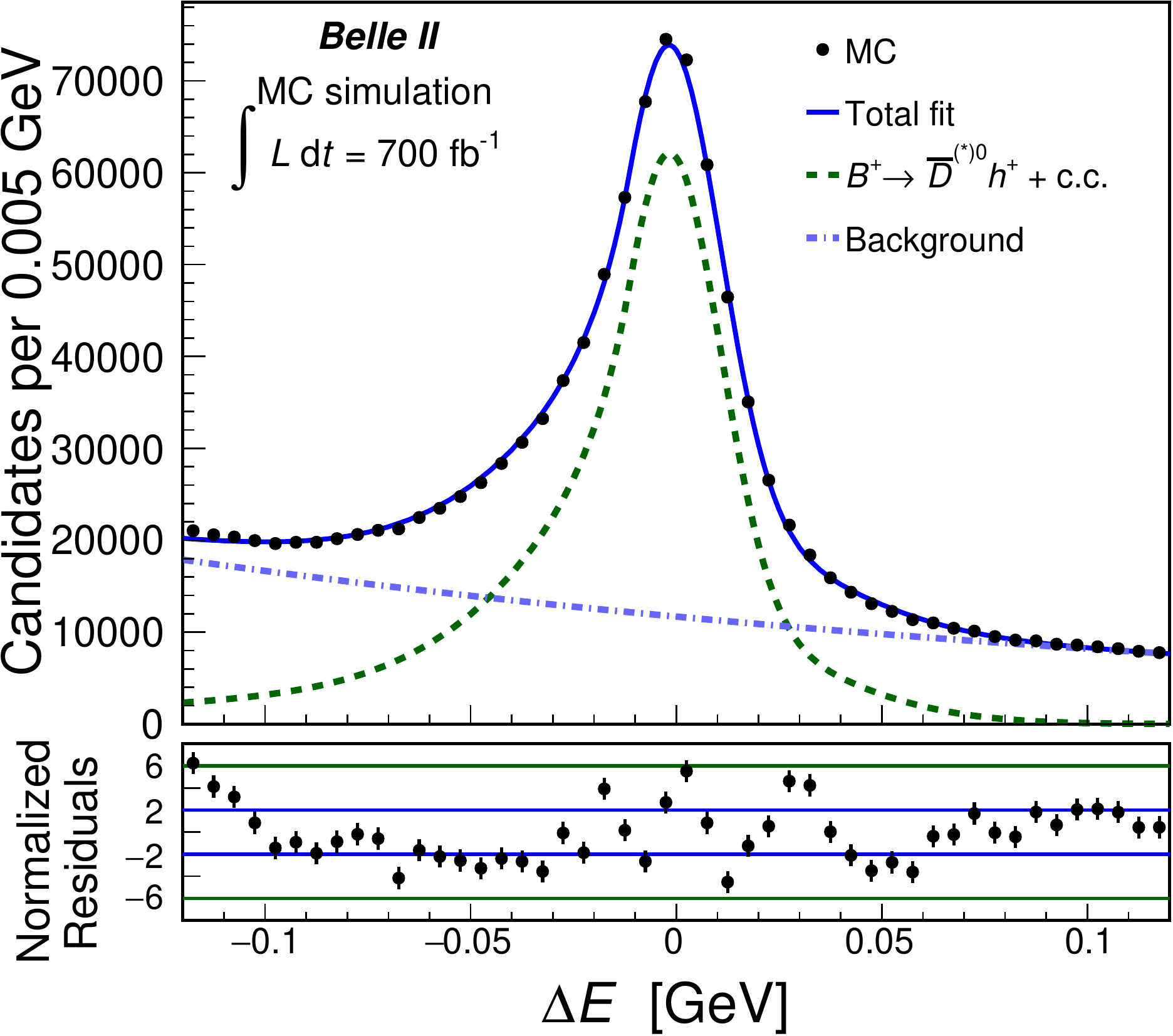}\hfill
 \includegraphics[width=0.475\textwidth]{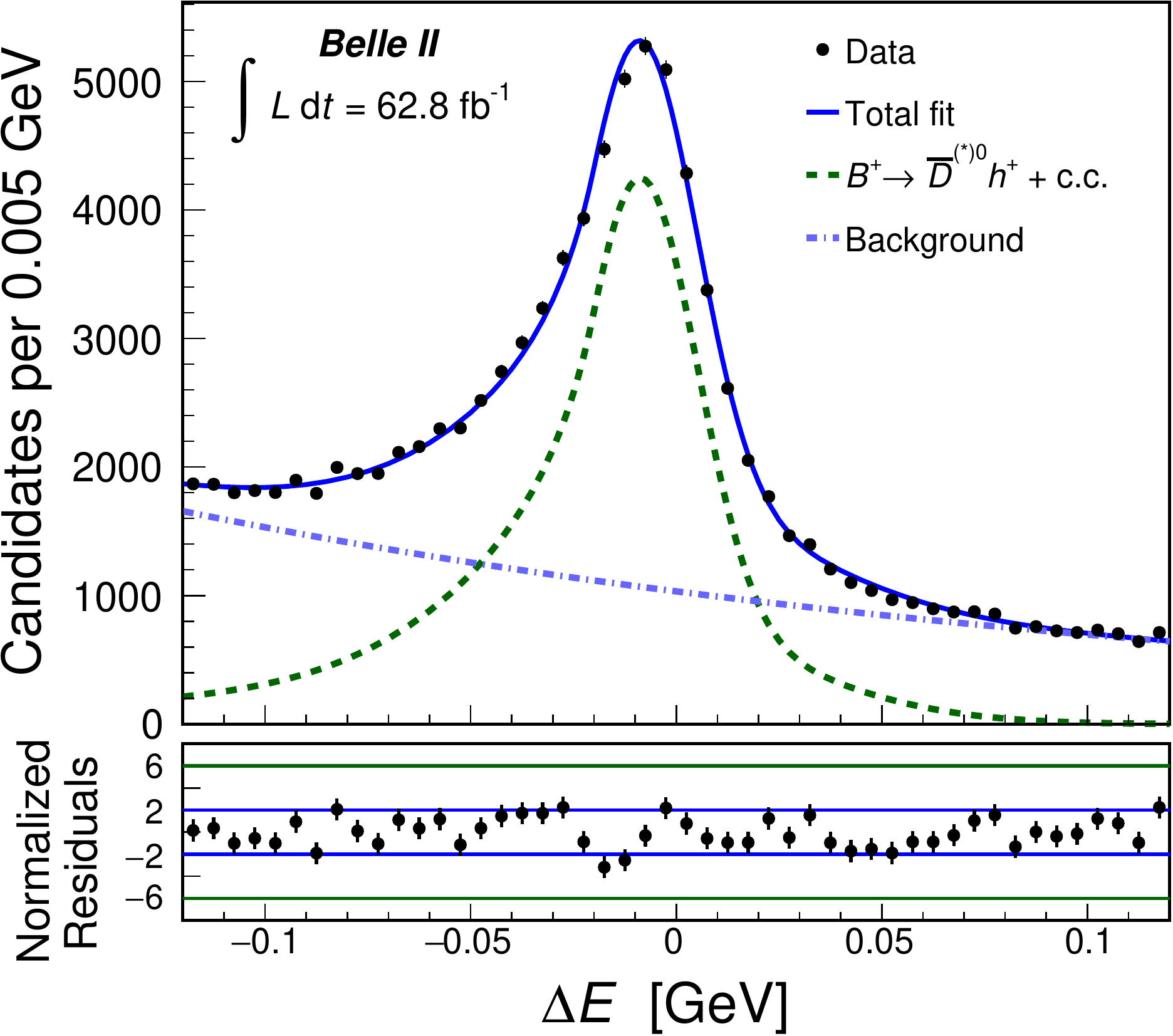}
 
 \vspace{0.3cm} \text{(b) $\PBplus\to\APD^{(*)0}h^{+}$.}\\\vspace{0.5cm} 
 
 \caption{Distributions of $\Delta E$ for (top)~neutral and (bottom)~charged $\PB\to\PD^{(*)}\Ph^{+}$ candidates reconstructed in (left)~simulation and (right)~data, restricted to \mbox{$M_{\rm bc} > 5.27$\,GeV/$c^2$}. The fit projection of the maximum likelihood fit is overlaid. 
 }
 \label{fig:fit_dE_unbinned}
\end{figure*}

 To validate the $\Delta E$ model, we first perform an extended maximum likelihood fit to the unbinned distribution of $\Delta E$ (without flavor part) in simulation and data.
 Figure~\ref{fig:fit_dE_unbinned} shows the $\Delta E$ fit projections in data and simulation for charged and neutral $\PB\to\PD^{(*)}h^{+}$ candidates. Table~\ref{tab:yield_summary} summarizes the yields obtained from the fits. We observe a relatively good agreement between data and simulation, but a tendency to lower yields with respect to the expectation, especially for charged signal \PB~candidates. 
 
To determine the partial efficiencies~$\varepsilon_i$ and the wrong-tag fractions~$w_i$, we perform a fit of the full model in a single step. 
%There are in total $31$ free parameters. 
For neutral candidates, we constrain the value of the signal $\chi_d^{\rm sig}$~parameter via a Gaussian constraint,
\begin{equation*}
   \hspace{2.1cm}  \mathcal{L}\; \Rightarrow\; {\rm G}(\chi_d^{\rm sig} - \chi_d,\, \delta\chi_d)\cdot\mathcal{L} \text{,}  
\end{equation*}
where $\chi_d$ and $\delta \chi_d$ are the central value and the uncertainty of the world average. For charged \PB~mesons,  $\chi_d$ is zero since there is no flavor mixing. % due to electric charge conservation. 

\vspace{1cm}

\clearpage

\begin{table}[!hb]
    \centering
    \caption{Summary of yields and yields per integrated luminosity obtained from the fit to MC~simulation and  data. The uncertainties are only statistical.}     
\begin{tabular}{l@{\hskip 5pt}  r@{\hskip 5pt}  r@{\hskip 5pt}  r@{\hskip 5pt}  r@{\hskip 5pt} }
\multicolumn{5}{l}{ $\PBzero\to\PD^{(*)-}h^{+}$}\\
\hline\hline
\multicolumn{1}{c}{} & 
\multicolumn{2}{c}{Yield} &   
\multicolumn{2}{c}{Yield/$\si{fb^{-1}}$} \\\hline
& \multicolumn{1}{c}{MC} & \multicolumn{1}{c}{Data} & \multicolumn{1}{c}{MC} & \multicolumn{1}{c}{Data} \\\hline
   Signal      &	$375105 \pm  1084$ &	$31423 \pm  360$ &	$536 \pm  2$ &	$ 500 \pm  6$ \\
   Background   &	$397275 \pm  1094$ &	$33764 \pm  364$ &	$568 \pm  2$ &	$ 538 \pm  6$ \\\hline\\
\multicolumn{5}{l}{ $\PBplus\to\APD^{(*)0}h^{+}$}   \\
 \hline\hline
\multicolumn{1}{c}{} & 
\multicolumn{2}{c}{Yield} &   
\multicolumn{2}{c}{Yield/$\si{fb^{-1}}$} \\\hline 
  & \multicolumn{1}{c}{MC} & \multicolumn{1}{c}{Data} & \multicolumn{1}{c}{MC} & \multicolumn{1}{c}{Data} \\\hline
   Signal      &	$610304 \pm  1461$ &	$ 46530 \pm  482$   &	$ 872 \pm  2$ & 	$  741 \pm  8$ \\
   Background   &	$587618 \pm  1453$ &	$ 51420 \pm  487$  &	$840 \pm  2$ & 	$ 819 \pm  8$ \\ 
\hline
\end{tabular}
    \label{tab:yield_summary}
\end{table}

\section{Comparison of performance in data and simulation}

\label{sec:splot}

We check the agreement between data and MC distributions of the flavor-tagger output by performing an $s\mathcal{P}lot$~\cite{Pivk:2004ty} analysis using $\Delta E$ as the control variable. We determine $s\mathcal{P}lot$ weights using the $\Delta E$~fit model introduced in the previous section. We weight the data with the $s\mathcal{P}lot$ weights to obtain the individual distributions of the signal and background components in data and compare them with MC~simulation. We normalize the simulated samples by scaling the total number of events to those observed in data. The procedure is validated by performing the $s\mathcal{P}lot$~analysis using MC~simulation and verifying that the obtained signal and background distributions correspond to the distributions obtained using the MC~truth.   

\begin{figure*}
\centering
\includegraphics[width=0.475\linewidth]{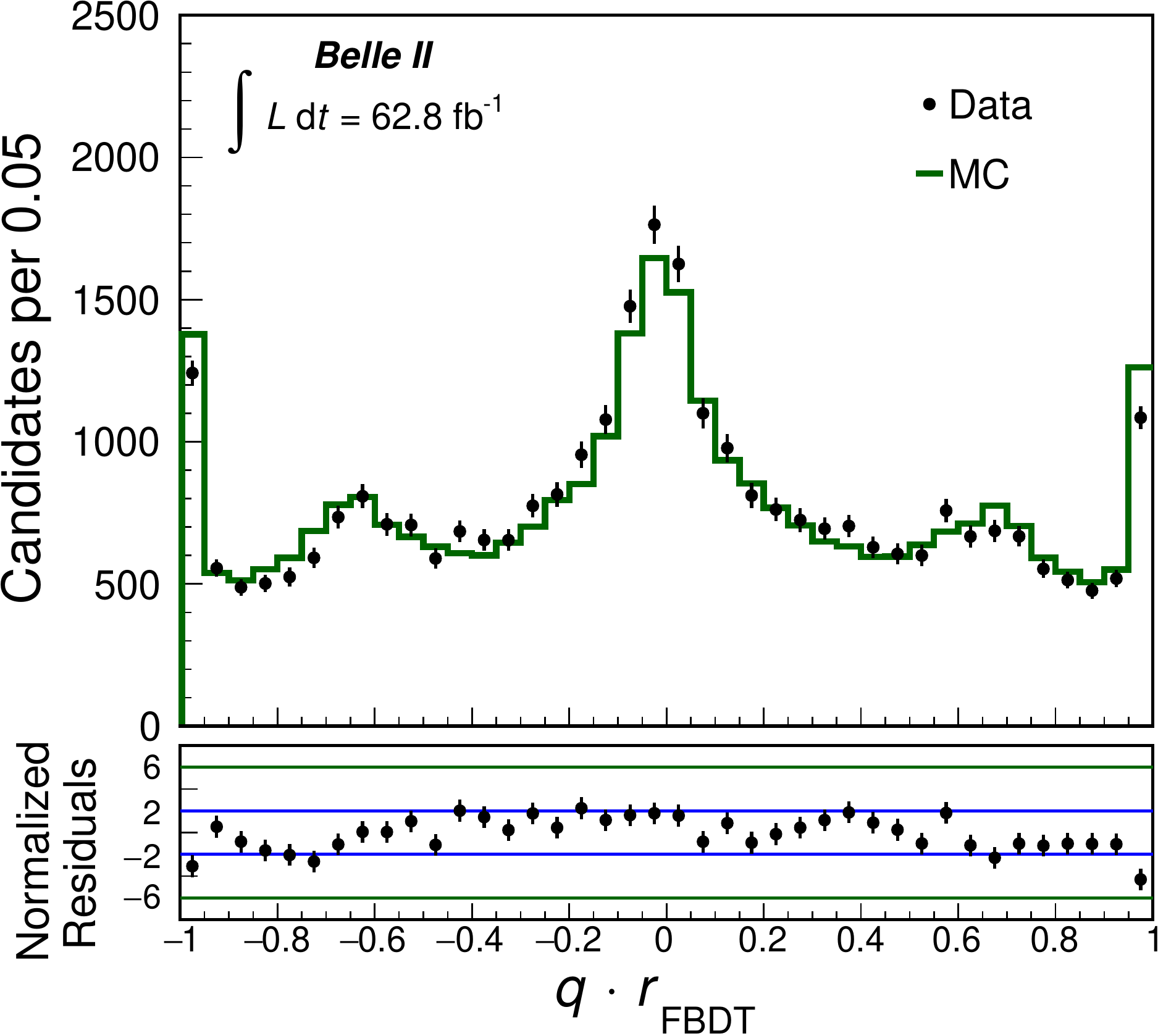}\hfill
\includegraphics[width=0.475\linewidth]{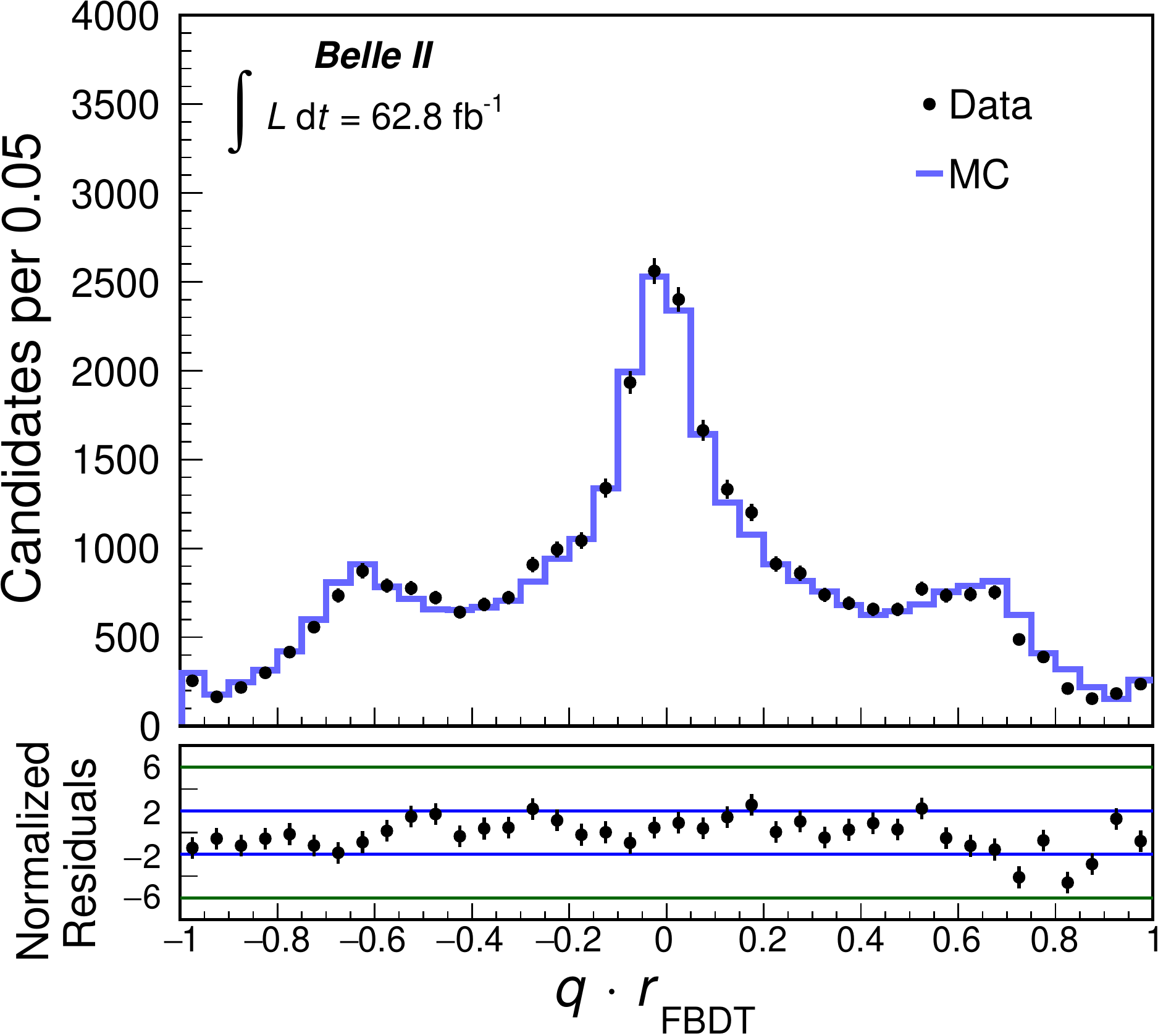}\\
    
\vspace{0.3cm} \text{(a) $\PBzero\to\PD^{(*)-}h^{+}$.}\\\vspace{1.0cm}
        
\includegraphics[width=0.475\linewidth]{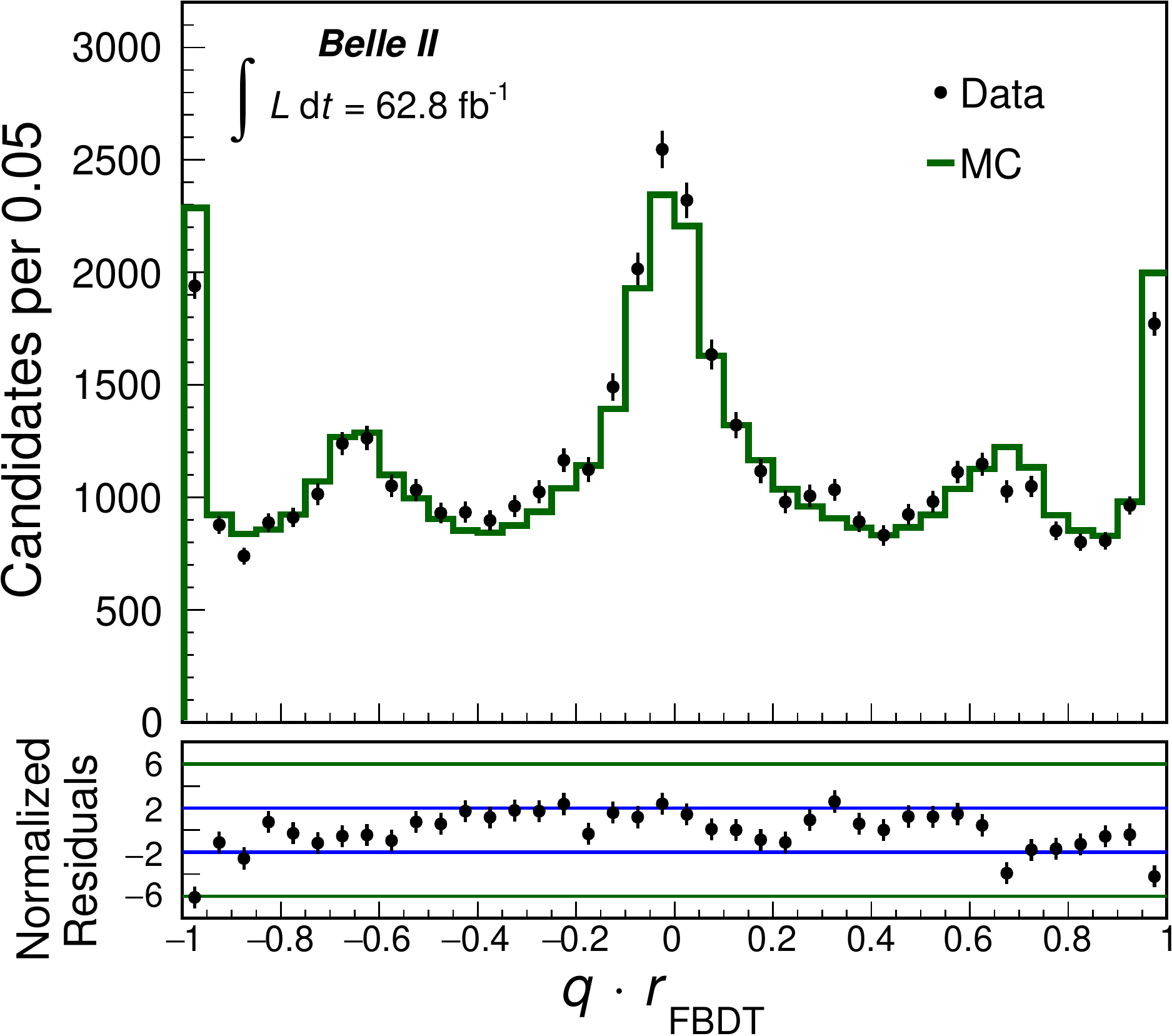}\hfill
\includegraphics[width=0.475\linewidth]{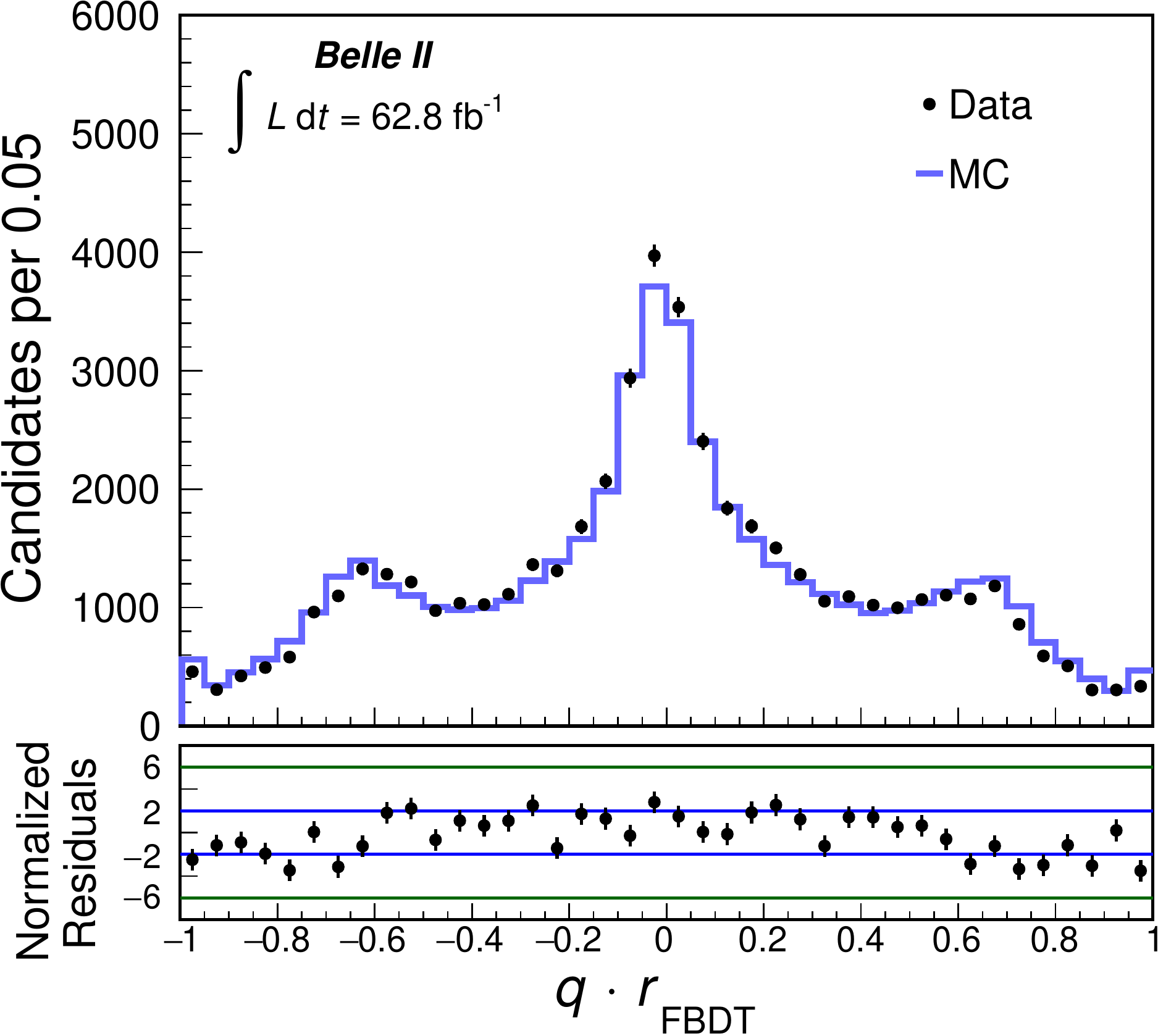}

\vspace{0.3cm} \text{(b) $\PBplus\to\APD^{(*)0}h^{+}$.}\\\vspace{0.5cm} 

\caption{\label{fig:qrSigDistDataFBDT} Normalized $q\cdot r$ distributions obtained with the category-based tagger in data and MC~simulation. 
The contribution (left) from the signal component in data is compared with correctly associated signal~MC events and (right)  from the background component in data is compared with sideband MC~events for (top)~neutral and (bottom)~charged $\PB\to\PD^{(*)}\Ph^{+}$ candidates.
}
\end{figure*}

\begin{figure*}
\centering
\includegraphics[width=0.475\linewidth]{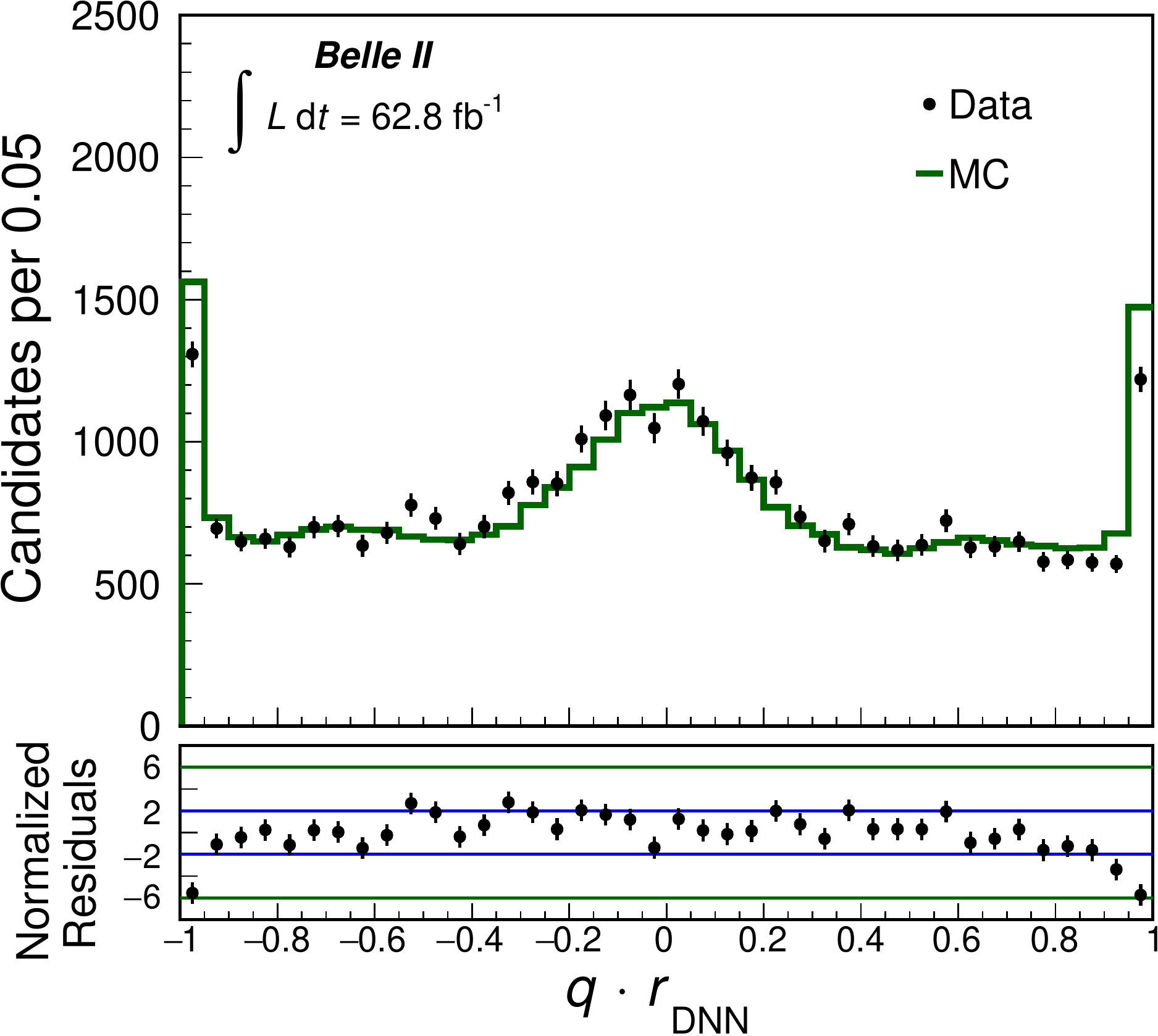}\hfill
\includegraphics[width=0.475\linewidth]{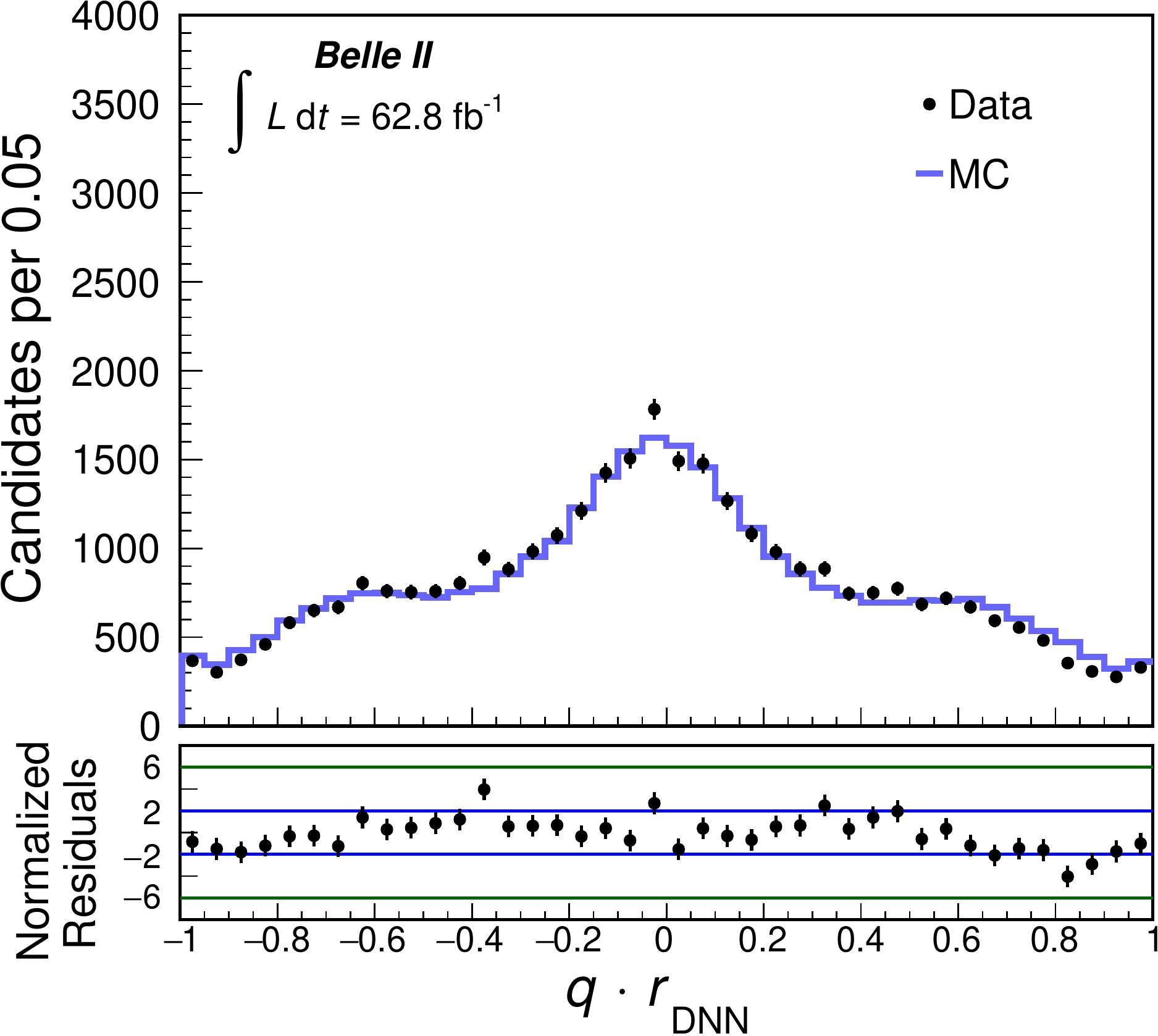}\\

\vspace{0.3cm} \text{(a) $\PBzero\to\PD^{(*)-}h^{+}$.}\\\vspace{1.0cm}

\includegraphics[width=0.475\linewidth]{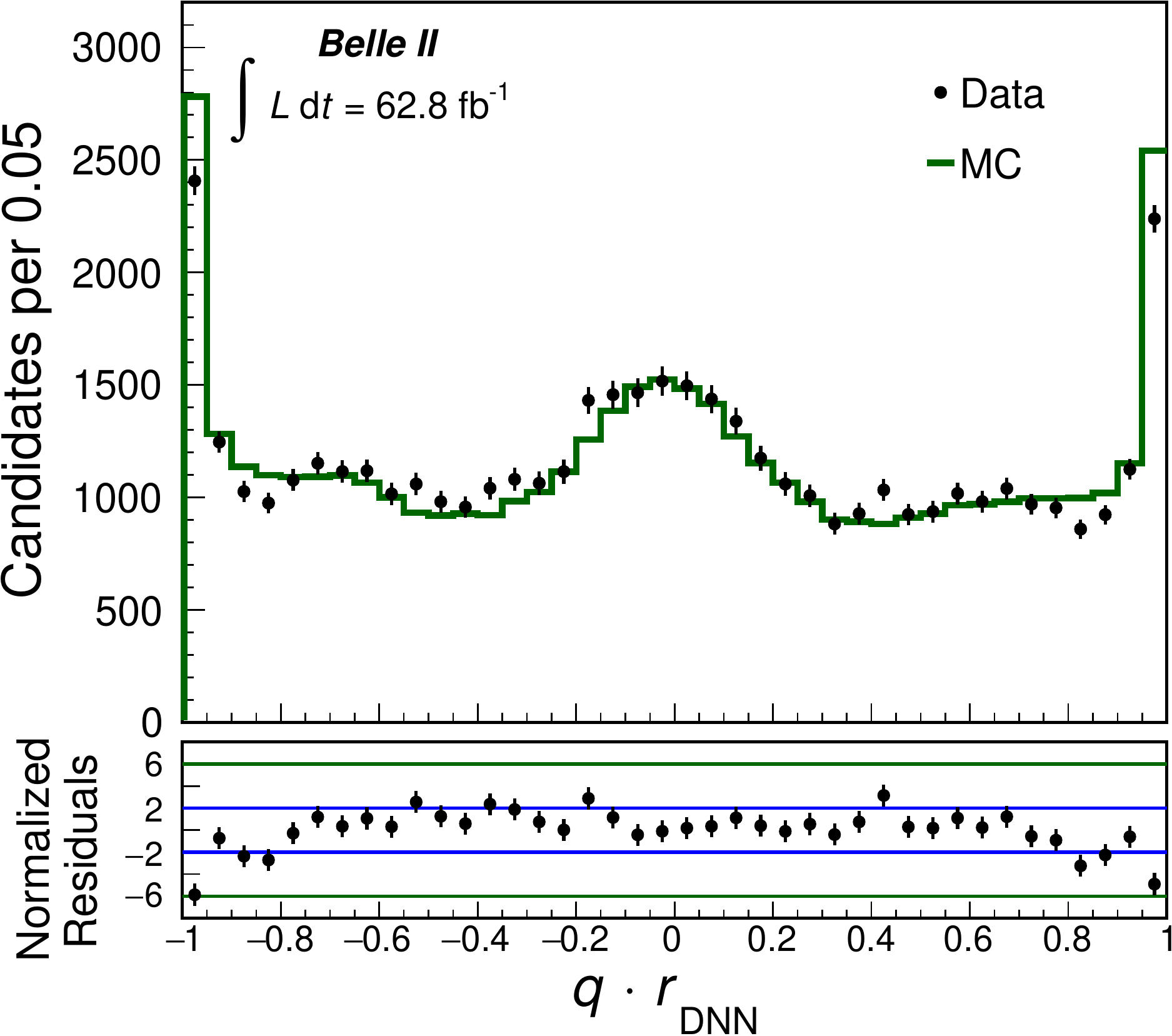}\hfill
\includegraphics[width=0.475\linewidth]{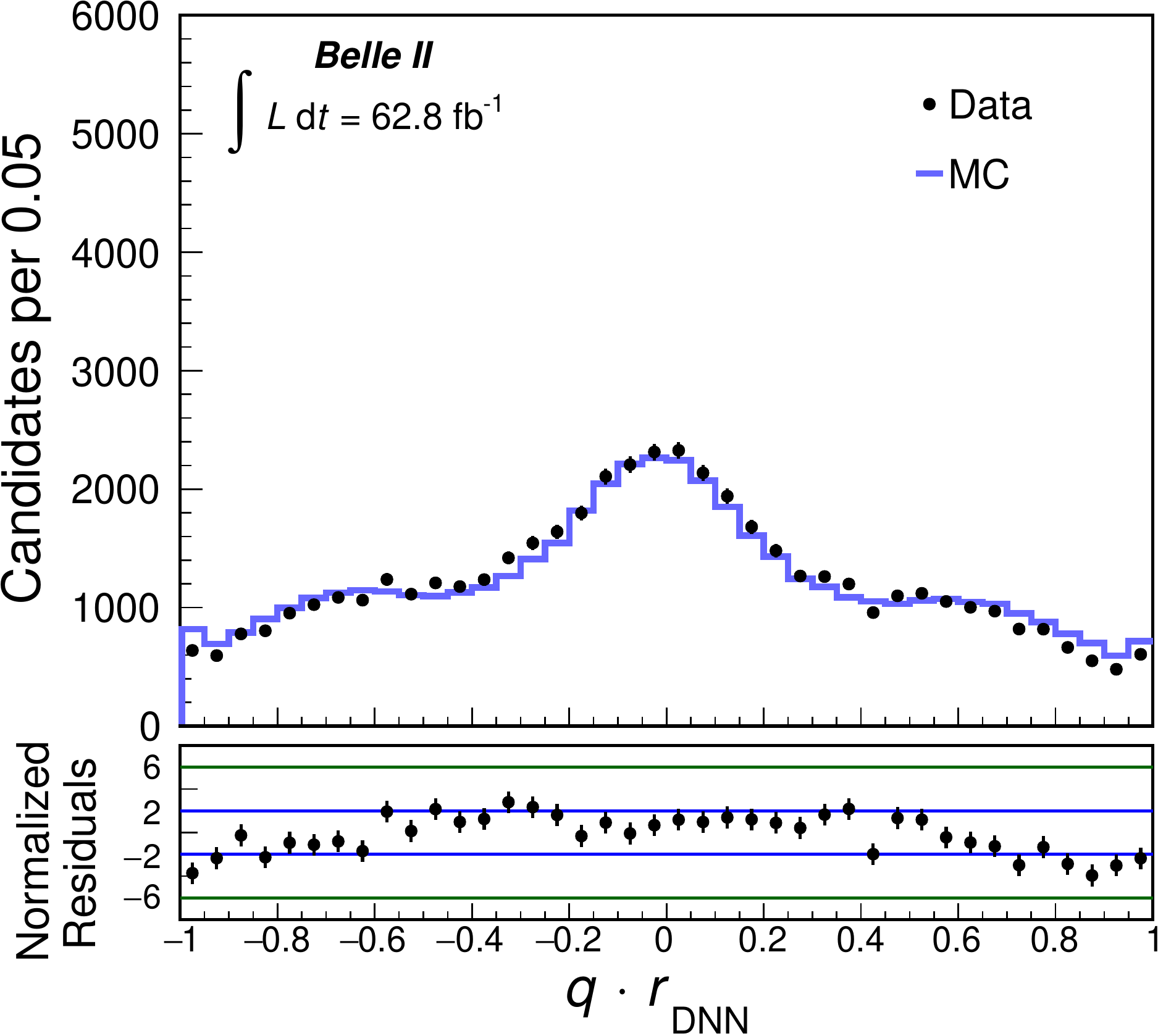}

\vspace{0.3cm} \text{(b) $\PBplus\to\APD^{(*)0}h^{+}$.}\\\vspace{0.5cm} 

\caption{\label{fig:qrSigDistDataDNN} Normalized $q\cdot r$ distributions obtained with the DNN tagger in data and MC~simulation. 
The contribution (left) from the signal component in data is compared with correctly associated signal~MC events and (right)  from the background component in data is compared with sideband MC~events for (top)~neutral and (bottom)~charged $\PB\to\PD^{(*)}\Ph^{+}$ candidates.
}
\end{figure*}

\begin{figure*}
    \centering
    \subfigure{\includegraphics[width=0.46\textwidth]{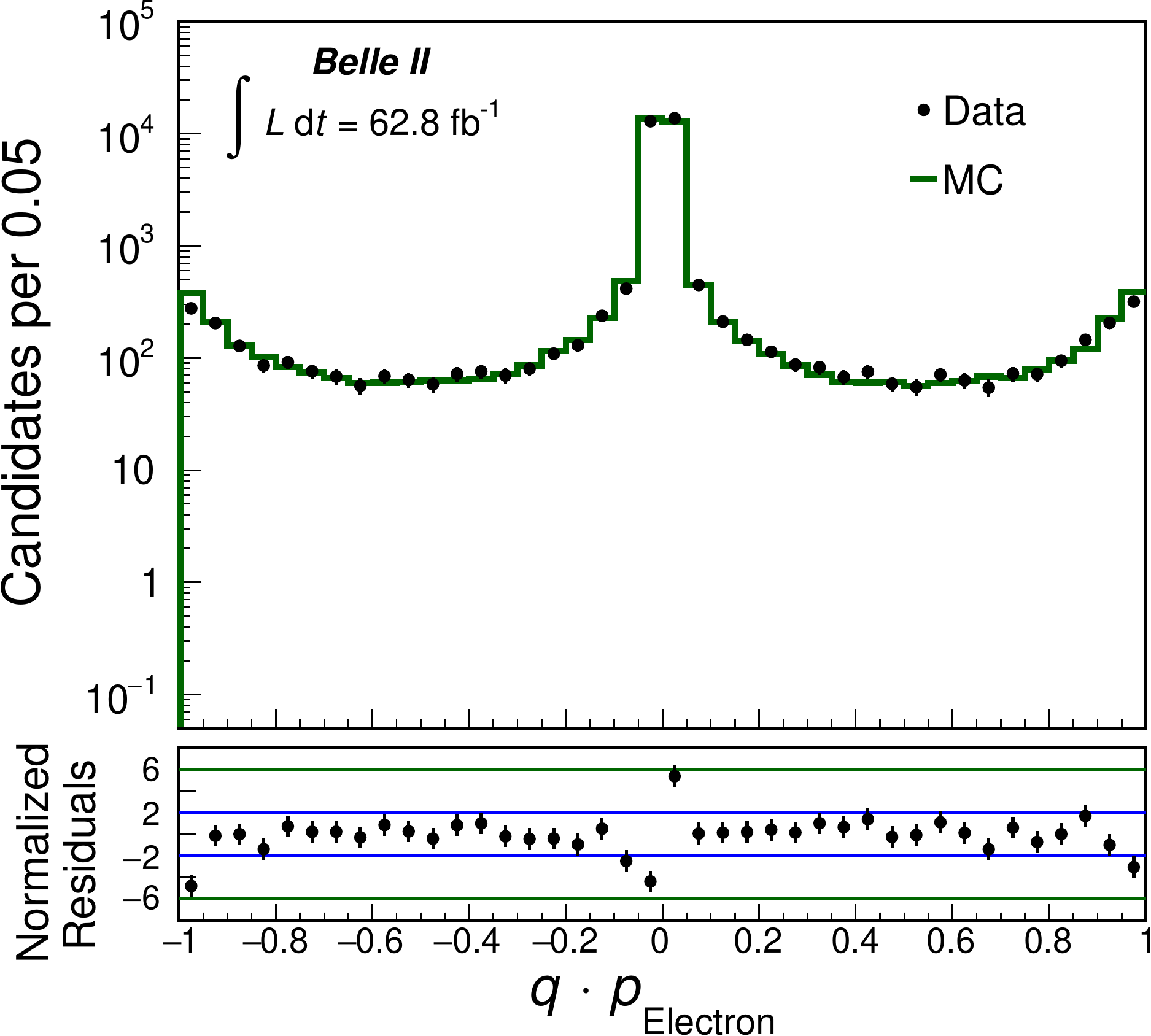}}\hfill 
    \subfigure{\includegraphics[width=0.46\textwidth]{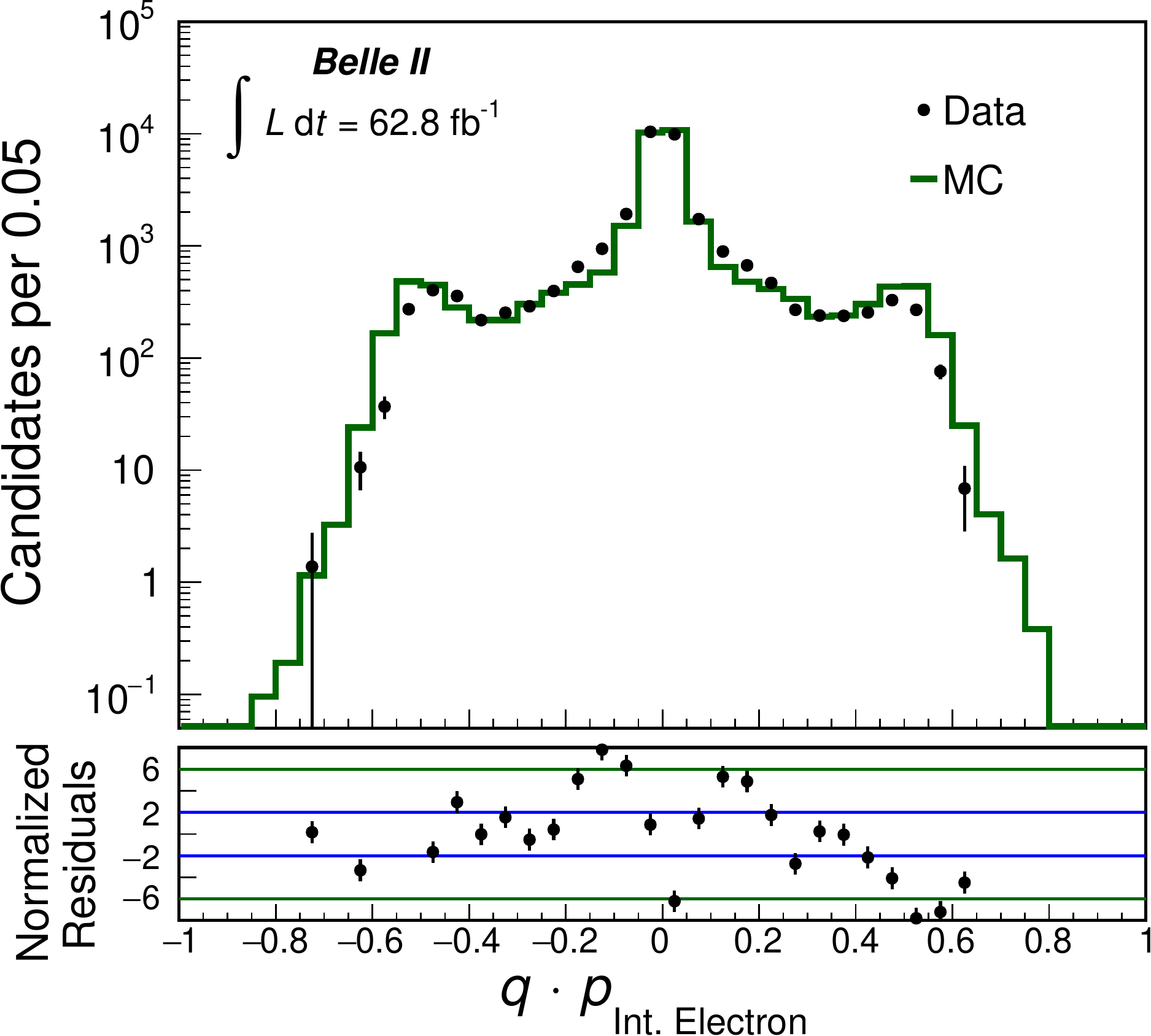}} 
    \subfigure{\includegraphics[width=0.46\textwidth]{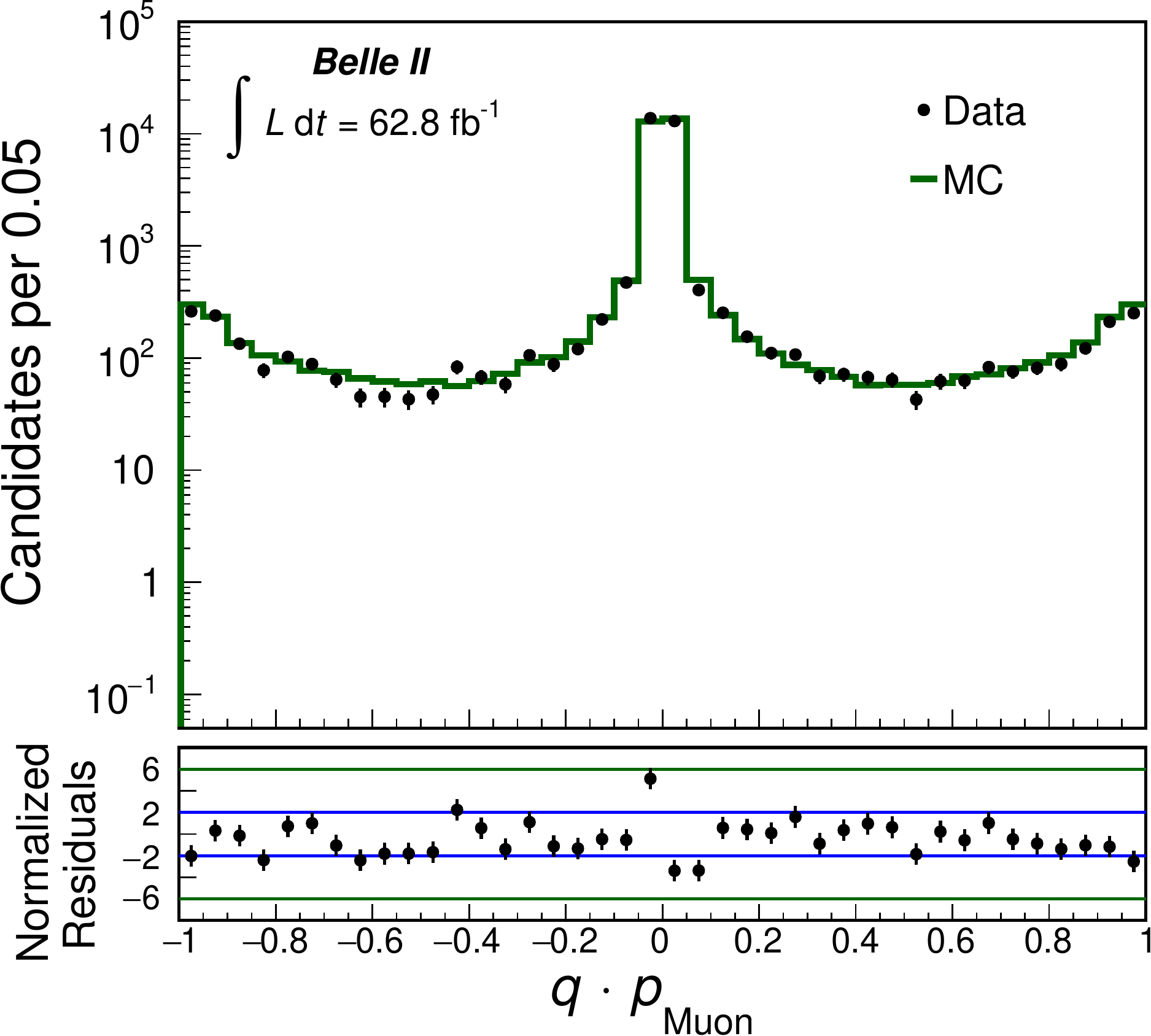}} \hfill 
    \subfigure{\includegraphics[width=0.46\textwidth]{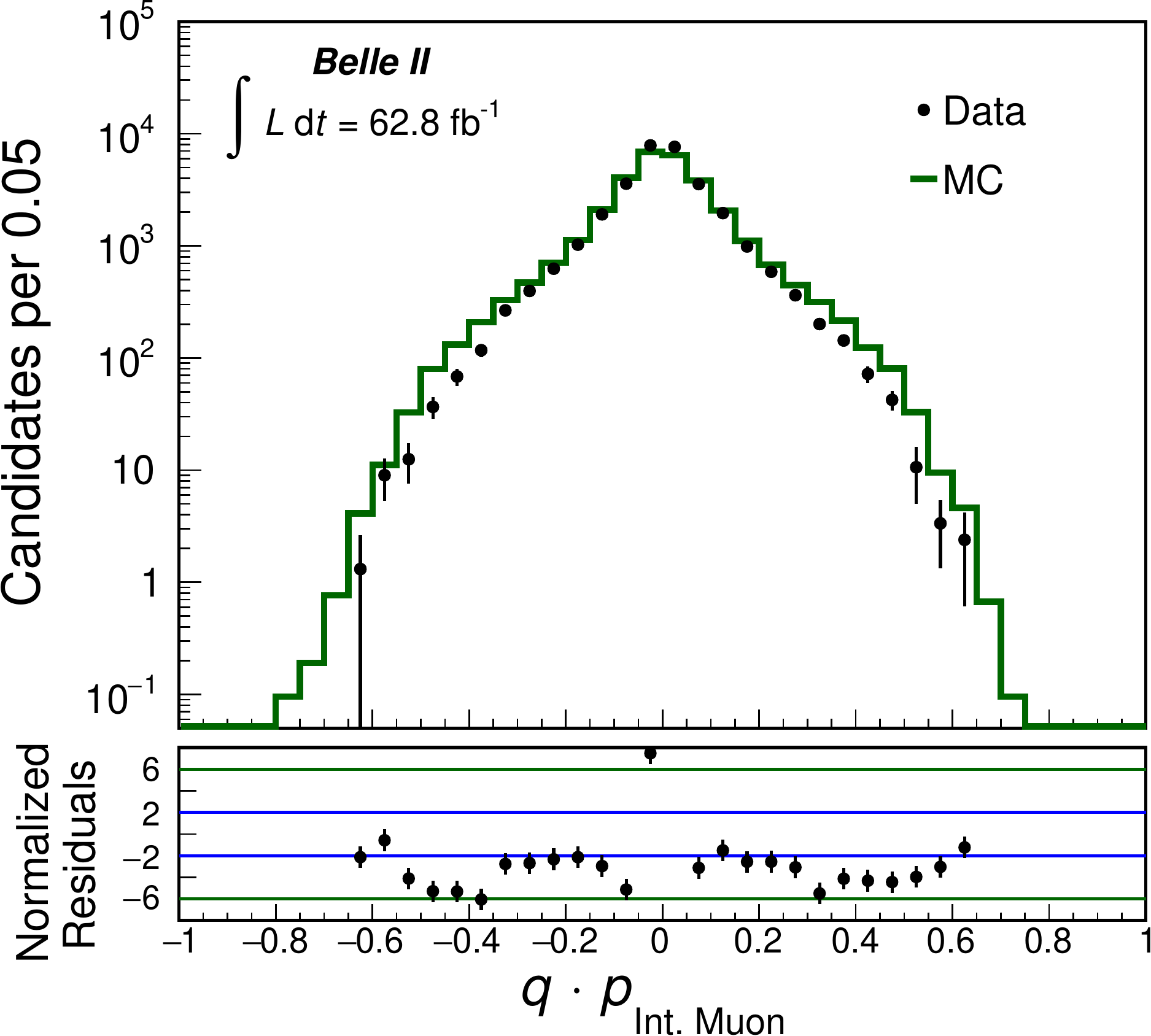}} 
    \subfigure{\includegraphics[width=0.46\textwidth]{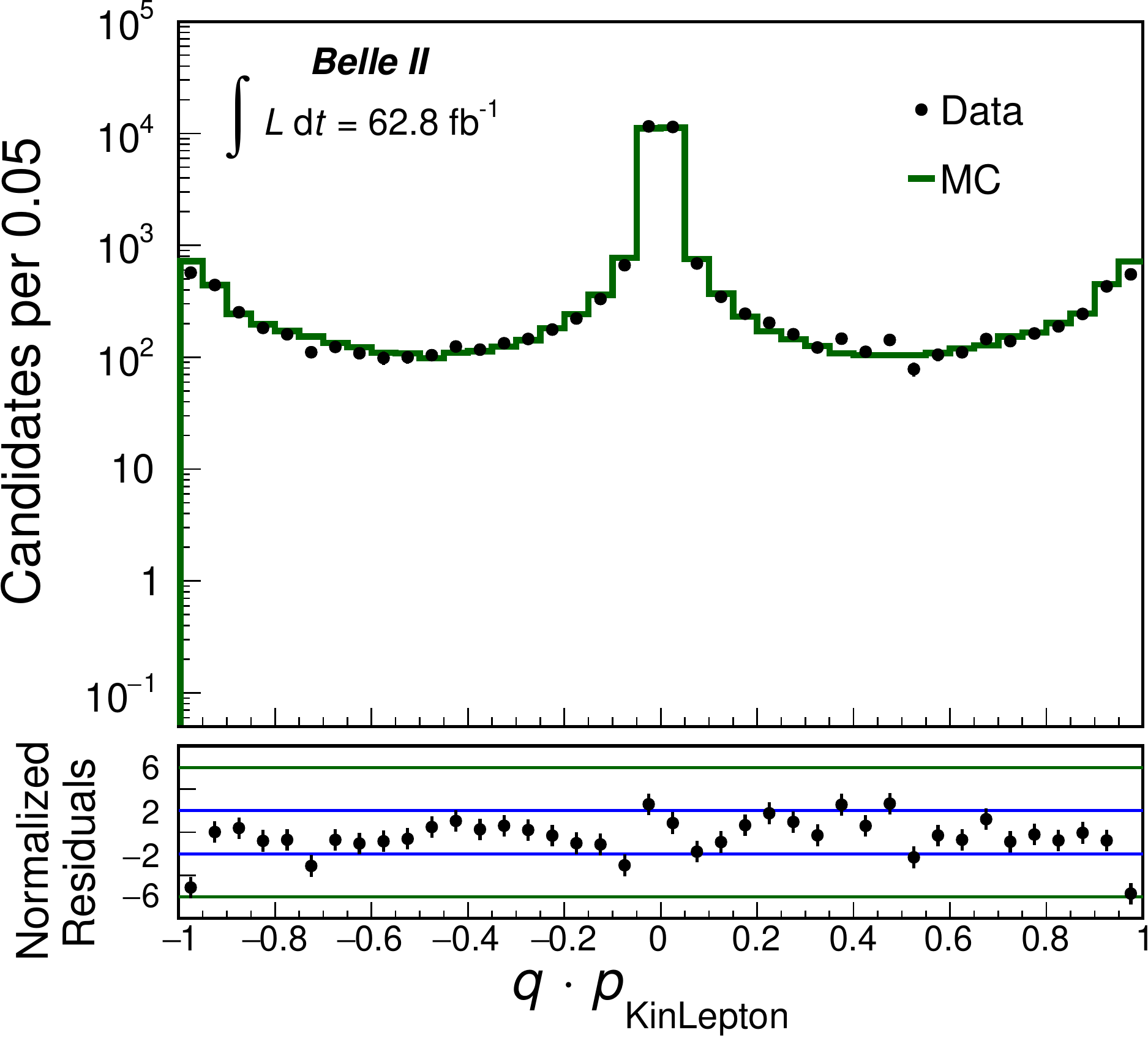}} \hfill
    \subfigure{\includegraphics[width=0.46\textwidth]{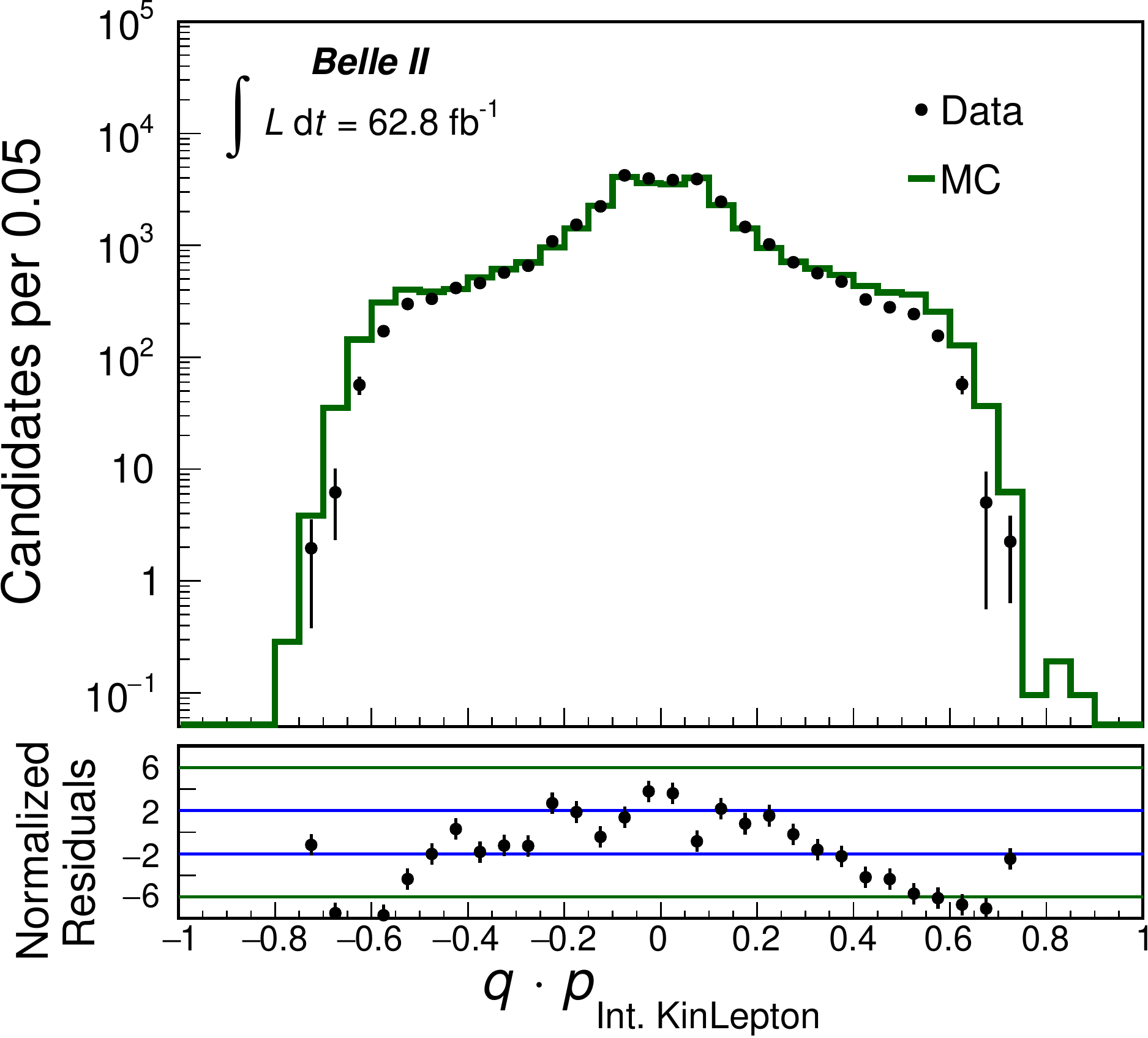}} 
    \caption{
    Normalized output distributions of the Electron, Intermediate Electron, Muon, Intermediate Muon, Kinetic Lepton, and Intermediate Kinetic Lepton categories in data and MC~simulation for $\PBzero\to\PD^{(*)-}\Ph^{+}$ candidates. The contribution from the signal component in data is compared with correctly associated signal MC~events.}
    \label{fig:QP_data_splotMC_1}
\end{figure*}

\begin{figure*}
    \centering
    \subfigure{\includegraphics[width=0.46\textwidth]{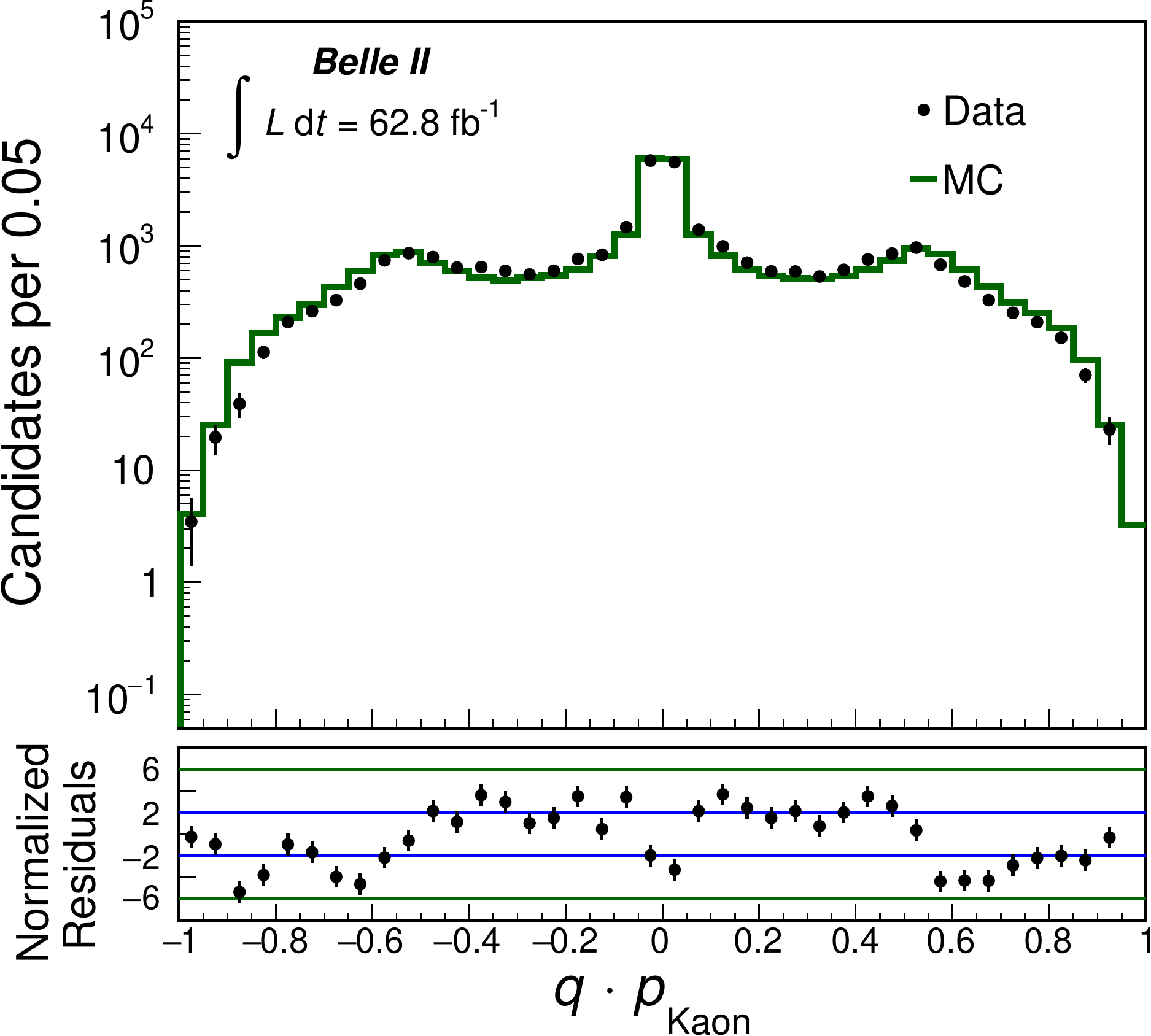}}\hfill 
    \subfigure{\includegraphics[width=0.46\textwidth]{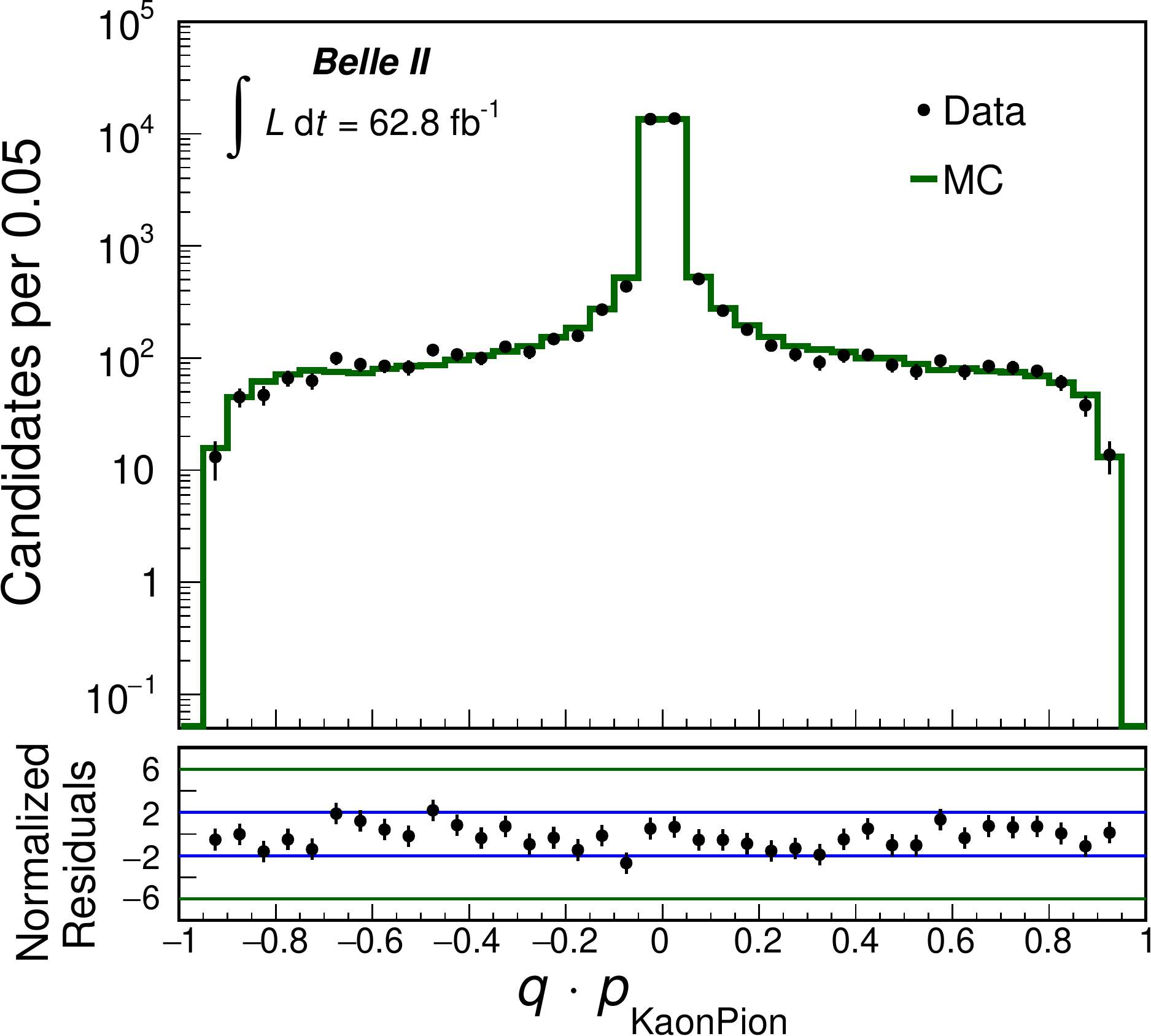}} 
    \subfigure{\includegraphics[width=0.46\textwidth]{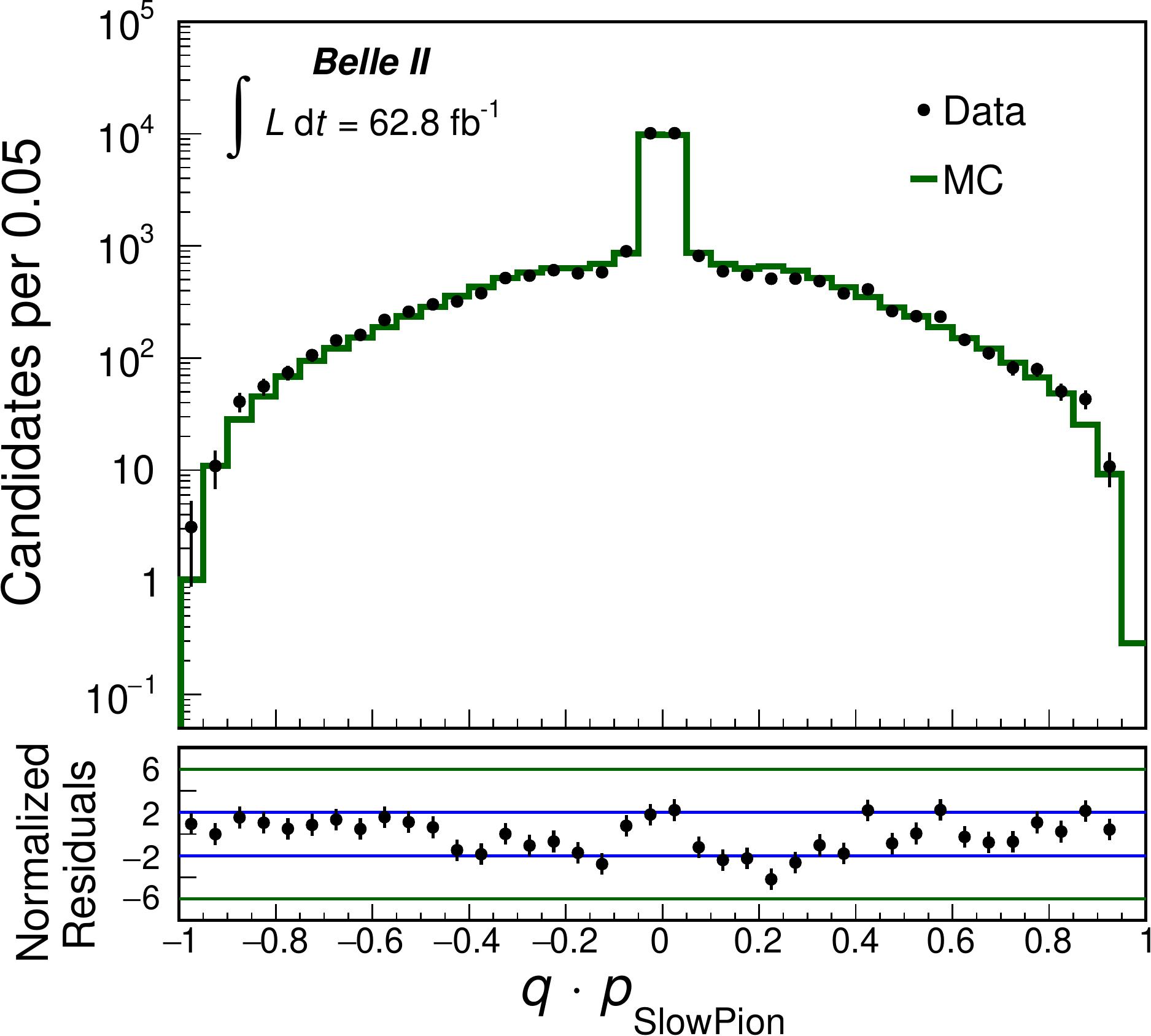}} \hfill 
    \subfigure{\includegraphics[width=0.46\textwidth]{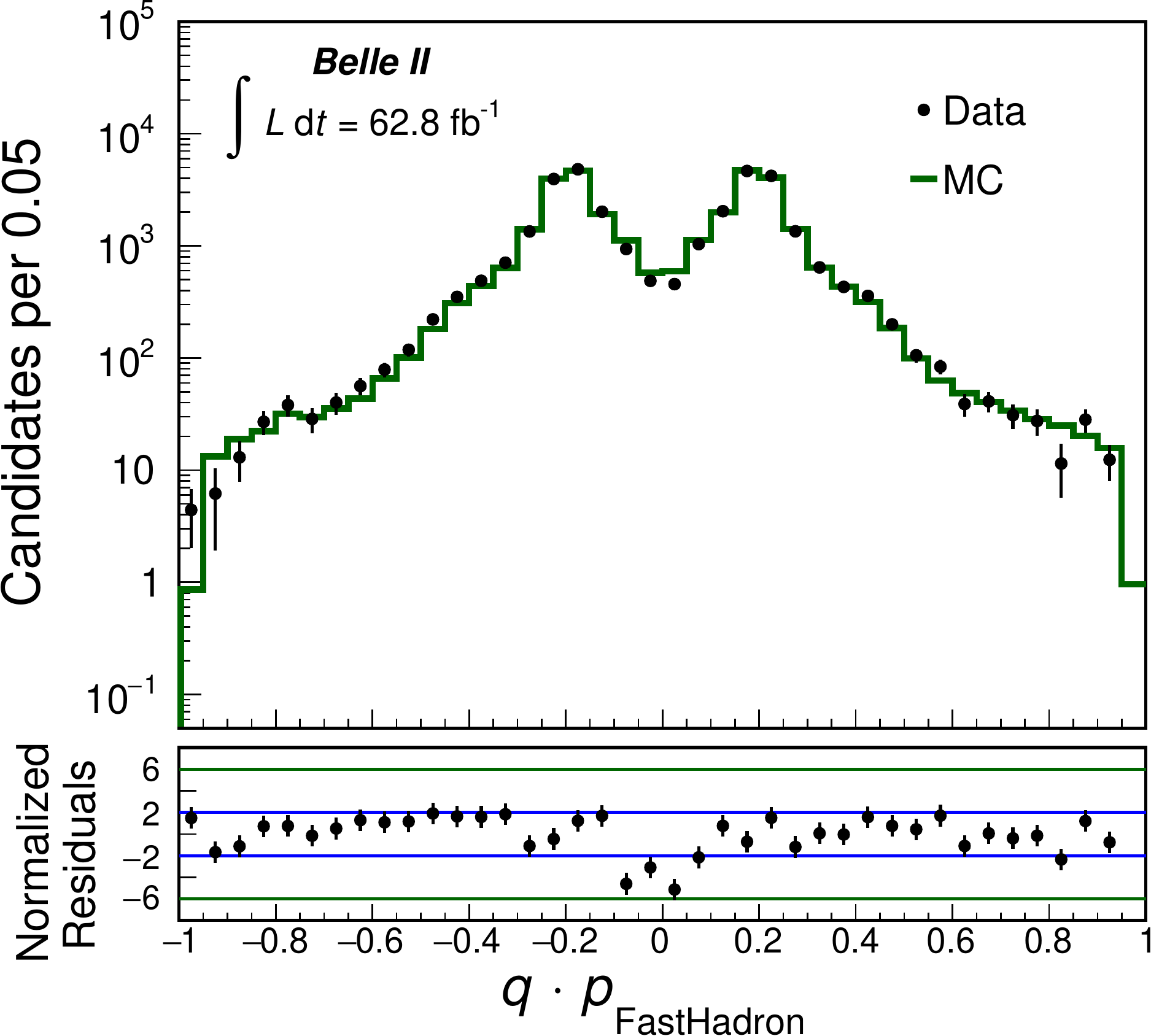}} 
    \subfigure{\includegraphics[width=0.46\textwidth]{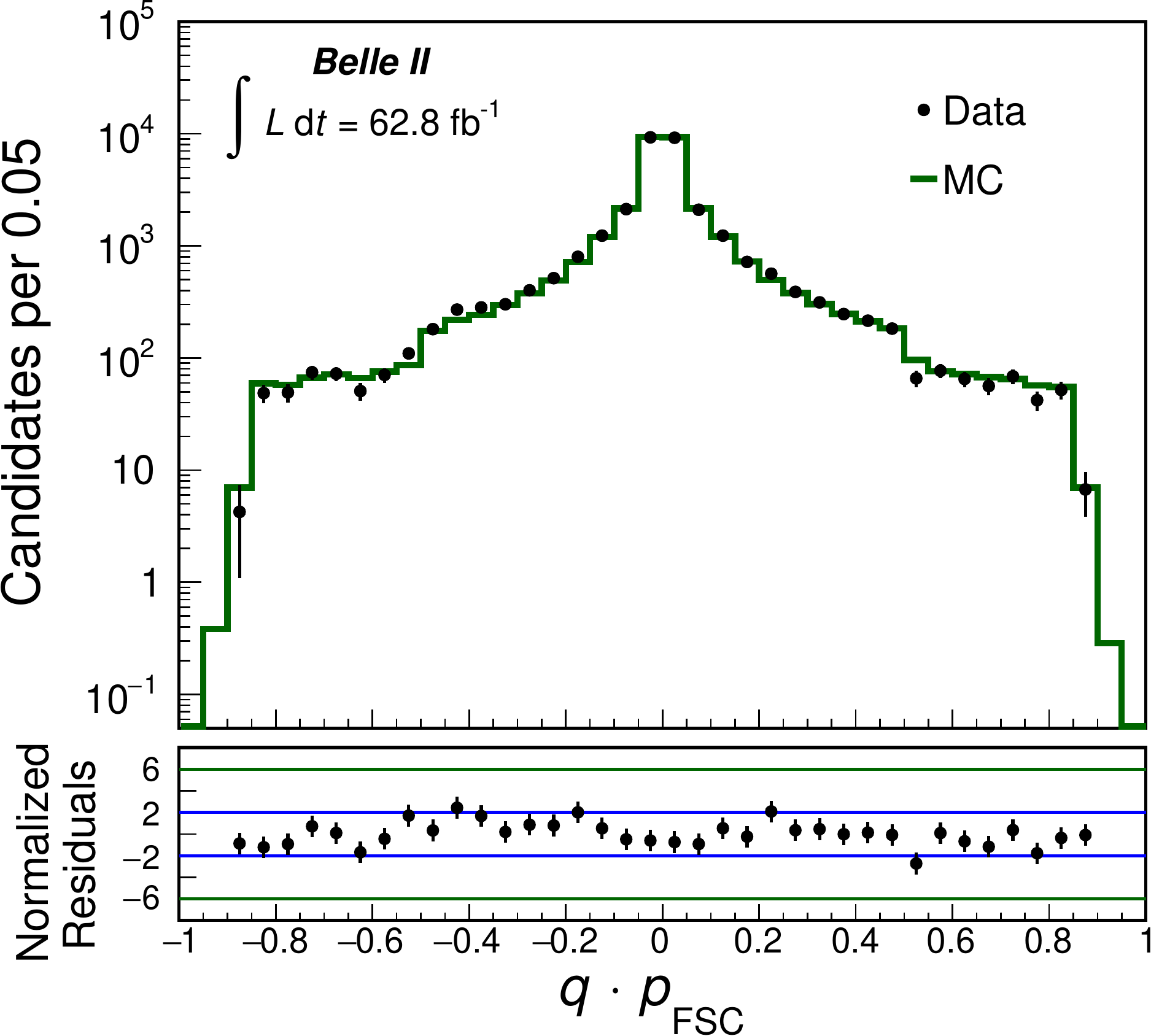}} \hfill
    \subfigure{\includegraphics[width=0.46\textwidth]{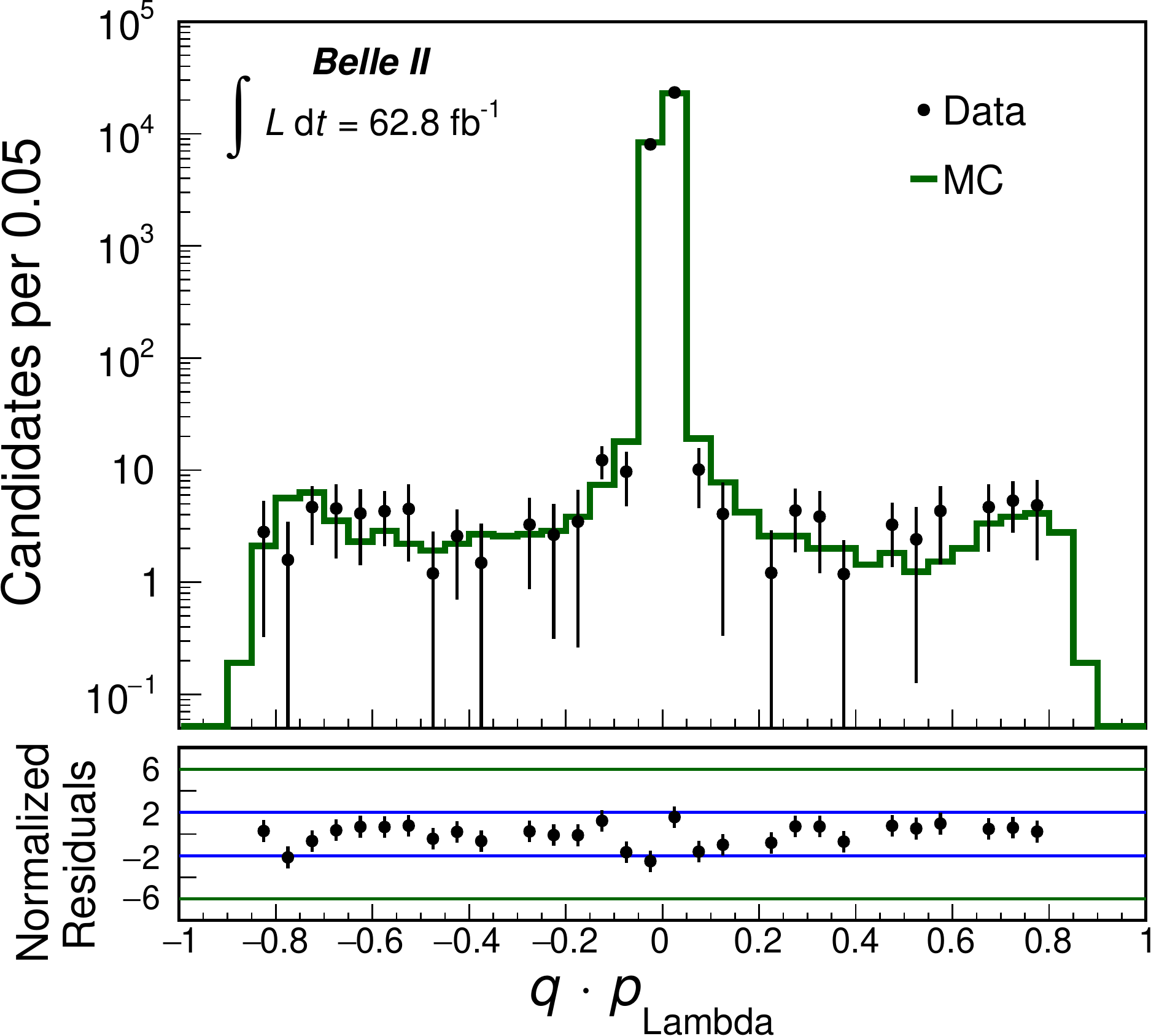}} 
    \caption{
    Normalized output distributions of the Kaon, Kaon-Pion, Slow Pion, Fast Hadron, Fast-Slow-Correlated, and Lambda categories  in data and MC~simulation for $\PBzero\to\PD^{(*)-}\Ph^{+}$ candidates. The contribution from the signal component in data is compared with correctly associated signal MC~events.}
    \label{fig:QP_data_splotMC_2}
\end{figure*}

\begin{figure}[!ht]
    \centering
    \subfigure{\includegraphics[width=0.46\textwidth]{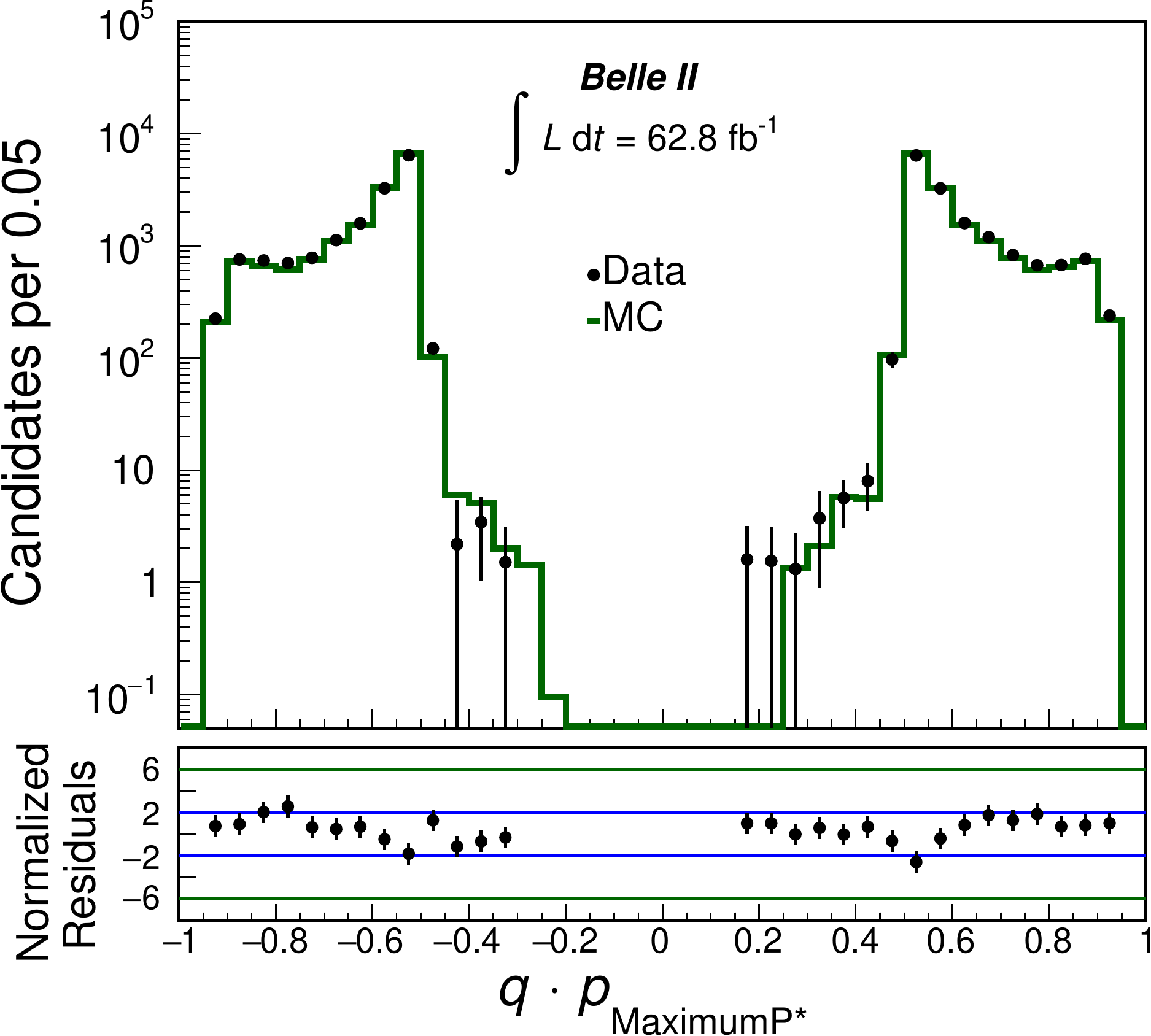}}
    \caption{ Normalized output distributions of the Maximum~$p^*$ category in data and MC~simulation for $\PBzero\to\PD^{(*)-}\Ph^{+}$ candidates. The contribution from the signal component in data is compared with correctly associated signal MC~events.}
    \label{fig:QP_data_splotMC_3}
\end{figure}

Figures~\ref{fig:qrSigDistDataFBDT} and~\ref{fig:qrSigDistDataDNN} show the $q\cdot r$~distributions provided by the FBDT and by the DNN flavor tagger; the signal and background distributions for neutral and charged $\PB\to\PD^{(*)}\Ph^{+}$ candidates are shown separately. 
%We use the subscript ${\rm FBDT}$ to label the dilution provided by the category-based flavor tagging algorithm.  
We compare the distribution of the signal component in data  with the distribution of correctly associated MC~events, and the distribution of the background component in data with the distribution of sideband MC~events (\mbox{$M_{\rm bc} < \SI{5.27}{GeV}/c^2$} 
and same fit range
\mbox{$\vert \Delta E \vert < \SI{0.12}{GeV}$}). We also compare the distributions of the signal component in data  with the distribution of correctly associated MC~events for the individual tagging categories~\mbox{(Figs.~\ref{fig:QP_data_splotMC_1}--\ref{fig:QP_data_splotMC_3})}. On the MC distributions, the statistical uncertainties are very small and thus not visible.  

In general, the results show a good consistency between data and simulation, with a slightly worse performance in data. In the signal $q\cdot r$~distributions, we observe some considerable differences around $\vert q\cdot r\vert\approx1$. We attribute these differences to discrepancies between data and simulation for some of the discriminating input variables, in particular for the electron and muon PID likelihoods.  Some differences are observed for categories associated with intermediate slow particles. However, these categories provide only marginal tagging power without degrading the overall tagging performance.

\clearpage

\section{Results}

\label{sec:results}

We obtain the partial tagging efficiencies $\varepsilon_i$, the wrong-tag fractions $w_i$, the asymmetries $\mu_i$ and $\Delta w_i$ and the correlation coefficients between them from the maximum-likelihood fit of the full model to data. To evaluate the tagging performance, we calculate the total effective efficiency as
\begin{equation*}
\hspace{1.7cm}    \varepsilon_{\rm eff} =  \sum_{i} \varepsilon_{{\rm eff}, i} = \sum_{i} \varepsilon_i \cdot (1-2w_i)^2\text{,}
\end{equation*}
where $\varepsilon_{{\rm eff}, i}$ is the partial effective efficiency in the $i$-th $r$~bin. The effective tagging efficiency is a measure for the effective reduction of events due to the flavor dilution $r$. In \CP~violation analyses, the statistical uncertainty of measured \CP~asymmetries is approximately proportional to \mbox{$1/\sqrt{N_{\rm eff}}=1/\sqrt{N\cdot \varepsilon_{\rm eff}}$}, where $N_{\rm eff}$ is the number of effectively tagged events. Thus, one would obtain the same statistical precision for $N_{\rm eff}$ perfectly tagged events or for $N$ events tagged with an effective efficiency $\varepsilon_{\rm eff}$.

\begin{figure*}[!h]
    \centering

    FBDT \hspace{8cm} DNN\\
    \includegraphics[width=0.475\textwidth]{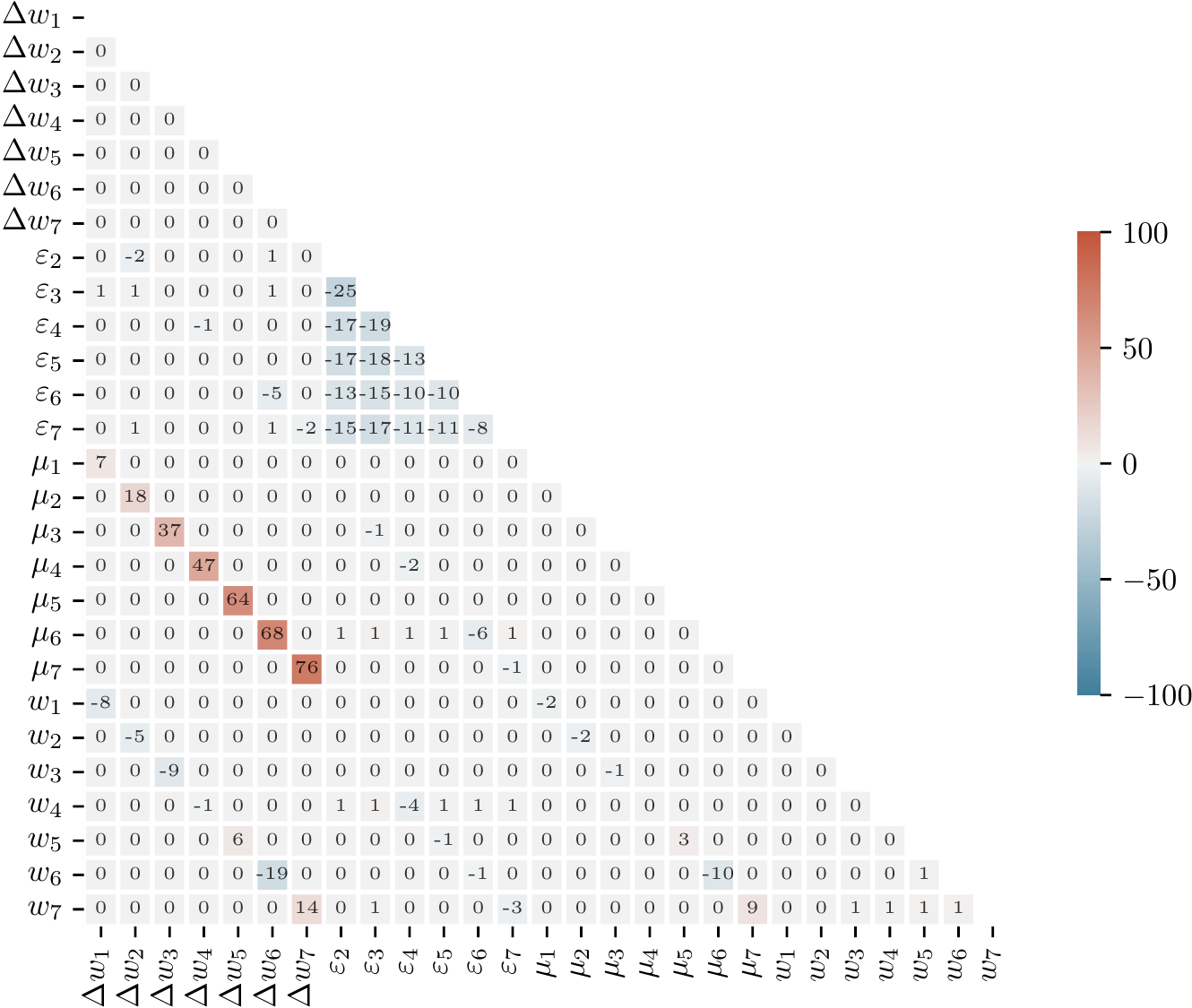} \hfill
    \includegraphics[width=0.475\textwidth]{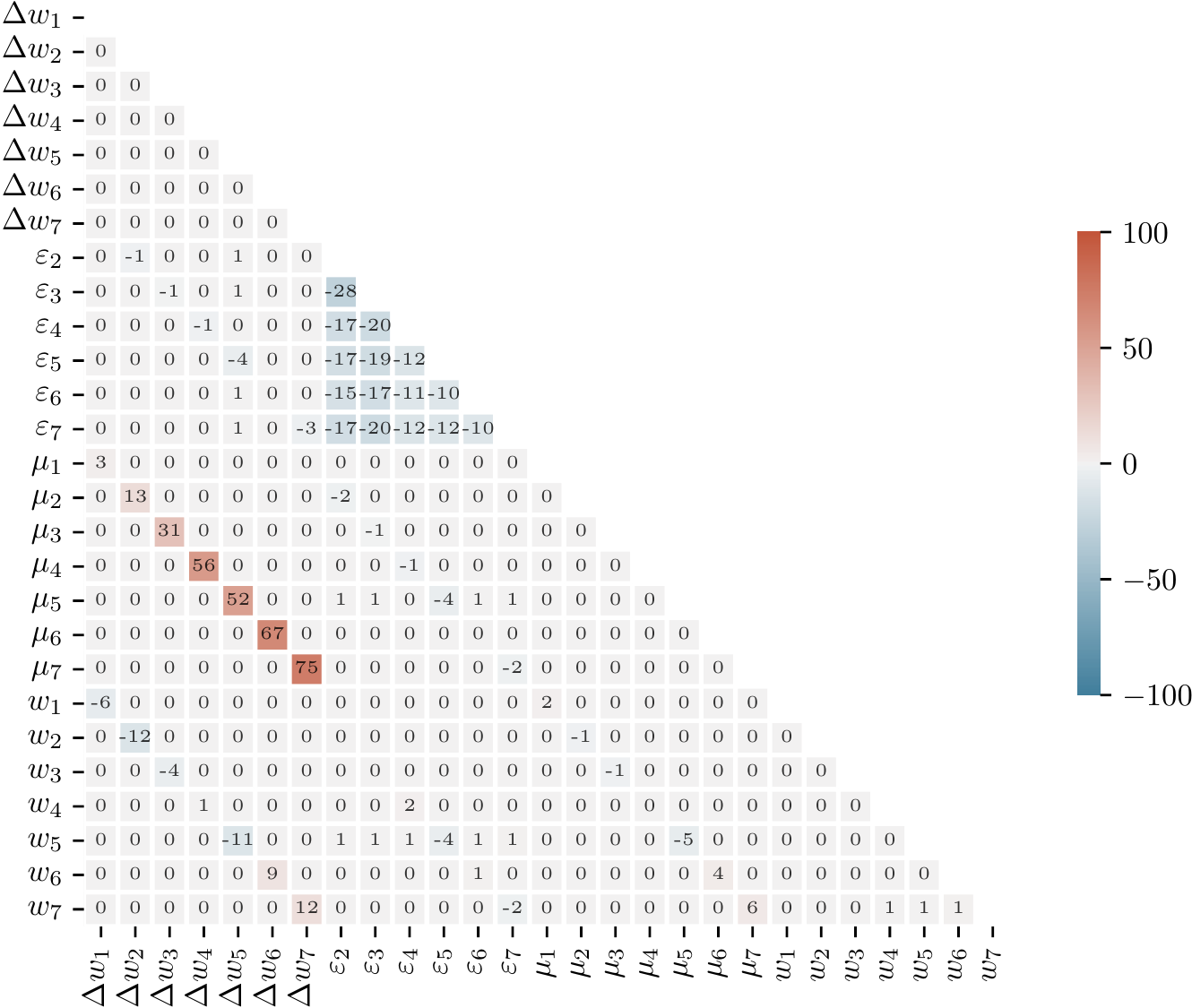}\\
        \vspace{0.3cm} \text{(a) $\PBzero\to\PD^{(*)-}h^{+}$.}\\\vspace{0.5cm}

    FBDT \hspace{8cm} DNN\\
    \includegraphics[width=0.475\textwidth]{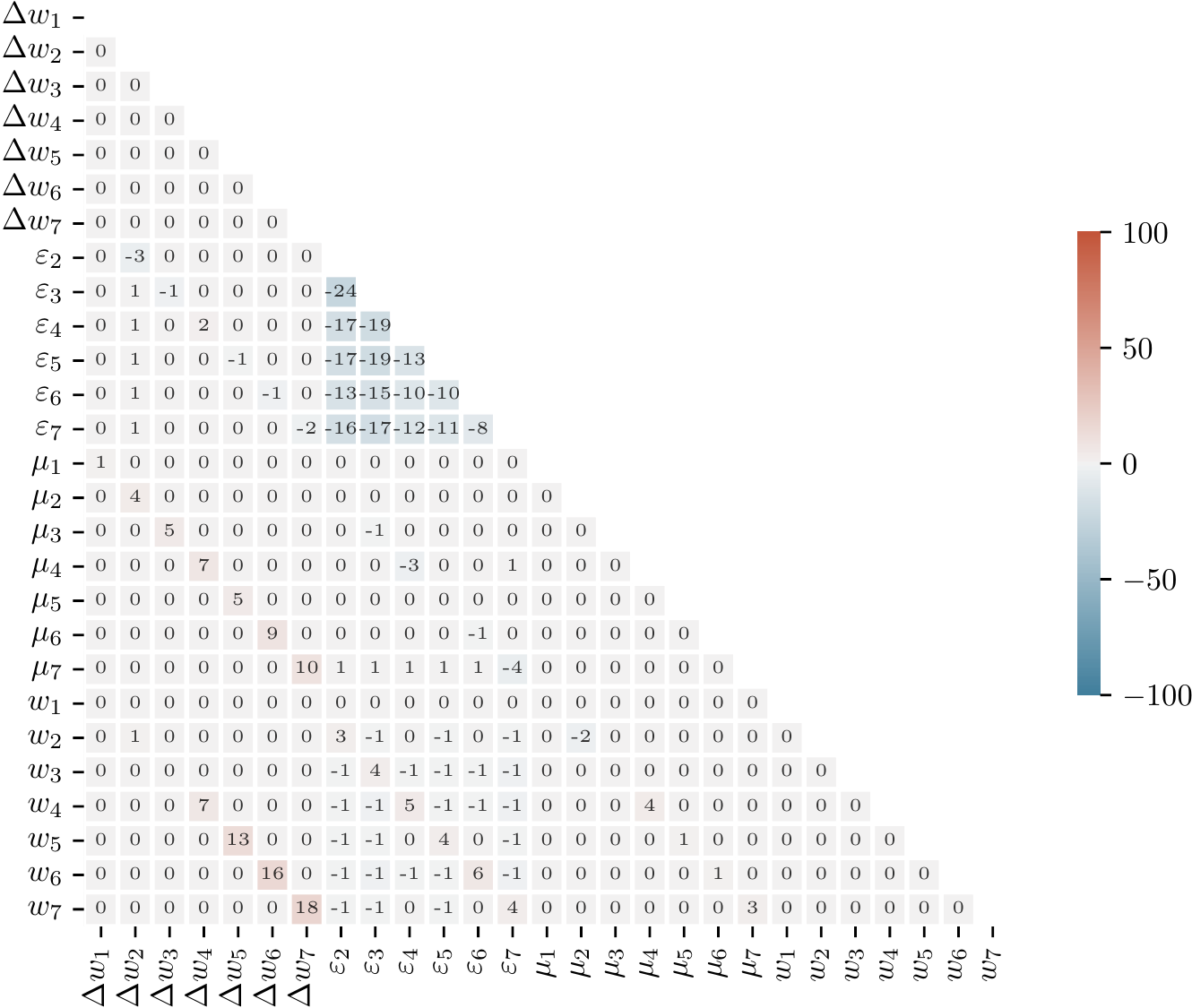}  \hfill
    \includegraphics[width=0.475\textwidth]{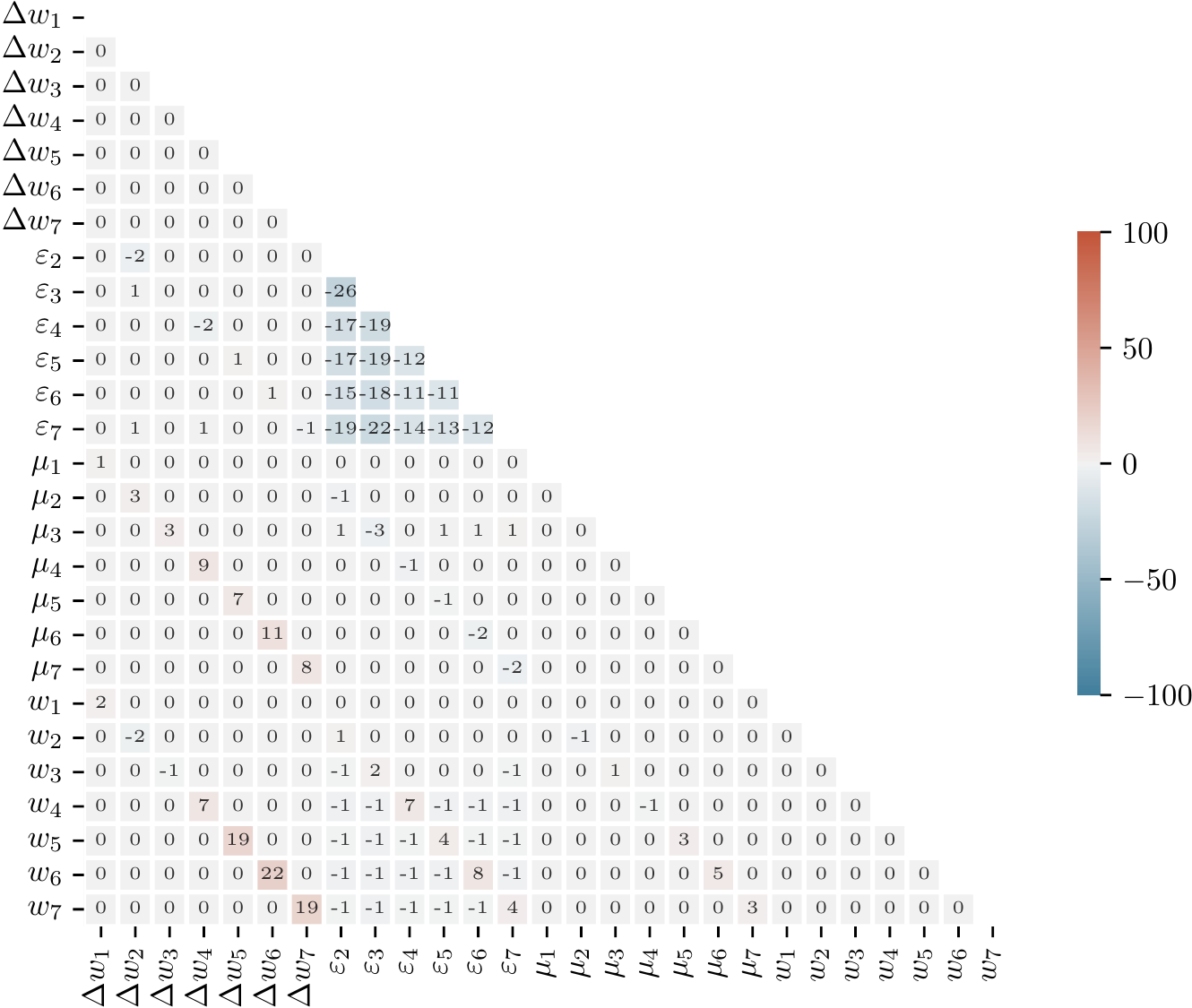}\\
        \vspace{0.3cm} \text{(b) $\PBplus\to\APD^{(*)0}h^{+}$.}\\\vspace{0.5cm}    
    
    \caption{{\color[rgb]{0,0.5,0}} Correlation coefficients between $\varepsilon_i$, $\Delta w_i$, $\Delta w_i$, and $\mu_i$ for the (left) category-based and (right) DNN flavor tagger in data. The results are shown for (top)~neutral and (bottom)~charged \mbox{$\PB\to\PD^{(*)}\Ph^{+}$} candidates.}
    \label{fig:sig_corrs}
\end{figure*}

Tables~\ref{tab:sys_fitres_data} and~\ref{tab:sys_fitres_dataDNN} show the fit results for the category-based and the DNN flavor taggers. The respective effective efficiencies for both flavor taggers are shown in 
Tables~\ref{tab:sys_effeff_data} and~\ref{tab:sys_effeff_dataDNN}. Figure~\ref{fig:sig_corrs} shows the Pearson correlation coefficients obtained from the Hessian matrix determined by the fit. We observe considerable dependencies among the $\varepsilon_i$ efficiencies for both charged and neutral \PB~candidates, and among the asymmetries $\Delta w_i$ and $\mu_i$ for neutral \PB~candidates.

\subsection{Systematic uncertainties}

We consider the systematic uncertainties associated with the $\Delta E$~PDF parametrization, 
%the $\Delta E$~fit range, 
the flavor mixing of the background,  
the fit bias, and 
 the eventual bias introduced by model assumptions.

\textbf{$\Delta E$~PDF parametrization:} we perform simplified simulated experiments using an alternative model with a different $\Delta E$ parametrization. We perform 
fits to simulated data samples bootstrapped~(sampled with replacement)~\cite{Efron:1979bxm} from the generic MC~simulation. We fit 
using default and alternative models and calculate for each fit parameter~$x_i$ the difference~$\delta x_i$ between the results obtained with the alternative model and the results obtained with the default model. We obtain the mean difference $\delta \hat{x}_i$ by fitting a Gaussian function to the distribution of $\delta x_i$ and take the full mean $\delta \hat{x}_i$ as systematic uncertainty.
%For the alternative signal $\Delta E$ PDF, we use a double Gaussian function determined using truth-matched MC~events, with the additional flexibility of a global shift of peak position and a global scaling factor for the width. For the alternative background $\Delta E$ PDF, we use a first order polynomial function. The alternative model has the same amount of free fit parameters as the default one.

For the alternative signal $\Delta E$ PDF, we use a triple Gaussian function with the additional flexibility of a global shift of peak position and a global scaling factor for the width. For the alternative background $\Delta E$ PDF, we use a second-order Chebyshev polynomial function; we determine the coefficient of the quadratic term by fitting to the distribution of the generic simulation without signal~MC events, and then leave the coefficient of the linear term free, in order to have the same degrees of freedom as the default model. 

We also check whether the signal $\Delta E$~PDF shape changes as a function of $r$ and \PB~flavor and find no significant dependences.

\textbf{Background mixing:} our fit takes into account the uncertainty on the world average for the signal $\chi_d$ in the Gaussian constraint. However, 
we assume that there is no mixing in the background~($\chi_d^{\rm bkg}=0$). Since the background includes $\PBzero\APBzero$ events, we study the effect of flavor mixing in the background by varying the value of the background $\chi_d^{\rm bkg}$ by a small amount~$\pm \delta\chi_d^{\rm bkg}$, corresponding to the statistical uncertainty when $\chi_d^{\rm bkg}$ is a free parameter in the fit. 
%We then take for each fit parameter~$x_i$ half the difference between the results for $\chi_d^{\rm bkg} + \delta\chi_d^{\rm bkg}$ and for  $\chi_d^{\rm bkg} - \delta\chi_d^{\rm bkg}$ as systematic uncertainty. 
%We obtain the value of $\delta\chi_d^{\rm bkg}$ by performing bootstrapped pseudo-experiments in which we leave $\chi_d^{\rm bkg}$ free to float and fix the signal $\chi_d$ to the true MC~value and the signal $\varepsilon_i$ and $w_i$ to the generated values~(count method). We take the width $\sigma$ of the residuals for $\chi_d^{\rm bkg}$ as $\delta\chi_d^{\rm bkg}$. 
 We find that the difference between the results for $\chi_d^{\rm bkg} + \delta\chi_d^{\rm bkg}$ and for $\chi_d^{\rm bkg} - \delta\chi_d^{\rm bkg}$ are below~$1\%$ of the statistical uncertainty for all fit parameters, which is negligible.

\textbf{Fit\; bias:}  for each fit parameter $x_i$, we determine the fit bias using the residuals from bootstrapped simulated experiments. The residuals are the differences between the fit results for the individual bootstrapped data samples and the fit results for the parent MC~sample. We take the full bias as systematic uncertainty.

\textbf{Fit model:} before performing the fit to data, we check that the results of the fit to the full MC~sample and the true values determined using MC~information agree within about one standard deviation for all fit parameters without tendency of over or underestimation across neighboring $r$ bins. We consider possible bias that cannot be resolved with the current sample sizes and that cannot be attributed to a single effect, for example bias due to fit model or due to MC association, by assigning the uncertainty of the fit to the full MC~sample as systematic uncertainty.
   
We further study possible asymmetries in the reconstruction of signal \PB~candidates that might cause bias in the measurement of the flavor tagging parameters as they are neglected in the fit model. We observe small reconstruction asymmetries between signal $\PBzero$($\PBplus$) and $\APBzero$($\PBminus$) in some individual $r$-bins below or around~$5\%$. However, we find that they do not cause statistically significant bias for samples up to $700\,\si{fb^{-1}}$ by performing fits to generic simulation.

% \clearpage

\begin{table*}[p]
    \centering
    \caption{Results for $\varepsilon_i$, $w_i$, $\mu_i$, and $\Delta w_i$ for the category-based~(FBDT) tagger: systematic uncertainties associated with the $\Delta E$ parametrization, fit bias, and fit model, and total systematic uncertainty are shown together with the fit results~(with stat. and syst. uncertainty). The results are given in percent.}    
\scriptsize
    \begin{tabular}{c r c c c r }
\multicolumn{6}{l}{ $\PBzero\to\PD^{(*)-}h^{+}$}\\
    \hline\hline
    \multicolumn{1}{c}{Parameter} & 
    \multicolumn{1}{c}{$\Delta E$ PDF} &
    \multicolumn{1}{c}{Fit bias}  &  
    \multicolumn{1}{c}{Model} &
    \multicolumn{1}{c}{Total syst. uncty.}  & \multicolumn{1}{c}{Fit results}\\\hline 
$\varepsilon_1$ & $0.01$ & $0.00$ & $0.09$ & $0.1$ & $19.0 \pm 0.3 \pm 0.1$\\
$\varepsilon_2$ & $0.01$ & $0.00$ & $0.09$ & $0.1$ & $17.1 \pm 0.3 \pm 0.1$\\
$\varepsilon_3$ & $0.01$ & $0.03$ & $0.09$ & $0.1$ & $21.3 \pm 0.3 \pm 0.1$\\
$\varepsilon_4$ & $0.01$ & $0.01$ & $0.07$ & $0.1$ & $11.3 \pm 0.3 \pm 0.1$\\
$\varepsilon_5$ & $0.01$ & $0.01$ & $0.07$ & $0.1$ & $10.7 \pm 0.3 \pm 0.1$\\
$\varepsilon_6$ & $0.00$ & $0.01$ & $0.06$ & $0.1$ & $ 8.2 \pm 0.2 \pm 0.1$\\
$\varepsilon_7$ & $0.02$ & $0.00$ & $0.07$ & $0.1$ & $12.4 \pm 0.2 \pm 0.1$
\\\hline 
$w_1$ & $0.03$ & $0.03$ & $0.45$ & $0.5$ & $47.1 \pm 1.6 \pm 0.5$\\
$w_2$ & $0.01$ & $0.01$ & $0.46$ & $0.5$ & $41.3 \pm 1.7 \pm 0.5$\\
$w_3$ & $0.07$ & $0.04$ & $0.40$ & $0.4$ & $30.3 \pm 1.4 \pm 0.4$\\
$w_4$ & $0.17$ & $0.18$ & $0.53$ & $0.6$ & $22.9 \pm 1.8 \pm 0.6$\\
$w_5$ & $0.01$ & $0.01$ & $0.49$ & $0.5$ & $12.4 \pm 1.8 \pm 0.5$\\
$w_6$ & $0.02$ & $0.02$ & $0.50$ & $0.5$ & $ 9.4 \pm 1.9 \pm 0.5$\\
$w_7$ & $0.01$ & $0.01$ & $0.37$ & $0.4$ & $ 2.3 \pm 1.3 \pm 0.4$\\
\hline 
$\mu_1$ & $0.01$ & $0.03$ & $0.91$ & $0.9$ &  $4.4 \pm 3.2 \pm 0.9$\\
$\mu_2$ & $0.15$ & $0.13$ & $0.93$ & $0.9$ &  $3.9 \pm 3.3 \pm 0.9$\\
$\mu_3$ & $0.05$ & $0.00$ & $0.82$ & $0.8$ &  $6.8 \pm 2.9 \pm 0.8$\\
$\mu_4$ & $0.04$ & $0.01$ & $1.12$ & $1.1$ &  $3.2 \pm 4.0 \pm 1.1$\\
$\mu_5$ & $0.16$ & $0.23$ & $1.06$ & $1.1$ & -$0.5 \pm 4.1 \pm 1.1$\\
$\mu_6$ & $0.02$ & $0.04$ & $1.14$ & $1.1$ & $10.8 \pm 4.3 \pm 1.1$\\
$\mu_7$ & $0.37$ & $0.27$ & $0.86$ & $1.0$ & -$3.7 \pm 3.2 \pm 1.0$\\\hline
$\Delta w_1$ & $0.16$ & $0.12$ & $0.57$ & $0.6$ & $8.8 \pm 2.0 \pm 0.6$\\
$\Delta w_2$ & $0.12$ & $0.15$ & $0.59$ & $0.6$ & $6.1 \pm 2.1 \pm 0.6$\\
$\Delta w_3$ & $0.12$ & $0.11$ & $0.54$ & $0.6$ & $2.7 \pm 1.9 \pm 0.6$\\
$\Delta w_4$ & $0.05$ & $0.01$ & $0.77$ & $0.8$ & $5.5 \pm 2.6 \pm 0.8$\\
$\Delta w_5$ & $0.05$ & $0.03$ & $0.74$ & $0.7$ & $0.7 \pm 2.9 \pm 0.7$\\
$\Delta w_6$ & $0.08$ & $0.07$ & $0.84$ & $0.9$ & $7.7 \pm 3.2 \pm 0.9$\\
$\Delta w_7$ & $0.19$ & $0.17$ & $0.66$ & $0.7$ & $0.6 \pm 2.4 \pm 0.7$\\\hline

\multicolumn{4}{l}{ }\\
\multicolumn{4}{l}{ $\PBplus\to\APD^{(*)0}h^{+}$}\\\hline\hline
    \multicolumn{1}{c}{Parameter} & 
    \multicolumn{1}{c}{$\Delta E$ PDF} &
    \multicolumn{1}{c}{Fit bias}  &  
    \multicolumn{1}{c}{Model} &
    \multicolumn{1}{c}{Total syst. uncty.}  & \multicolumn{1}{c}{Fit results}\\\hline 
$\varepsilon_1$ & $0.04$ & $0.00$ & $0.07$ & $0.1$ & $18.3 \pm 0.3 \pm 0.1$\\
$\varepsilon_2$ & $0.02$ & $0.00$ & $0.07$ & $0.1$ & $15.4 \pm 0.3 \pm 0.1$\\
$\varepsilon_3$ & $0.02$ & $0.01$ & $0.07$ & $0.1$ & $20.2 \pm 0.3 \pm 0.1$\\
$\varepsilon_4$ & $0.00$ & $0.01$ & $0.06$ & $0.1$ & $11.5 \pm 0.2 \pm 0.1$\\
$\varepsilon_5$ & $0.00$ & $0.00$ & $0.06$ & $0.1$ & $11.9 \pm 0.2 \pm 0.1$\\
$\varepsilon_6$ & $0.02$ & $0.00$ & $0.05$ & $0.1$ & $ 8.9 \pm 0.2 \pm 0.1$\\
$\varepsilon_7$ & $0.07$ & $0.01$ & $0.06$ & $0.1$ & $13.8 \pm 0.2 \pm 0.1$\\\hline 
$w_1$ & $0.01$ & $0.01$ & $0.23$ & $0.2$ & $48.2 \pm 0.9 \pm 0.2$\\
$w_2$ & $0.01$ & $0.01$ & $0.24$ & $0.2$ & $40.9 \pm 0.9 \pm 0.2$\\
$w_3$ & $0.05$ & $0.00$ & $0.19$ & $0.2$ & $28.3 \pm 0.7 \pm 0.2$\\
$w_4$ & $0.04$ & $0.01$ & $0.21$ & $0.2$ & $15.6 \pm 0.8 \pm 0.2$\\
$w_5$ & $0.03$ & $0.01$ & $0.17$ & $0.2$ & $11.9 \pm 0.7 \pm 0.2$\\
$w_6$ & $0.02$ & $0.01$ & $0.15$ & $0.2$ & $ 5.4 \pm 0.6 \pm 0.2$\\
$w_7$ & $0.00$ & $0.00$ & $0.06$ & $0.1$ & $ 1.2 \pm 0.2 \pm 0.1$\\
\hline
$\mu_1$ & $0.03$ & $0.02$ & $0.46$ & $0.5$ &  $0.2 \pm 1.8 \pm 0.5$\\
$\mu_2$ & $0.03$ & $0.05$ & $0.49$ & $0.5$ &  $0.3 \pm 1.9 \pm 0.5$\\
$\mu_3$ & $0.00$ & $0.02$ & $0.43$ & $0.4$ &  $1.5 \pm 1.6 \pm 0.4$\\
$\mu_4$ & $0.10$ & $0.09$ & $0.55$ & $0.6$ & -$3.0 \pm 2.1 \pm 0.6$\\
$\mu_5$ & $0.03$ & $0.06$ & $0.51$ & $0.5$ & -$3.1 \pm 2.0 \pm 0.5$\\
$\mu_6$ & $0.09$ & $0.10$ & $0.56$ & $0.6$ & -$0.6 \pm 2.2 \pm 0.6$\\
$\mu_7$ & $0.08$ & $0.07$ & $0.42$ & $0.4$ & -$0.6 \pm 1.6 \pm 0.4$\\\hline
$\Delta w_1$ & $0.10$ & $0.09$ & $0.46$ & $0.5$ &  $7.1 \pm 1.8 \pm 0.5$\\
$\Delta w_2$ & $0.08$ & $0.07$ & $0.48$ & $0.5$ &  $5.8 \pm 1.9 \pm 0.5$\\
$\Delta w_3$ & $0.06$ & $0.05$ & $0.39$ & $0.4$ &  $1.3 \pm 1.5 \pm 0.4$\\
$\Delta w_4$ & $0.04$ & $0.05$ & $0.43$ & $0.4$ & -$1.2 \pm 1.6 \pm 0.4$\\
$\Delta w_5$ & $0.06$ & $0.04$ & $0.35$ & $0.4$ &  $2.0 \pm 1.4 \pm 0.4$\\
$\Delta w_6$ & $0.06$ & $0.05$ & $0.31$ & $0.3$ &  $1.8 \pm 1.2 \pm 0.3$\\
$\Delta w_7$ & $0.01$ & $0.01$ & $0.12$ & $0.1$ &  $0.5 \pm 0.5 \pm 0.1$
\\\hline
    \end{tabular}

    \label{tab:sys_fitres_data}
\end{table*}

% \clearpage

\begin{table*}[p]
    \centering
    \caption{Results for $\varepsilon_i$, $w_i$, $\mu_i$, and $\Delta w_i$ for the DNN tagger: systematic uncertainties associated with the $\Delta E$ parametrization, fit bias, and fit model, and total systematic uncertainty  are shown together with the fit results~(with stat. and syst. uncertainty). The results are given in percent.}    
 \scriptsize   
    \begin{tabular}{c r c c c r }
\multicolumn{6}{l}{ $\PBzero\to\PD^{(*)-}h^{+}$}\\
    \hline\hline
    \multicolumn{1}{c}{Parameter} & 
    \multicolumn{1}{c}{$\Delta E$ PDF} &
    \multicolumn{1}{c}{Fit bias}  &  
    \multicolumn{1}{c}{Model} &
    \multicolumn{1}{c}{Total syst. uncty.}  & \multicolumn{1}{c}{Fit results}\\\hline 
$\varepsilon_1$ & $0.02$ & $0.01$ & $0.09$ & $0.1$ & $14.3 \pm 0.3 \pm 0.1$\\
$\varepsilon_2$ & $0.00$ & $0.01$ & $0.09$ & $0.1$ & $17.9 \pm 0.3 \pm 0.1$\\
$\varepsilon_3$ & $0.01$ & $0.02$ & $0.10$ & $0.1$ & $22.5 \pm 0.4 \pm 0.1$\\
$\varepsilon_4$ & $0.00$ & $0.00$ & $0.07$ & $0.1$ & $11.0 \pm 0.3 \pm 0.1$\\
$\varepsilon_5$ & $0.00$ & $0.00$ & $0.07$ & $0.1$ & $10.4 \pm 0.3 \pm 0.1$\\
$\varepsilon_6$ & $0.01$ & $0.00$ & $0.07$ & $0.1$ & $ 9.6 \pm 0.2 \pm 0.1$\\
$\varepsilon_7$ & $0.02$ & $0.01$ & $0.08$ & $0.1$ & $14.2 \pm 0.3 \pm 0.1$ \\\hline
$w_1$ & $0.05$ & $0.07$ & $0.51$ & $0.5$ & $48.2 \pm 1.9 \pm 0.5$\\
$w_2$ & $0.02$ & $0.02$ & $0.46$ & $0.5$ & $43.6 \pm 1.6 \pm 0.5$\\
$w_3$ & $0.03$ & $0.02$ & $0.40$ & $0.4$ & $33.9 \pm 1.4 \pm 0.4$\\
$w_4$ & $0.08$ & $0.06$ & $0.54$ & $0.6$ & $19.3 \pm 1.9 \pm 0.6$\\
$w_5$ & $0.05$ & $0.04$ & $0.52$ & $0.5$ & $19.7 \pm 1.9 \pm 0.5$\\
$w_6$ & $0.07$ & $0.04$ & $0.49$ & $0.5$ & $10.8 \pm 1.8 \pm 0.5$\\
$w_7$ & $0.02$ & $0.04$ & $0.36$ & $0.4$ & $ 3.5 \pm 1.2 \pm 0.4$ \\
\hline 
$\mu_1$ & $0.35$ & $0.24$ & $1.02$ & $1.1$ &  $3.7 \pm 3.7 \pm 1.1$\\
$\mu_2$ & $0.12$ & $0.08$ & $0.92$ & $0.9$ &  $7.3 \pm 3.2 \pm 0.9$\\
$\mu_3$ & $0.00$ & $0.02$ & $0.81$ & $0.8$ &  $4.6 \pm 2.9 \pm 0.8$\\
$\mu_4$ & $0.39$ & $0.35$ & $1.14$ & $1.3$ &  $2.2 \pm 4.0 \pm 1.3$\\
$\mu_5$ & $0.12$ & $0.16$ & $1.11$ & $1.1$ &  $7.4 \pm 4.1 \pm 1.1$\\
$\mu_6$ & $0.16$ & $0.06$ & $1.10$ & $1.1$ &  $1.5 \pm 4.1 \pm 1.1$\\
$\mu_7$ & $0.12$ & $0.07$ & $0.80$ & $0.8$ & -$2.5 \pm 3.1 \pm 0.8$\\
\hline    
$\Delta w_1$ & $0.13$ & $0.11$ & $0.64$ & $0.7$ & -$1.1 \pm 2.3 \pm 0.7$\\
$\Delta w_2$ & $0.18$ & $0.16$ & $0.58$ & $0.6$ &  $5.6 \pm 2.1 \pm 0.6$\\
$\Delta w_3$ & $0.01$ & $0.01$ & $0.53$ & $0.5$ &  $7.1 \pm 1.8 \pm 0.5$\\
$\Delta w_4$ & $0.08$ & $0.05$ & $0.77$ & $0.8$ &  $4.5 \pm 2.8 \pm 0.8$\\
$\Delta w_5$ & $0.06$ & $0.03$ & $0.78$ & $0.8$ &  $7.9 \pm 2.7 \pm 0.8$\\
$\Delta w_6$ & $0.09$ & $0.12$ & $0.80$ & $0.8$ &  $5.7 \pm 3.0 \pm 0.8$\\
$\Delta w_7$ & $0.12$ & $0.09$ & $0.61$ & $0.6$ &  $3.4 \pm 2.3 \pm 0.6$ \\\hline
\multicolumn{4}{l}{ }\\
\multicolumn{4}{l}{ $\PBplus\to\APD^{(*)0}h^{+}$}\\\hline\hline
    \multicolumn{1}{c}{Parameter} & 
    \multicolumn{1}{c}{$\Delta E$ PDF} &
    \multicolumn{1}{c}{Fit bias}  &  
    \multicolumn{1}{c}{Model} &
    \multicolumn{1}{c}{Total syst. uncty.}  & \multicolumn{1}{c}{Fit results}\\\hline
$\varepsilon_1$ & $0.01$ & $0.01$ & $0.07$ & $0.1$ & $12.7 \pm 0.3 \pm 0.1$\\
$\varepsilon_2$ & $0.04$ & $0.00$ & $0.07$ & $0.1$ & $16.2 \pm 0.3 \pm 0.1$\\
$\varepsilon_3$ & $0.02$ & $0.01$ & $0.08$ & $0.1$ & $21.2 \pm 0.3 \pm 0.1$\\
$\varepsilon_4$ & $0.01$ & $0.00$ & $0.06$ & $0.1$ & $11.0 \pm 0.2 \pm 0.1$\\
$\varepsilon_5$ & $0.01$ & $0.01$ & $0.06$ & $0.1$ & $11.3 \pm 0.2 \pm 0.1$\\
$\varepsilon_6$ & $0.01$ & $0.01$ & $0.06$ & $0.1$ & $10.3 \pm 0.2 \pm 0.1$\\
$\varepsilon_7$ & $0.08$ & $0.01$ & $0.07$ & $0.1$ & $17.3 \pm 0.2 \pm 0.1$
 \\\hline
$w_1$ & $0.01$ & $0.00$ & $0.27$ & $0.3$ & $47.2 \pm 1.1 \pm 0.3$\\
$w_2$ & $0.01$ & $0.00$ & $0.24$ & $0.2$ & $40.9 \pm 0.9 \pm 0.2$\\
$w_3$ & $0.06$ & $0.02$ & $0.19$ & $0.2$ & $29.8 \pm 0.7 \pm 0.2$\\
$w_4$ & $0.02$ & $0.02$ & $0.23$ & $0.2$ & $17.2 \pm 0.9 \pm 0.2$\\
$w_5$ & $0.06$ & $0.02$ & $0.19$ & $0.2$ & $11.9 \pm 0.7 \pm 0.2$\\
$w_6$ & $0.01$ & $0.02$ & $0.14$ & $0.1$ & $ 4.6 \pm 0.5 \pm 0.1$\\
$w_7$ & $0.01$ & $0.00$ & $0.06$ & $0.1$ & $ 1.8 \pm 0.2 \pm 0.1$\\
\hline
$\mu_1$ & $0.08$ & $0.10$ & $0.54$ & $0.6$ &  $0.4 \pm 2.1 \pm 0.6$\\
$\mu_2$ & $0.11$ & $0.08$ & $0.49$ & $0.5$ &  $3.2 \pm 1.9 \pm 0.5$\\
$\mu_3$ & $0.04$ & $0.02$ & $0.42$ & $0.4$ &  $4.4 \pm 1.6 \pm 0.4$\\
$\mu_4$ & $0.01$ & $0.04$ & $0.57$ & $0.6$ & -$1.4 \pm 2.2 \pm 0.6$\\
$\mu_5$ & $0.02$ & $0.02$ & $0.54$ & $0.5$ & -$4.1 \pm 2.1 \pm 0.5$\\
$\mu_6$ & $0.07$ & $0.05$ & $0.53$ & $0.5$ & -$5.4 \pm 2.1 \pm 0.5$\\
$\mu_7$ & $0.04$ & $0.03$ & $0.37$ & $0.4$ & -$4.5 \pm 1.5 \pm 0.4$\\
\hline 
$\Delta w_1$ & $0.02$ & $0.04$ & $0.54$ & $0.5$ & $0.8 \pm 2.1 \pm 0.5$\\
$\Delta w_2$ & $0.00$ & $0.03$ & $0.48$ & $0.5$ & $6.4 \pm 1.8 \pm 0.5$\\
$\Delta w_3$ & $0.05$ & $0.08$ & $0.39$ & $0.4$ & $5.3 \pm 1.5 \pm 0.4$\\
$\Delta w_4$ & $0.11$ & $0.12$ & $0.46$ & $0.5$ & $3.4 \pm 1.8 \pm 0.5$\\
$\Delta w_5$ & $0.14$ & $0.11$ & $0.38$ & $0.4$ & $2.5 \pm 1.5 \pm 0.4$\\
$\Delta w_6$ & $0.05$ & $0.03$ & $0.29$ & $0.3$ & $0.5 \pm 1.1 \pm 0.3$\\
$\Delta w_7$ & $0.01$ & $0.02$ & $0.11$ & $0.1$ & $0.4 \pm 0.5 \pm 0.1$\\
\hline
    \end{tabular}
    \label{tab:sys_fitres_dataDNN}
\end{table*}

\begin{table*}[!h]
    \centering
    \caption{Effective efficiencies for the category-based~(FBDT) tagger: systematic uncertainties associated with the $\Delta E$ parametrization, fit bias, and fit model, and total systematic uncertainty  are shown together with the fit results~(with stat. and syst. uncertainty). The results are given in percent.}    
\scriptsize    
    \begin{tabular}{c r c c c r }
\multicolumn{6}{l}{ $\PBzero\to\PD^{(*)-}h^{+}$}\\
    \hline\hline
    \multicolumn{1}{c}{Parameter} & 
    \multicolumn{1}{c}{$\Delta E$ PDF} &
    \multicolumn{1}{c}{Fit bias}  &  
    \multicolumn{1}{c}{Model} &
    \multicolumn{1}{c}{Total syst. uncty.}  & \multicolumn{1}{c}{Fit results}\\\hline 
$\varepsilon_{\rm{eff}, 1}$ & $0.02$ & $0.02$ & $0.02$ & $0.0$ & $ 0.1 \pm 0.1 \pm 0.0$\\
$\varepsilon_{\rm{eff}, 2}$ & $0.02$ & $0.02$ & $0.06$ & $0.1$ & $ 0.5 \pm 0.2 \pm 0.1$\\
$\varepsilon_{\rm{eff}, 3}$ & $0.03$ & $0.01$ & $0.13$ & $0.1$ & $ 3.3 \pm 0.5 \pm 0.1$\\
$\varepsilon_{\rm{eff}, 4}$ & $0.04$ & $0.04$ & $0.14$ & $0.2$ & $ 3.3 \pm 0.5 \pm 0.2$\\
$\varepsilon_{\rm{eff}, 5}$ & $0.01$ & $0.00$ & $0.16$ & $0.2$ & $ 6.1 \pm 0.6 \pm 0.2$\\
$\varepsilon_{\rm{eff}, 6}$ & $0.00$ & $0.00$ & $0.16$ & $0.2$ & $ 5.4 \pm 0.5 \pm 0.2$\\
$\varepsilon_{\rm{eff}, 7}$ & $0.03$ & $0.01$ & $0.20$ & $0.2$ & $11.3 \pm 0.6 \pm 0.2$
 \\\hline
 Total & $0.06$ & $0.05$ & $0.36$ & $0.4$ & $30.0 \pm 1.2 \pm 0.4$\\\hline 
 \multicolumn{4}{l}{ }\\
\multicolumn{4}{l}{ $\PBplus\to\APD^{(*)0}h^{+}$}\\\hline\hline
    \multicolumn{1}{c}{Parameter} & 
    \multicolumn{1}{c}{$\Delta E$ PDF} &
    \multicolumn{1}{c}{Fit bias}  &  
    \multicolumn{1}{c}{Model} &
    \multicolumn{1}{c}{Total syst. uncty.}  & \multicolumn{1}{c}{Fit results}\\\hline 
$\varepsilon_{\rm{eff}, 1}$ & $0.00$ & $0.00$ & $0.01$ & $0.0$ & $ 0.0 \pm 0.0 \pm 0.0$\\
$\varepsilon_{\rm{eff}, 2}$ & $0.00$ & $0.00$ & $0.03$ & $0.0$ & $ 0.5 \pm 0.1 \pm 0.0$\\
$\varepsilon_{\rm{eff}, 3}$ & $0.01$ & $0.00$ & $0.07$ & $0.1$ & $ 3.8 \pm 0.3 \pm 0.1$\\
$\varepsilon_{\rm{eff}, 4}$ & $0.01$ & $0.00$ & $0.07$ & $0.1$ & $ 5.4 \pm 0.3 \pm 0.1$\\
$\varepsilon_{\rm{eff}, 5}$ & $0.01$ & $0.01$ & $0.08$ & $0.1$ & $ 7.0 \pm 0.3 \pm 0.1$\\
$\varepsilon_{\rm{eff}, 6}$ & $0.02$ & $0.00$ & $0.06$ & $0.1$ & $ 7.1 \pm 0.2 \pm 0.1$\\
$\varepsilon_{\rm{eff}, 7}$ & $0.07$ & $0.01$ & $0.06$ & $0.1$ & $13.2 \pm 0.2 \pm 0.1$
\\\hline
Total & $0.08$ & $0.02$ & $0.16$ & $0.2$ & $37.0 \pm 0.6 \pm 0.2$ \\\hline
    \end{tabular}

    \label{tab:sys_effeff_data}
\end{table*}

\begin{table*}[h!]
    \centering
    \caption{Effective efficiencies for the DNN tagger: systematic uncertainties associated with the $\Delta E$ parametrization, fit bias, and fit model, and total systematic uncertainty are shown together with the fit results~(with stat. and syst. uncertainty). The results are given in percent.}    
\scriptsize    
    \begin{tabular}{c r c c c r }
\multicolumn{6}{l}{ $\PBzero\to\PD^{(*)-}h^{+}$}\\
    \hline\hline
    \multicolumn{1}{c}{Parameter} & 
    \multicolumn{1}{c}{$\Delta E$ PDF} &
    \multicolumn{1}{c}{Fit bias}  &  
    \multicolumn{1}{c}{Model} &
    \multicolumn{1}{c}{Total syst. uncty.}  & \multicolumn{1}{c}{Fit results}\\\hline 
$\varepsilon_{\rm{eff}, 1}$ & $0.02$ & $0.02$ & $0.01$ & $<0.1$ & $ 0.0 \pm 0.0 \pm 0.0$\\
$\varepsilon_{\rm{eff}, 2}$ & $0.01$ & $0.01$ & $0.05$ & $0.1$ & $ 0.3 \pm 0.1 \pm 0.1$\\
$\varepsilon_{\rm{eff}, 3}$ & $0.00$ & $0.00$ & $0.13$ & $0.1$ & $ 2.3 \pm 0.4 \pm 0.1$\\
$\varepsilon_{\rm{eff}, 4}$ & $0.00$ & $0.00$ & $0.12$ & $0.1$ & $ 4.2 \pm 0.5 \pm 0.1$\\
$\varepsilon_{\rm{eff}, 5}$ & $0.01$ & $0.01$ & $0.15$ & $0.2$ & $ 3.8 \pm 0.5 \pm 0.2$\\
$\varepsilon_{\rm{eff}, 6}$ & $0.02$ & $0.02$ & $0.16$ & $0.2$ & $ 5.9 \pm 0.6 \pm 0.2$\\
$\varepsilon_{\rm{eff}, 7}$ & $0.03$ & $0.01$ & $0.23$ & $0.2$ & $12.3 \pm 0.7 \pm 0.2$\\\hline
Total & $0.05$ & $0.03$ & $0.37$ & $0.4$ & $28.8 \pm 1.2 \pm 0.4$\\\hline \multicolumn{4}{l}{ }\\
\multicolumn{4}{l}{ $\PBplus\to\APD^{(*)0}h^{+}$}\\\hline\hline
    \multicolumn{1}{c}{Parameter} & 
    \multicolumn{1}{c}{$\Delta E$ PDF} &
    \multicolumn{1}{c}{Fit bias}  &  
    \multicolumn{1}{c}{Model} &
    \multicolumn{1}{c}{Total syst. uncty.}  & \multicolumn{1}{c}{Fit results}\\\hline
$\varepsilon_{\rm{eff}, 1}$ & $0.00$ & $0.00$ & $0.01$ & $<0.1$ & $ 0.0 \pm 0.0 \pm 0.0$\\
$\varepsilon_{\rm{eff}, 2}$ & $0.00$ & $0.00$ & $0.03$ & $<0.1$ & $ 0.5 \pm 0.1 \pm 0.0$\\
$\varepsilon_{\rm{eff}, 3}$ & $0.02$ & $0.01$ & $0.07$ & $0.1$ & $ 3.5 \pm 0.3 \pm 0.1$\\
$\varepsilon_{\rm{eff}, 4}$ & $0.01$ & $0.00$ & $0.07$ & $0.1$ & $ 4.7 \pm 0.3 \pm 0.1$\\
$\varepsilon_{\rm{eff}, 5}$ & $0.01$ & $0.00$ & $0.07$ & $0.1$ & $ 6.6 \pm 0.3 \pm 0.1$\\
$\varepsilon_{\rm{eff}, 6}$ & $0.01$ & $0.01$ & $0.07$ & $0.1$ & $ 8.5 \pm 0.3 \pm 0.1$\\
$\varepsilon_{\rm{eff}, 7}$ & $0.09$ & $0.00$ & $0.07$ & $0.1$ & $16.1 \pm 0.3 \pm 0.1$\\\hline
                 Total & $0.09$ & $0.01$ & $0.16$ & $0.2$ & $39.9 \pm 0.6 \pm 0.2$\\\hline
    \end{tabular}

    \label{tab:sys_effeff_dataDNN}
\end{table*}

\clearpage

\clearpage

\section{Linearity check}

\label{sec:linearity}

\begin{figure*}
\centering
\includegraphics[width=0.475\linewidth]{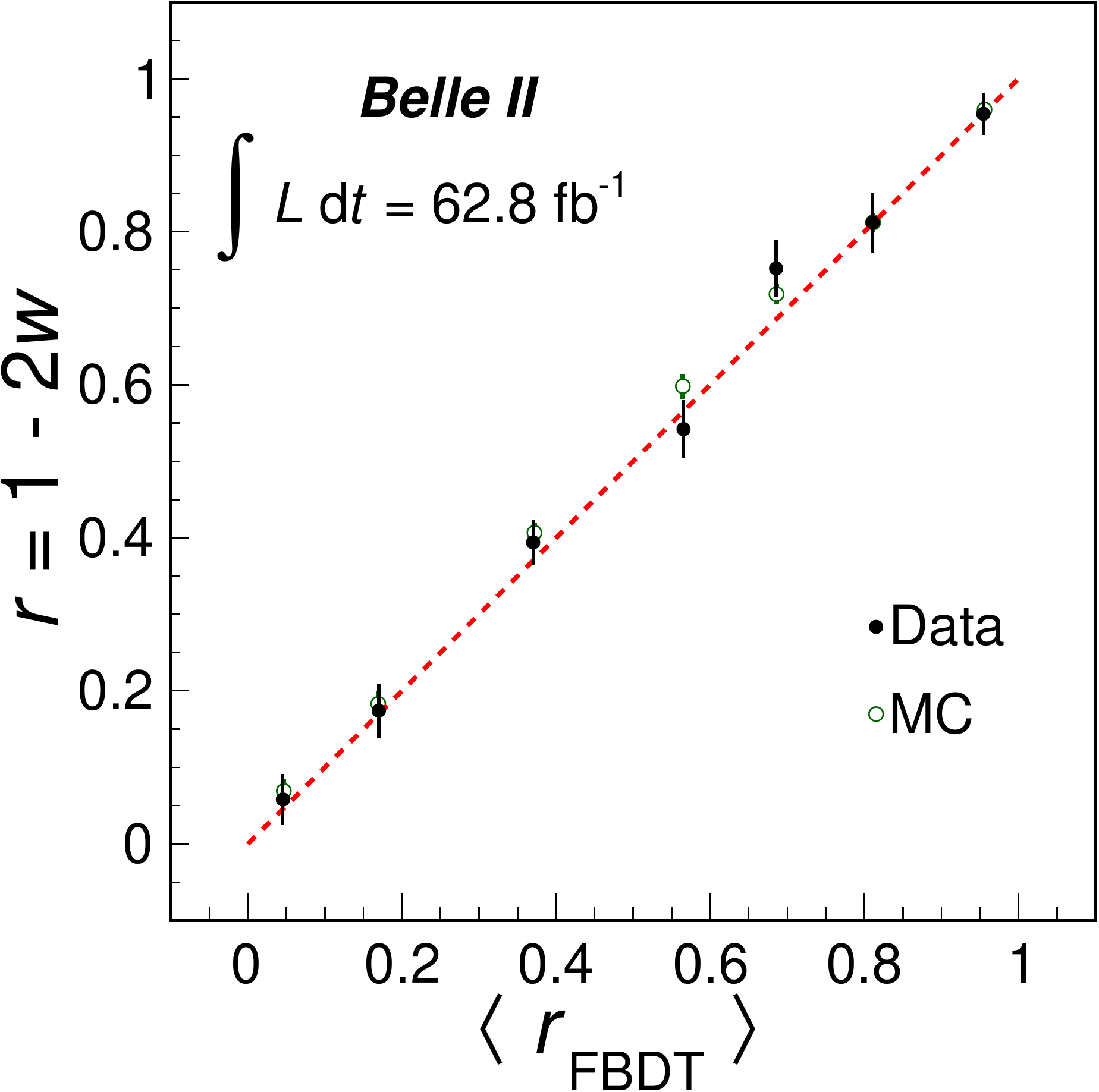}\hfill
\includegraphics[width=0.475\linewidth]{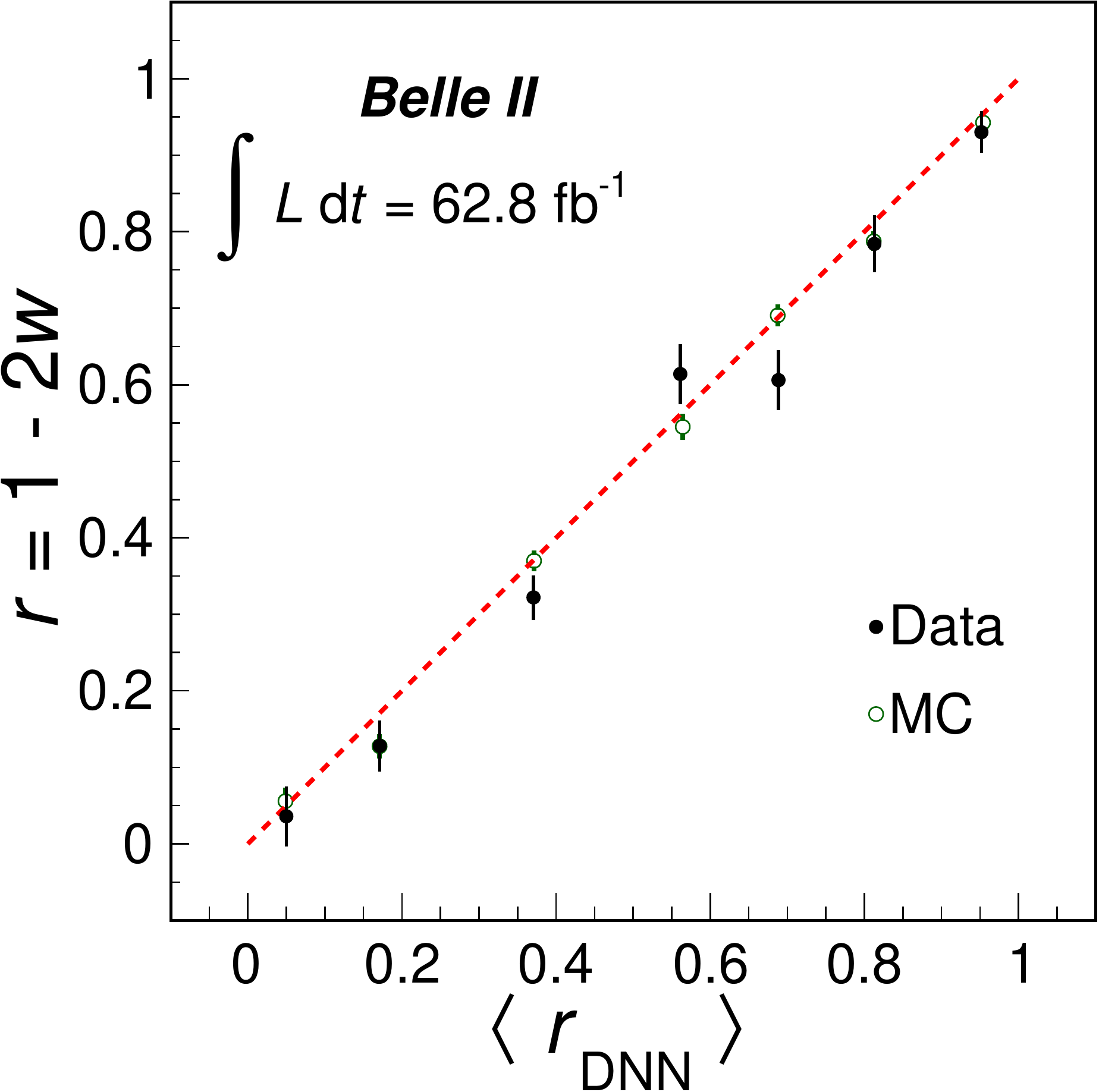}\\
        \vspace{0.3cm} \text{(a) $\PBzero\to\PD^{(*)-}h^{+}$.}\\\vspace{1.0cm}
        
\includegraphics[width=0.475\linewidth]{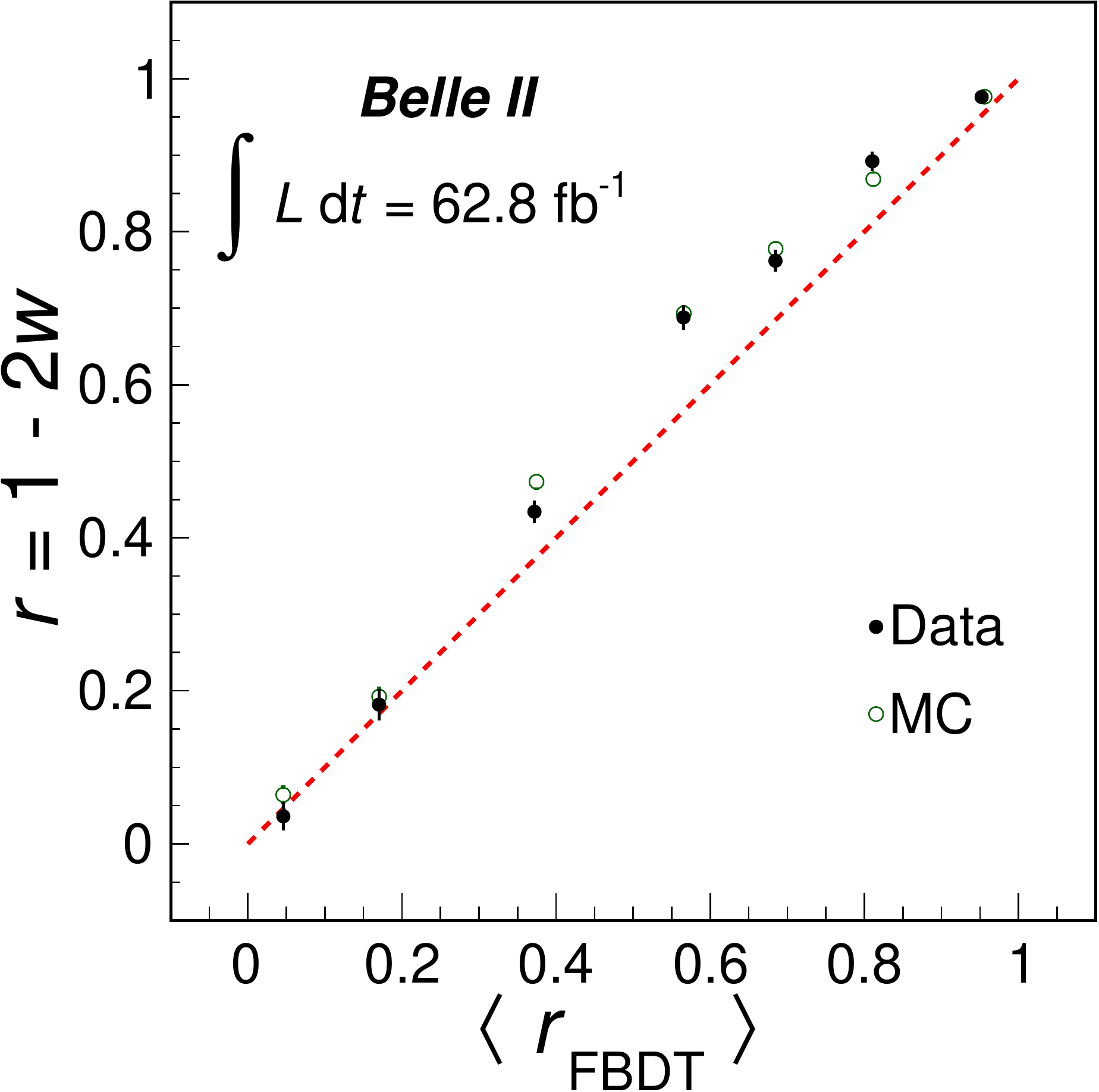}\hfill
\includegraphics[width=0.475\linewidth]{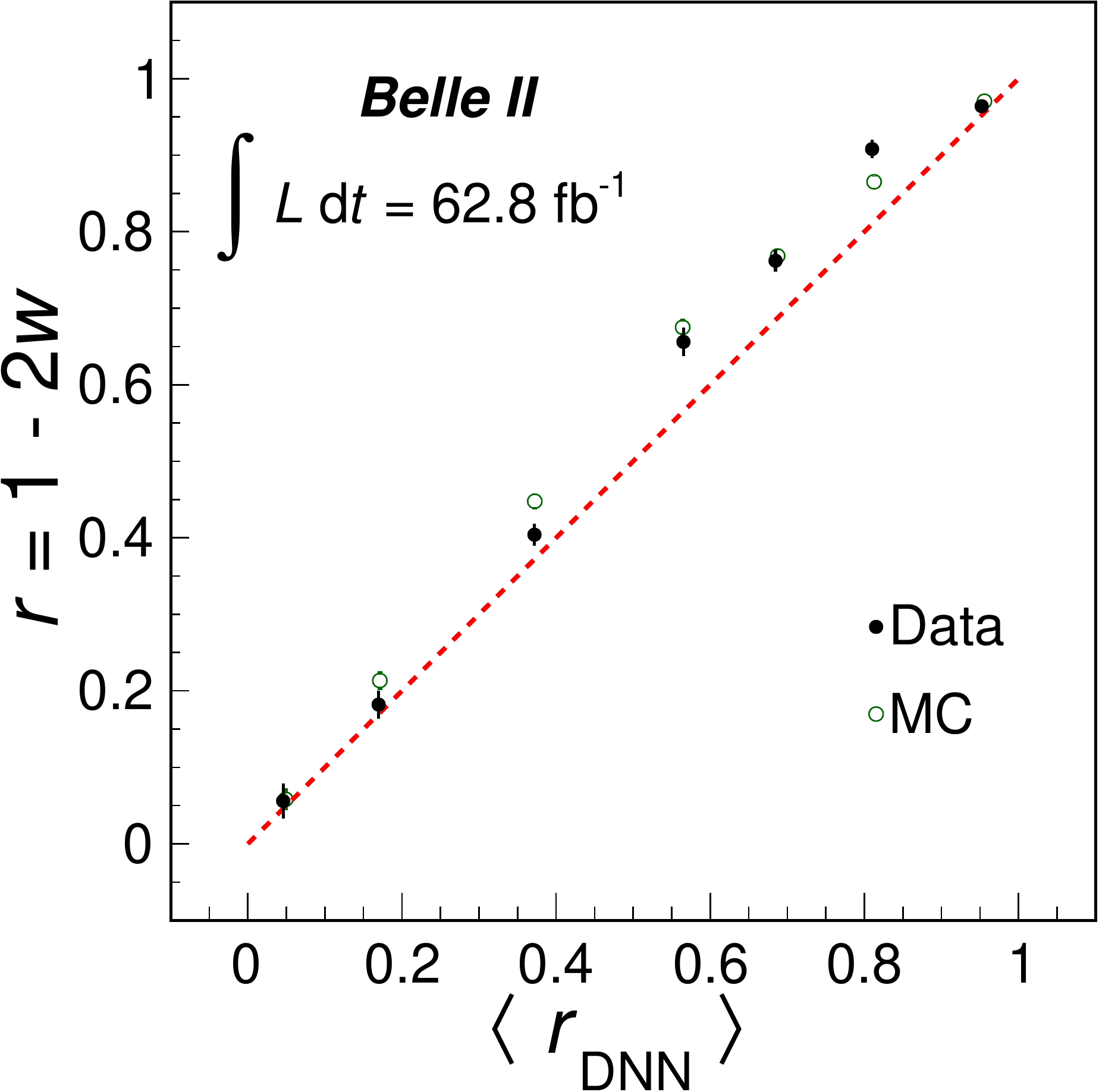}\\
        \vspace{0.3cm} \text{(b) $\PBplus\to\APD^{(*)0}h^{+}$.}\\\vspace{0.5cm}  
        
\caption{\label{fig:cal_plot_data} Dilution factor $r = 1-2w$ as a function of the mean dilution~\mbox{$\langle \vert q\cdot r \vert \rangle$} provided by the (left) category-based and (right) DNN flavor tagger in data and MC~simulation for (top)~neutral and (bottom)~charged \mbox{$\PB\to\PD^{(*)}\Ph^{+}$}~candidates. The red dashed lines correspond to a linear function with an intercept at zero and a slope of one, corresponding to a perfect agreement between predicted and measured dilution. }
\end{figure*}

 We check whether the dilution $r$ provided by the flavor tagger corresponds to the actual definition $r \coloneqq 1-2w$ by performing a linearity check.  Figure~\ref{fig:cal_plot_data} shows the linearity check for both flavor taggers in simulation and data.

For simulation, we determine the true wrong-tag fraction $w_{\rm MC}$ by comparing the MC~truth with the flavor-tagger output, and calculate the true dilution $r_{\rm MC} = 1 -2w_{\rm MC}$. The mean dilution $\langle r_{\rm FBDT}\rangle$ is simply the mean of $\vert q\cdot r_{\rm FBDT}\vert$ for correctly associated MC~events in each $r$~bin.  For data, we obtain the mean \mbox{$\langle r_{\rm FBDT} \rangle = \langle\vert q\cdot r_{\rm FBDT} \vert\rangle$} values from the signal $q\cdot r_{\rm FBDT}$ distribution provided by the $s\mathcal{P}lot$ analysis in Sec.~\ref{sec:splot}. The dilution \mbox{$r=1-2\cdot w$} in data is obtained from the fit results for $w$. The linearity verifies the equivalence on average between the dilution provided by the flavor tagger and the measured one within the uncertainties.  For charged $\PB$~candidates, we observe a slightly non-linear behavior, which is attributed to the fact that both flavor taggers are optimized and trained only for neutral \PB~mesons. However, we observe a good agreement between data and simulation for both neutral and charged $\PB$~candidates.

\clearpage

\clearpage

\section{Comparison with the previous Belle algorithm}

\label{sec:compBelle}

\begin{figure*}[h!]
 
     \centering
 
 \vspace{0.5cm}
 
     \includegraphics[width=0.8\textwidth]{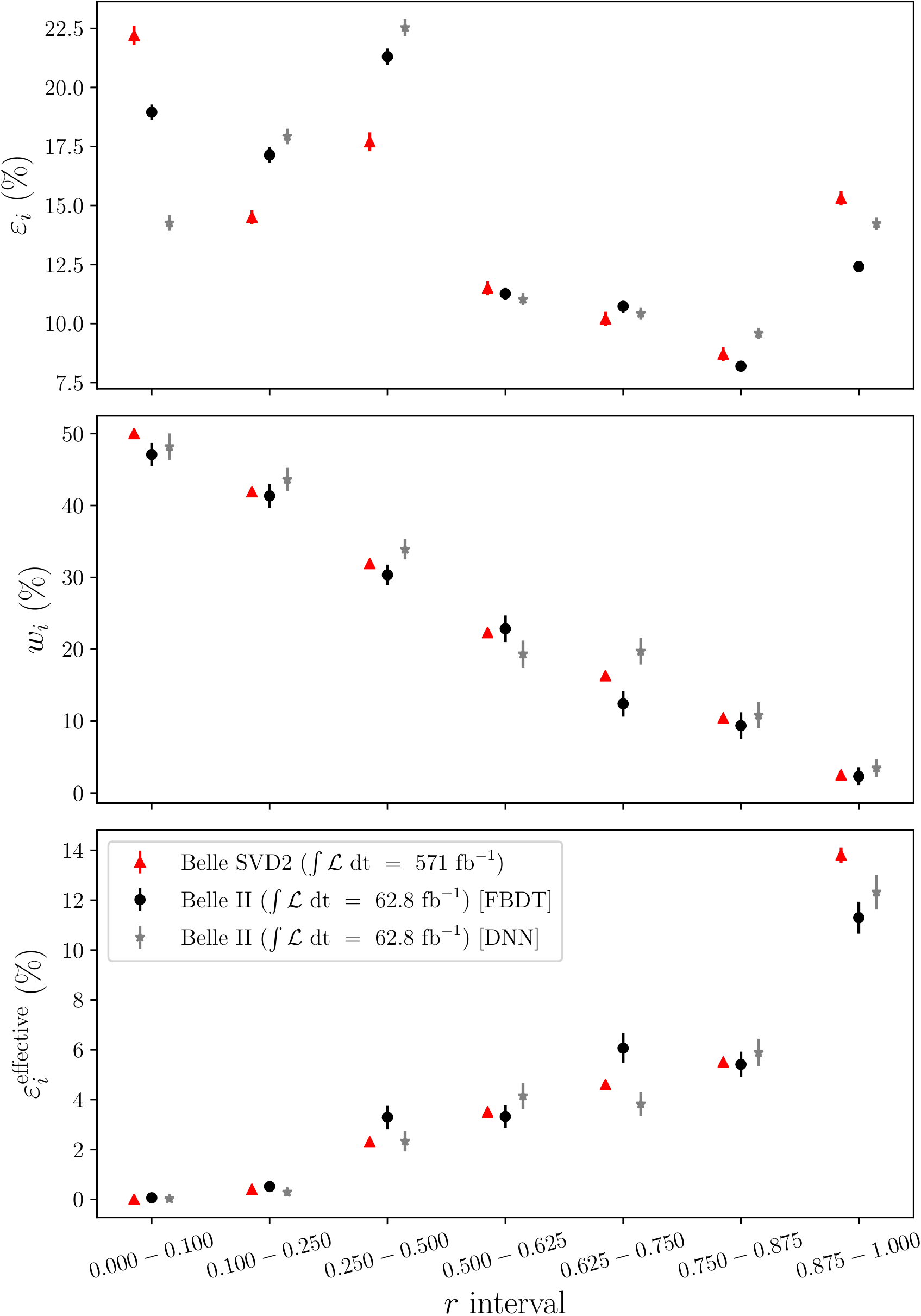}
     
\vspace{0.5cm}
     
     \caption{ Performance of the Belle~II category-based~(FBDT) and deep-learning~(DNN) flavor taggers in 2019-2020 Belle~II data and of the Belle~flavor tagger in 2003-2010 Belle data~\cite{Bevan:2014iga}  taken with the second silicon-vertex detector configuration~(SVD2).}
     
\vspace{0.5cm}
     
     \label{fig:Belle2-vs-Belle_FBDT+DNN}
 \end{figure*}

\begin{figure*}[h!]
\centering

 \vspace{0.5cm}

    \includegraphics[width=0.76\textwidth]{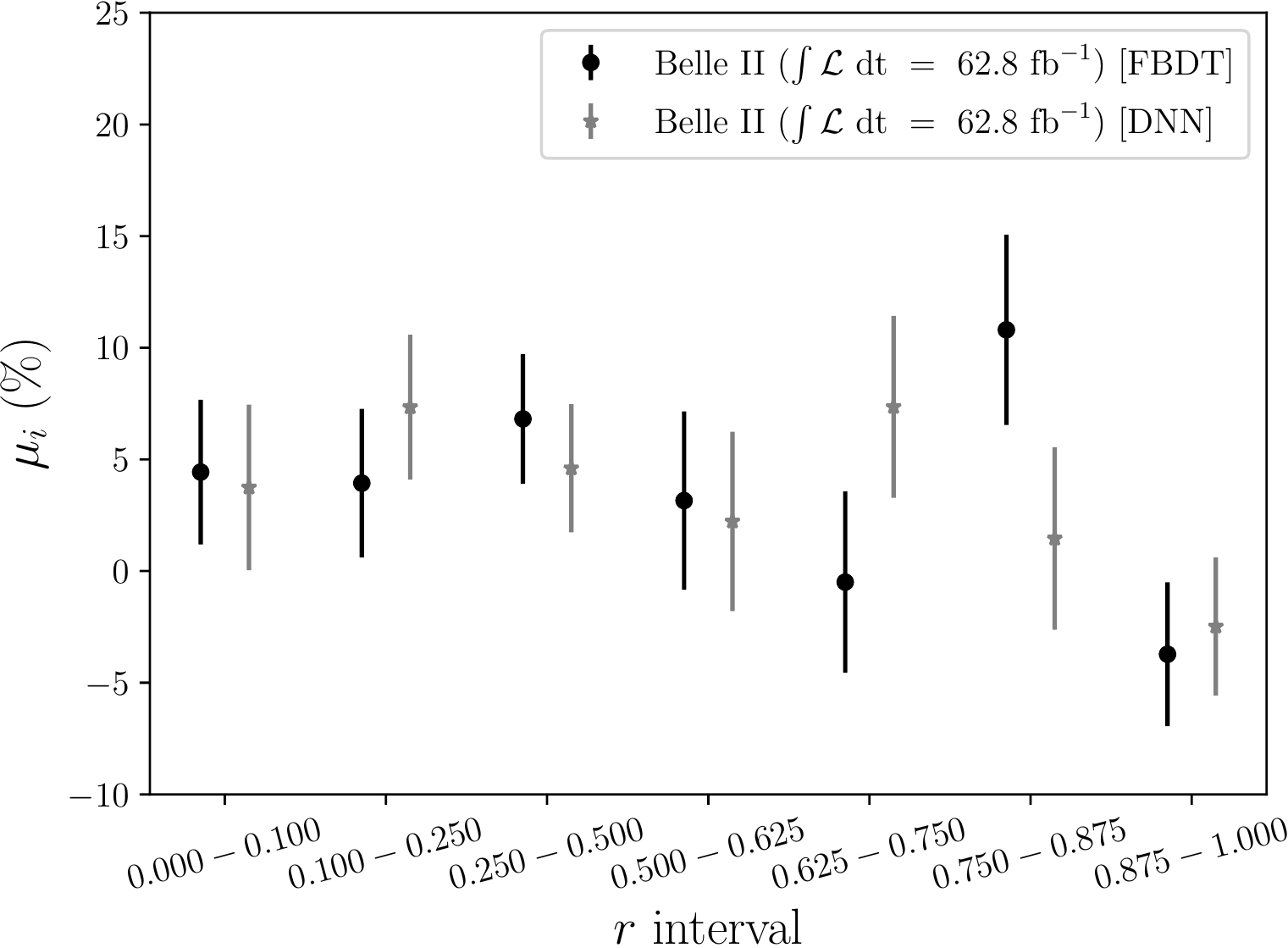}\\\vspace{0.5cm}
    
    \includegraphics[width=0.76\textwidth]{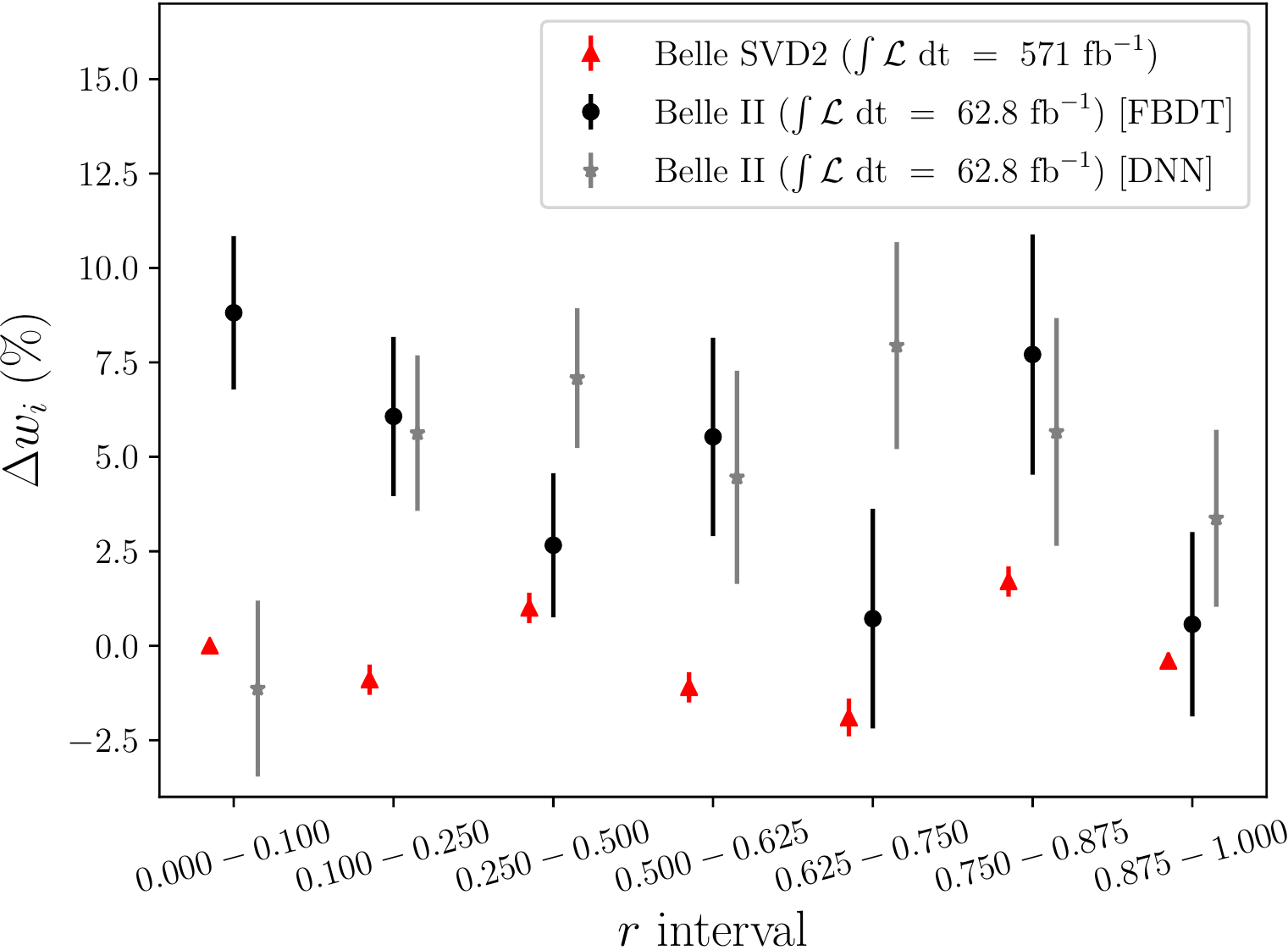}

 \vspace{0.5cm}
    
    \caption{ Comparison of (top) tagging efficiency asymmetries $\mu_i$ and (bottom)  wrong-tag fraction asymmetries $\Delta w_i$ for the Belle~II category-based~(FBDT) and deep-learning~(DNN) flavor taggers in 2019-2020  Belle~II data and for the Belle~flavor tagger in 2003-2010 Belle data~\cite{Bevan:2014iga}  taken with the second silicon-vertex detector configuration~(SVD2). There are no available Belle results for $\mu$.}
    \label{fig:Belle2-vs-Belle_FBDT+DNN_dw_mu}
\end{figure*}

A comparison of the current results with the latest results on flavor tagging obtained by Belle~\cite{Bevan:2014iga} provides interesting insight about the current and projected performance of Belle~II. We compare partial efficiencies, wrong-tag fractions, total effective efficiencies, and wrong-tag asymmetries in each $r$-bin. Table~\ref{tab:Belle2-vs-Belle_FBDT+DNN}, and Figs.~\ref{fig:Belle2-vs-Belle_FBDT+DNN} and~\ref{fig:Belle2-vs-Belle_FBDT+DNN_dw_mu} compare Belle~II and Belle results. 

The Belle flavor tagger, which was a category-based algorithm, reached a total effective efficiency of $(30.1\pm 0.4)\%$ on Belle data~\cite{Bevan:2014iga}. We observe about the same or slightly better performance than Belle in all bins except in the highest $r$~bin, for which we observe a smaller partial efficiency and also a slightly worse performance with respect to expectations~(see Figs.~\ref{fig:qrSigDistDataFBDT} and~\ref{fig:qrSigDistDataDNN}).  For the wrong-tag fractions, we observe larger asymmetries between $\PBzero$ and $\APBzero$ than Belle.

In comparison with the Belle algorithm~\cite{Kakuno:2004cf,Bevan:2014iga}, the Belle~II category-based flavor tagger considers more flavor signatures and more input variables, and is fully based on multivariate methods avoiding the cut-based identification of decay products and exploiting the correlations between input variables and between flavor signatures. 

% For the current version of the Belle~II flavor taggers, we avoid the use of track impact parameters~(displacement from nominal interaction point), which are not yet well simulated for small displacements below $0.1\,\si{cm}$. 
For the current version of the category-based and the DNN flavor taggers, we obtain a total effective efficiency around $32.5\%$ in simulation. Previous studies~\cite{Abudinen:2018,Gemmler:2020,Kou:2018nap} show that using track impact parameters as additional input variables potentially improves the total effective tagging efficiency by about $2$ to $3\%$ in its absolute value. Nonetheless, the current results show that Belle~II can reach a tagging performance comparable with the one obtained by Belle even with a not yet fully-optimized calibration of the tracking and PID systems, and operating in harsher background conditions than those experienced by Belle.

{\renewcommand{\arraystretch}{1.1}
 \begin{table*}[h!]
 \centering
 \caption{\label{tab:Belle2-vs-Belle_FBDT+DNN}Partial efficiencies $\varepsilon_i$, wrong-tag fractions $w_i$, total effective efficiencies $\varepsilon_{\text{eff},i}$, tagging efficiency asymmetries $\mu = \Delta\varepsilon/(2\varepsilon)$, and wrong-tag fraction asymmetries $\Delta w$ obtained with the Belle~II category-based~(FBDT) and deep-learning~(DNN) flavor taggers in 2019-2020 Belle~II data and with the Belle flavor tagger in 2003-2010~Belle data~\cite{Bevan:2014iga}  taken with the second silicon-vertex detector configuration~(SVD2). There are no available Belle results for $\mu$. Statistical and systematical uncertainties are added in quadrature. All values are given in percent.}
 
 \vspace{5mm}

 \makebox[\textwidth]{\footnotesize\begin{tabular}{ l  r  r  r  r  r  r }
 &
 \multicolumn{3}{c}{$\varepsilon_i \pm \delta\varepsilon_{i}$} &
 \multicolumn{3}{c}{$w_i \pm \delta w_i\;\, $} \\\hline
$r$- Interval & \multicolumn{1}{c}{FBDT} & \multicolumn{1}{c}{DNN} & \multicolumn{1}{c}{Belle}  &
\multicolumn{1}{c}{FBDT} & \multicolumn{1}{c}{DNN} & \multicolumn{1}{c}{Belle} \\ \hline\hline
$ 0.000 - 0.100$ & $19.0  \pm 0.3$ & $14.3  \pm 0.3$ & $22.2 \pm 0.4$ & $47.1 \pm 1.7$ & $48.2 \pm 2.0$ & $50.0 \quad $   \\
$ 0.100 - 0.250$ & $17.1  \pm 0.3$ & $17.9  \pm 0.3$ & $14.5 \pm 0.3$ & $41.3 \pm 1.8$ & $43.6 \pm 1.7$ & $41.9 \pm 0.4$  \\
$ 0.250 - 0.500$ & $21.3  \pm 0.3$ & $22.5  \pm 0.4$ & $17.7 \pm 0.4$ & $30.3 \pm 1.5$ & $33.9 \pm 1.5$ & $31.9 \pm 0.3$  \\
$ 0.500 - 0.625$ & $11.3  \pm 0.3$ & $11.0  \pm 0.3$ & $11.5 \pm 0.3$ & $22.9 \pm 2.0$ & $19.3 \pm 2.0$ & $22.3 \pm 0.4$  \\
$ 0.625 - 0.750$ & $10.7  \pm 0.3$ & $10.4  \pm 0.3$ & $10.2 \pm 0.3$ & $12.4 \pm 1.9$ & $19.7 \pm 2.0$ & $16.3 \pm 0.4$  \\
$ 0.750 - 0.875$ & $ 8.2  \pm 0.2$ & $ 9.6  \pm 0.2$ & $ 8.7 \pm 0.3$ & $ 9.4 \pm 2.0$ & $10.8 \pm 1.9$ & $10.4 \pm 0.4$  \\
$ 0.875 - 1.000$ & $12.4  \pm 0.2$ & $14.2  \pm 0.3$ & $15.3 \pm 0.3$ & $ 2.3 \pm 1.4$ & $ 3.5 \pm 1.4$ & $ 2.5 \pm 0.3$  \\
\hline\hline
 \end{tabular}}
 
  \vspace{5mm}
  
  \makebox[\textwidth]{\footnotesize\begin{tabular}{ l  r  r  r}
 &
 \multicolumn{3}{c}{$\varepsilon_{\text{eff}, i} \pm \delta\varepsilon_{\text{eff}, i}\enskip\,$}\\\hline
$r$- Interval & 
\multicolumn{1}{c}{FBDT} & \multicolumn{1}{c}{DNN} & \multicolumn{1}{c}{Belle}\\ \hline\hline
$ 0.000 - 0.100$ &  $ 0.1 \pm 0.1$ & $ 0.0 \pm 0.1$ & $ 0.0 \quad$ \\
$ 0.100 - 0.250$ &  $ 0.5 \pm 0.2$ & $ 0.3 \pm 0.1$ & $ 0.4 \pm 0.1$ \\
$ 0.250 - 0.500$ &  $ 3.3 \pm 0.5$ & $ 2.3 \pm 0.4$ & $ 2.3 \pm 0.1$ \\
$ 0.500 - 0.625$ &  $ 3.3 \pm 0.5$ & $ 4.2 \pm 0.5$ & $ 3.5 \pm 0.1$ \\
$ 0.625 - 0.750$ &  $ 6.1 \pm 0.6$ & $ 3.8 \pm 0.5$ & $ 4.6 \pm 0.2$ \\
$ 0.750 - 0.875$ &  $ 5.4 \pm 0.5$ & $ 5.9 \pm 0.6$ & $ 5.5 \pm 0.1$ \\
$ 0.875 - 1.000$ &  $11.3 \pm 0.6$ & $12.3 \pm 0.7$ & $13.8 \pm 0.3$ \\
\hline\hline
\multicolumn{1}{c}{Total} & $30.0 \pm 1.3$ & $28.8 \pm 1.3$  & $30.1\pm 0.4$\\
 \hline
 \end{tabular}}
 
  \vspace{5mm} 
  
 \makebox[\textwidth]{\footnotesize\begin{tabular}{ l  r  r  c  r  r  r  }
 &
 \multicolumn{3}{c}{$\mu_i \pm \delta\mu_{i}$} &
 \multicolumn{3}{c}{$\Delta w_i \pm \delta \Delta w_i\;\, $} \\\hline
$r$- Interval & \multicolumn{1}{c}{FBDT} & \multicolumn{1}{c}{DNN} & \multicolumn{1}{c}{Belle}  &
\multicolumn{1}{c}{FBDT} & \multicolumn{1}{c}{DNN} & \multicolumn{1}{c}{Belle} \\ \hline\hline
$ 0.000 - 0.100$ &  $4.4 \pm 3.4$ &  $3.7 \pm 3.9$  & - & $ 8.8 \pm 2.1 $ & -$1.1 \pm 2.4$ &  $0.0 \quad$\\   
$ 0.100 - 0.250$ &  $3.9 \pm 3.4$ &  $7.3 \pm 3.4$  & - & $ 6.1 \pm 2.2 $ & $ 5.6 \pm 2.2$ & -$0.9 \pm 0.4$\\
$ 0.250 - 0.500$ &  $6.8 \pm 3.0$ &  $4.6 \pm 3.0$  & - & $ 2.7 \pm 2.0 $ & $ 7.1 \pm 2.0$ &  $1.0 \pm 0.4$\\
$ 0.500 - 0.625$ &  $3.2 \pm 4.1$ &  $2.2 \pm 4.2$  & - & $ 5.5 \pm 2.8 $ & $ 4.5 \pm 2.9$ & -$1.1 \pm 0.4$\\
$ 0.625 - 0.750$ & -$0.5 \pm 4.2$ &  $7.4 \pm 4.2$  & - & $ 0.7 \pm 3.0 $ & $ 7.9 \pm 2.9$ & -$1.9 \pm 0.5$\\
$ 0.750 - 0.875$ & $10.8 \pm 4.4$ &  $1.5 \pm 4.2$  & - & $ 7.7 \pm 3.3 $ & $ 5.7 \pm 3.1$ &  $1.7 \pm 0.4$\\
$ 0.875 - 1.000$ & -$3.7 \pm 3.4$ & -$2.5 \pm 4.2$  & - & $ 0.6 \pm 2.5 $ & $ 3.4 \pm 2.5$ & -$0.4 \pm 0.2$\\
\hline\hline

\vspace{6cm}

 \end{tabular}}
 
 \end{table*}

 }

%\clearpage

\clearpage

\section{Summary}

\label{sec:summary}
We report on the performance of two new Belle~II~$\PB$-flavor tagging algorithms on Belle~II data collected at the \PUpsilonFourS~resonance between 2019 and 2020. 
The algorithms exploit modern machine-learning 
techniques to determine the quark-flavor content of neutral \PB~mesons from the kinematic, track-hit, and particle-identification information associated with the reconstructed decay products. 
We validate the algorithms in simulation and in data using samples containing one fully reconstructed signal \PB~decay. We reconstruct abundant signal \PB~decays to flavor-specific hadronic final states and then use the remaining tracks and neutral clusters in each event as input for the flavor taggers. 

We use the $\Delta E$ distribution of the fully reconstructed \PB~candidates, restricted in $M_{\rm bc}$, to identify the \PB~signals and measure the tagging efficiencies, fractions of wrongly tagged events and related asymmetries from the flavor evolution of the signal $\PB\APB$~pairs in a time-integrated way.  Using a category-based flavor tagging algorithm, we obtain for neutral $\PB$~candidates the total effective efficiency
\begin{center}
\mbox{$\varepsilon_{\rm eff} = \big(30.0 \pm 1.2(\text{stat}) \pm 0.4(\text{syst})\big)\% $}, 
\end{center}
and for charged $\PB$~candidates 
\begin{center}
\mbox{$\varepsilon_{\rm eff} = \big(37.0 \pm 0.6(\text{stat}) \pm 0.2(\text{syst})\big)\% $}.  
\end{center}
Using a deep-learning-based flavor tagging algorithm, we obtain for neutral $\PB$~candidates the total effective efficiency
\begin{center}
\mbox{$\varepsilon_{\rm eff} = \big(28.8 \pm 1.2(\text{stat}) \pm 0.4(\text{syst})\big)\% $},
\end{center}
and for charged $\PB$~candidates
\begin{center}
\mbox{$\varepsilon_{\rm eff} = \big(39.9 \pm 0.6(\text{stat}) \pm 0.2(\text{syst})\big)\% $}. 
\end{center}

The performance of the flavor taggers is generally compatible with expectations from simulation and is comparable with the best performance obtained by the Belle experiment within the uncertainties. 
While both flavor taggers perform equally good in simulation, the deep-learning-based algorithm performs slightly worse than the category-based one in data. This is most likely due to current discrepancies between data and simulation since deep-learning methods heavily rely on a good description of the dependences among input variables. Thus we expect improvements in the future.

This work marks a milestone for future calibrations, which will play an essential role in measurements of \CP asymmetries at Belle~II and ultimately in the search for deviations from the Standard Model expectations.

\begin{acknowledgements}
%\input acknowledgements.tex
We thank the SuperKEKB group for the excellent operation of the
accelerator; the KEK cryogenics group for the efficient
operation of the solenoid; the KEK computer group for
on-site computing support; and the raw-data centers at
BNL, DESY, GridKa, IN2P3, and INFN for off-site computing support.
This work was supported by the following funding sources:
%Armenia
Science Committee of the Republic of Armenia Grant No. 20TTCG-1C010;
%Australia
Australian Research Council and research grant Nos.
DP180102629, 
DP170102389, 
DP170102204, 
DP150103061, 
FT130100303, 
FT130100018,
and
FT120100745;
%Austria
Austrian Federal Ministry of Education, Science and Research,
Austrian Science Fund No. P 31361-N36, and
Horizon 2020 ERC Starting Grant no. 947006 ``InterLeptons''; 
%Canada
Natural Sciences and Engineering Research
Council of Canada, Compute Canada and CANARIE;
%China
Chinese Academy of Sciences and research grant No. QYZDJ-SSW-SLH011,
National Natural Science Foundation of China and research grant Nos.
11521505,
11575017,
11675166,
11761141009,
11705209,
and
11975076,
LiaoNing Revitalization Talents Program under contract No. XLYC1807135,
Shanghai Municipal Science and Technology Committee under contract No. 19ZR1403000,
Shanghai Pujiang Program under Grant No. 18PJ1401000,
and the CAS Center for Excellence in Particle Physics (CCEPP);
%Czech Republic
the Ministry of Education, Youth and Sports of the Czech Republic under Contract 
No.~LTT17020 and 
Charles University grants SVV 260448 and GAUK 404316;
%EU
European Research Council, 7th Framework PIEF-GA-2013-622527, 
Horizon 2020 ERC Advanced Grants No. 267104 and 884719,
Horizon 2020 ERC Consolidator Grant No. 819127,
Horizon 2020 Marie Sklodowska-Curie grant agreement No. 700525 `NIOBE,' 
and
Horizon 2020 Marie Sklodowska-Curie RISE project JENNIFER2 grant agreement No. 822070 (European grants);
%France
L'Institut National de Physique Nucl\'{e}aire et de Physique des Particules (IN2P3) du CNRS (France);
%Germany
BMBF, DFG, HGF, MPG, and AvH Foundation (Germany);
%India
Department of Atomic Energy under Project Identification No. RTI 4002 and Department of Science and Technology (India);
%Israel
Israel Science Foundation grant No. 2476/17,
United States-Israel Binational Science Foundation grant No. 2016113, and
Israel Ministry of Science grant No. 3-16543;
%Italy
Istituto Nazionale di Fisica Nucleare and the research grants BELLE2;
%Japan
Japan Society for the Promotion of Science,  Grant-in-Aid for Scientific Research grant Nos.
16H03968, 
16H03993, 
16H06492,
16K05323, 
17H01133, 
17H05405, 
18K03621, 
18H03710, 
18H05226,
19H00682, % Niigata
26220706,
and
26400255,
the National Institute of Informatics, and Science Information NETwork 5 (SINET5), 
and
the Ministry of Education, Culture, Sports, Science, and Technology (MEXT) of Japan;  
%Korea
National Research Foundation (NRF) of Korea Grant Nos.
2016R1\-D1A1B\-01010135,
2016R1\-D1A1B\-02012900,
2018R1\-A2B\-3003643,
2018R1\-A6A1A\-06024970,
2018R1\-D1A1B\-07047294,
2019K1\-A3A7A\-09033840,
and
2019R1\-I1A3A\-01058933,
Radiation Science Research Institute,
Foreign Large-size Research Facility Application Supporting project,
the Global Science Experimental Data Hub Center of the Korea Institute of Science and Technology Information
and
KREONET/GLORIAD;
%Malaysia
Universiti Malaya RU grant, Akademi Sains Malaysia and Ministry of Education Malaysia;
%Mexico
% CINVESTAV-IPN, UNAM, UAS, BUAP and CONACYT are funded under
Frontiers of Science Program contracts
FOINS-296,
CB-221329,
CB-236394,
CB-254409,
and
CB-180023, and SEP-CINVESTAV research grant 237 (Mexico);
%Poland
the Polish Ministry of Science and Higher Education and the National Science Center;
%Russia
the Ministry of Science and Higher Education of the Russian Federation,
Agreement 14.W03.31.0026, and
the HSE University Basic Research Program, Moscow;
%Saudi Arabia
University of Tabuk research grants
S-0256-1438 and S-0280-1439 (Saudi Arabia);
%Slovenia
Slovenian Research Agency and research grant Nos.
J1-9124
and
P1-0135; 
%Spain
Agencia Estatal de Investigacion, Spain grant Nos.
FPA2014-55613-P
and
FPA2017-84445-P,
and
CIDEGENT/2018/020 of Generalitat Valenciana;
%Taiwan
Ministry of Science and Technology and research grant Nos.
MOST106-2112-M-002-005-MY3
and
MOST107-2119-M-002-035-MY3, 
and the Ministry of Education (Taiwan);
%Thailand
Thailand Center of Excellence in Physics;
%Turkey
TUBITAK ULAKBIM (Turkey);
%Ukraine
Ministry of Education and Science of Ukraine;
%USA
the US National Science Foundation and research grant Nos.
PHY-1807007 % Luther
and
PHY-1913789, % Indiana CEEM
and the US Department of Energy and research grant Nos.
DE-AC06-76RLO1830, % PNNL
DE-SC0007983, % Wayne State
DE-SC0009824, % Florida
DE-SC0009973, % VPI
DE-SC0010007, % Frank
DE-SC0010073, % South Carolina
DE-SC0010118, % Carnegie Mellon
DE-SC0010504, % Hawaii
DE-SC0011784, % Cincinnati
DE-SC0012704, % BNL
DE-SC0021274; % Mississippi
%last group
and
%Vietnam
the Vietnam Acad. of Sci. and Technology (VAST) under grant DL0000.05/21-23.

\end{acknowledgements}


\bibliography{ftbibfile}

% % Non-BibTeX users please use
% \begin{thebibliography}{}
% %
% % and use \bibitem to create references. Consult the Instructions
% % for authors for reference list style.
% % Format for Journal Reference
% Author, Article title, Journal, Volume, page numbers (year)
% Author, Book title, page numbers. Publisher, place (year)
% \end{thebibliography}

\end{document}